\documentclass[ALICE,manyauthors]{cernphprep}
\usepackage[comma,square,numbers,sort&compress]{natbib}
\usepackage{hyperref}
\usepackage{lineno}
\usepackage{xspace}
\usepackage{color}
\usepackage{amssymb}
\usepackage{float}
\usepackage[T1]{fontenc}

\begin{document}
%

\newcommand{\pp}           {pp\xspace}
\newcommand{\ppbar}        {\mbox{$\mathrm {p\overline{p}}$}\xspace}
\newcommand{\XeXe}         {\mbox{Xe--Xe}\xspace}
\newcommand{\PbPb}         {\mbox{Pb--Pb}\xspace}
\newcommand{\pA}           {\mbox{pA}\xspace}
\newcommand{\pPb}          {\mbox{p--Pb}\xspace}
\newcommand{\AuAu}         {\mbox{Au--Au}\xspace}
\newcommand{\dAu}          {\mbox{d--Au}\xspace}

\newcommand{\s}            {\ensuremath{\sqrt{s}}\xspace}
\newcommand{\snn}          {\ensuremath{\sqrt{s_{\mathrm{NN}}}}\xspace}
\newcommand{\pt}           {\ensuremath{p_{\rm T}}\xspace}
\newcommand{\meanpt}       {$\langle p_{\mathrm{T}}\rangle$\xspace}
\newcommand{\ycms}         {\ensuremath{y_{\rm CMS}}\xspace}
\newcommand{\ylab}         {\ensuremath{y_{\rm lab}}\xspace}
\newcommand{\etarange}[1]  {\mbox{$\left | \eta \right |~<~#1$}}
\newcommand{\yrange}[1]    {\mbox{$\left | y \right |~<~#1$}}
\newcommand{\dndy}         {\ensuremath{\mathrm{d}N_\mathrm{ch}/\mathrm{d}y}\xspace}
\newcommand{\dndeta}       {\ensuremath{\mathrm{d}N_\mathrm{ch}/\mathrm{d}\eta}\xspace}
\newcommand{\avdndeta}     {\ensuremath{\langle\dndeta\rangle}\xspace}
\newcommand{\dNdy}         {\ensuremath{\mathrm{d}N_\mathrm{ch}/\mathrm{d}y}\xspace}
\newcommand{\Npart}        {\ensuremath{N_\mathrm{part}}\xspace}
\newcommand{\Ncoll}        {\ensuremath{N_\mathrm{coll}}\xspace}
\newcommand{\dEdx}         {\ensuremath{\textrm{d}E/\textrm{d}x}\xspace}
\newcommand{\RpPb}         {\ensuremath{R_{\rm pPb}}\xspace}

\newcommand{\nineH}        {$\sqrt{s}~=~0.9$~Te\kern-.1emV\xspace}
\newcommand{\seven}        {$\sqrt{s}~=~7$~Te\kern-.1emV\xspace}
\newcommand{\twoH}         {$\sqrt{s}~=~0.2$~Te\kern-.1emV\xspace}
\newcommand{\twosevensix}  {$\sqrt{s}~=~2.76$~Te\kern-.1emV\xspace}
\newcommand{\five}         {$\sqrt{s}~=~5.02$~Te\kern-.1emV\xspace}
\newcommand{\twosevensixnn}{$\sqrt{s_{\mathrm{NN}}}~=~2.76$~Te\kern-.1emV\xspace}
\newcommand{\fivenn}       {$\sqrt{s_{\mathrm{NN}}}~=~5.02$~Te\kern-.1emV\xspace}
\newcommand{\LT}           {L{\'e}vy-Tsallis\xspace}
\newcommand{\GeVc}         {Ge\kern-.1emV/$c$\xspace}
\newcommand{\MeVc}         {Me\kern-.1emV/$c$\xspace}
\newcommand{\TeV}          {Te\kern-.1emV\xspace}
\newcommand{\GeV}          {Ge\kern-.1emV\xspace}
\newcommand{\MeV}          {Me\kern-.1emV\xspace}
\newcommand{\GeVmass}      {Ge\kern-.2emV/$c^2$\xspace}
\newcommand{\MeVmass}      {Me\kern-.2emV/$c^2$\xspace}
\newcommand{\lumi}         {\ensuremath{\mathcal{L}}\xspace}

\newcommand{\ITS}          {\rm{ITS}\xspace}
\newcommand{\TOF}          {\rm{TOF}\xspace}
\newcommand{\ZDC}          {\rm{ZDC}\xspace}
\newcommand{\ZDCs}         {\rm{ZDCs}\xspace}
\newcommand{\ZNA}          {\rm{ZNA}\xspace}
\newcommand{\ZNC}          {\rm{ZNC}\xspace}
\newcommand{\SPD}          {\rm{SPD}\xspace}
\newcommand{\SDD}          {\rm{SDD}\xspace}
\newcommand{\SSD}          {\rm{SSD}\xspace}
\newcommand{\TPC}          {\rm{TPC}\xspace}
\newcommand{\TRD}          {\rm{TRD}\xspace}
\newcommand{\VZERO}        {\rm{V0}\xspace}
\newcommand{\VZEROA}       {\rm{V0A}\xspace}
\newcommand{\VZEROC}       {\rm{V0C}\xspace}
\newcommand{\Vdecay} 	   {\ensuremath{V^{0}}\xspace}

\newcommand{\ee}           {\ensuremath{e^{+}e^{-}}} 
\newcommand{\pip}          {\ensuremath{\pi^{+}}\xspace}
\newcommand{\pim}          {\ensuremath{\pi^{-}}\xspace}
\newcommand{\kap}          {\ensuremath{\rm{K}^{+}}\xspace}
\newcommand{\kam}          {\ensuremath{\rm{K}^{-}}\xspace}
\newcommand{\pbar}         {\ensuremath{\rm\overline{p}}\xspace}
\newcommand{\kzero}        {\ensuremath{{\rm K}^{0}_{\rm{S}}}\xspace}
\newcommand{\lmb}          {\ensuremath{\Lambda}\xspace}
\newcommand{\almb}         {\ensuremath{\overline{\Lambda}}\xspace}
\newcommand{\Om}           {\ensuremath{\Omega^-}\xspace}
\newcommand{\Mo}           {\ensuremath{\overline{\Omega}^+}\xspace}
\newcommand{\X}            {\ensuremath{\Xi^-}\xspace}
\newcommand{\Ix}           {\ensuremath{\overline{\Xi}^+}\xspace}
\newcommand{\Xis}          {\ensuremath{\Xi^{\pm}}\xspace}
\newcommand{\Oms}          {\ensuremath{\Omega^{\pm}}\xspace}
\newcommand{\degree}       {\ensuremath{^{\rm o}}\xspace}

\newcommand{\sqrtsnn} [1]  {\ensuremath{\sqrt{s_{\mathrm{NN}}} = #1}\xspace}
\newcommand {\y}         {\ensuremath{y} }
\newcommand {\sqrtSnn}   {\ensuremath{\sqrt{s_{_{\mathrm{NN}}}}}}
\newcommand {\sqrtSnnE}[2][TeV]  {$\sqrtSnn = #2\,\mathrm{#1}$}
\newcommand {\sqrtS}     {\ensuremath{\sqrt{s}}\,}
\newcommand {\sqrtSE}[2][TeV]  {$\sqrtS = #2\,\mathrm{#1}$}
\newcommand {\mumu}      {\ensuremath{\mu^+\mu^-}}
\newcommand {\mumuMB}    {\ensuremath{\mu \mu \mathrm{MB}}}
\newcommand {\Rpa}       {\ensuremath{R_\mathrm{pA}}}
\newcommand {\Rap}       {\ensuremath{R_\mathrm{Ap}} }
\newcommand {\Taa}       {\ensuremath{\langle T_\mathrm{AA} \rangle}}
\newcommand {\jpsi}      {\ensuremath{\mathrm{J}\kern-0.02em/\kern-0.05em\psi} }
\newcommand {\psip}      {\ensuremath{\psi\mathrm{(2S)}} }
\newcommand {\chic}      {\ensuremath{\chi_{\mathrm{c}}}}
\newcommand {\gev}       {\ensuremath{\,\mathrm{GeV}}}
\newcommand {\tev}       {\ensuremath{\,\mathrm{TeV}}}
\newcommand {\gevc}      {\ensuremath{\,\mathrm{GeV}\kern-0.05em/\kern-0.02em c}}
\newcommand {\gevcc}     {\ensuremath{\,\mathrm{GeV}\kern-0.05em/\kern-0.02em c^2}}
\newcommand{\RAA}{$R_{\mathrm{AA}}$}
\newcommand{\sqrts}{$\sqrt{s_{\mathrm{NN}}} = 8.16$ TeV}

\begin{titlepage}
\PHyear{2020}       
\PHnumber{133}      
\PHdate{14 July}  

\title{Centrality dependence of \jpsi and \psip production and nuclear modification in p--Pb collisions at \sqrtsnn{8.16}  TeV }
\ShortTitle{Centrality dependence of \jpsi and \psip production in p--Pb collisions}   

\Collaboration{ALICE Collaboration\thanks{See Appendix~\ref{app:collab} for the list of collaboration members}}
\ShortAuthor{ALICE Collaboration} 

\begin{abstract}
The inclusive production of the J/$\psi$ and $\psi$(2S) charmonium states is studied as a function of centrality in p--Pb collisions at a centre-of-mass energy per nucleon pair $\sqrt{s_{\rm NN}} = 8.16$ TeV at the LHC. The measurement is performed in the dimuon decay channel with the ALICE apparatus in the centre-of-mass rapidity intervals $-4.46 < y_{\rm cms} < -2.96$  (Pb-going direction) and $2.03 < y_{\rm cms} < 3.53$ (p-going direction), down to zero transverse momentum ($p_{\rm T}$). The J/$\psi$ and $\psi$(2S) production cross sections are evaluated as a function of the collision centrality, estimated through the energy deposited in the zero degree calorimeter located in the Pb-going direction. The $p_{\rm T}$-differential J/$\psi$ production cross section is measured at backward and forward rapidity for several centrality classes, together with the corresponding average $\langle p_{\rm T} \rangle$ and $\langle p^{2}_{\rm T} \rangle$ values. The nuclear effects affecting the production of both charmonium states are studied using the nuclear modification factor. In the p-going direction, a suppression of the production of both charmonium states is observed, which seems to increase from peripheral to central collisions. In the Pb-going direction, however, the centrality dependence is different for the two states: the nuclear modification factor of the J/$\psi$ increases from below unity in peripheral collisions to above unity in central collisions, while for the $\psi$(2S) it stays below or consistent with unity for all centralities with no significant centrality dependence. The results are compared with measurements in p--Pb collisions at  $\sqrt{s_{\rm NN}} = 5.02$ TeV and no significant dependence on the energy of the collision is observed. Finally, the results are compared with theoretical models implementing various nuclear matter effects.
\end{abstract}
\end{titlepage}

\setcounter{page}{2} 


\section{Introduction}
\label{Introduction}

Quarkonia, bound states of a heavy quark and its antiquark, are prominent probes of the properties of the strong interaction, which is described by quantum chromodynamics (QCD). 
In high-energy hadronic collisions, the production of quarkonia is usually factorised in a two-step process: the creation of a  heavy-quark pair, mainly by gluon fusion at LHC energies, followed by its evolution and binding into a colour-singlet state. 
The former is described using perturbative QCD calculations, while the latter involves non-perturbative processes and is described using effective models~\cite{Bodwin:1994jh,Paul:2014eka,Fritzsch:1977ay}. 

In high-energy heavy-ion collisions, the creation of a deconfined state of nuclear matter made of quarks and gluons, the so-called quark--gluon plasma (QGP), modifies the production rates of the various quarkonium states. 
On the one hand, the production of quarkonium states is expected to be suppressed by the large density of colour charges in the QGP~\cite{Matsui:1986dk}, with the suppression increasing with decreasing binding energy of the resonance~\cite{Digal:2001ue}. 
Such sequential suppression has been observed, most notably in the bottomonium (${\rm b\overline{b}}$) sector in Pb--Pb collisions at the LHC by the CMS~\cite{Chatrchyan:2012lxa,Khachatryan:2016xxp,Sirunyan:2017lzi,Sirunyan:2018nsz} and ALICE~\cite{Acharya:2018mni} collaborations.
On the other hand, quarkonia could also be regenerated during the QGP phase~\cite{Thews:2000rj} or at its late boundary~\cite{BraunMunzinger:2000px} by recombination of deconfined heavy quarks. 
Strong indications supporting such a regeneration mechanism, which (partially) compensates the aforementioned suppression, have been reported by the ALICE Collaboration in the charmonium (${\rm c\overline{c}}$) sector for the J/$\psi$ in Pb--Pb collisions at the LHC~\cite{Abelev:2012rv,Abelev:2013ila,Adam:2015isa,Adam:2016rdg,Acharya:2019iur}. 

However, to fully exploit those experimental results for the understanding of the inner-workings of the QGP, other nuclear effects, not related to the presence of the QGP, must be addressed. 
These are typically referred to as cold nuclear matter (CNM) effects, as opposed to those related to the hot medium, and include the effects described below. 
A significant contribution involves the nuclear modification of the parton distribution function (PDF) of the nucleons inside the nucleus~\cite{Eskola:2009uj}, i.e. the modification of the probability for a parton (quark or gluon) to carry a fraction $x$ of the momentum of the nucleon.
The gluon nuclear parton distribution function (nPDF) includes, most notably, a shadowing region at low $x$ ($x \lesssim 0.01$) corresponding to a suppression of gluons and an antishadowing region at intermediate $x$ ($0.01 \lesssim x \lesssim 0.3$) corresponding to an enhancement of gluons~\cite{Eskola:2009uj}.
The modification of the initial state of the nucleus with respect to an incoherent superposition of free nucleons can also be described in terms of the saturation of low-$x$ gluons as implemented in the Colour Glass Condensate (CGC) effective field theory~\cite{Fujii:2013gxa}.
In addition, coherent energy-loss effects involving the initial- and final-state partons can modify the production of heavy-quark pairs and thus of quarkonium states~\cite{Arleo:2014oha}.
The pre-resonant quarkonium state could also interact with the surrounding spectator nucleons. 
This nuclear absorption is expected to be negligible at LHC energies due to the short crossing time of the colliding nuclei~\cite{Ferreiro:2011xy}.
The CNM effects discussed above are expected to affect similarly all states of the same quarkonium family, as they act on the production cross section of heavy-quark pairs or on the pre-resonant quarkonium state.
On the contrary, final-state interactions with the co-moving medium~\cite{Ferreiro:2014bia} or with a medium including a short-lived QGP and a hadron resonance gas~\cite{Du:2015wha} could affect differently the various states of the same family.
Soft-color exchanges between the hadronising ${\rm c\overline{c}}$ pair and long-lived co-moving partons~\cite{Ma:2017rsu} could also affect differently the various charmonium states.

Cold nuclear matter effects are typically investigated using proton--nucleus collisions, where the formation of the QGP is not expected.
At the LHC, the production of quarkonia was extensively studied in p--Pb collisions at a centre-of-mass energy per nucleon pair $\sqrt{s_{\rm NN}} = 5.02$~TeV by the ALICE~\cite{Abelev:2013yxa,Abelev:2014zpa,Abelev:2014oea,Adam:2015iga,Adam:2015jsa,Adam:2016ohd,Acharya:2018yud}, ATLAS~\cite{Aaboud:2017cif}, CMS~\cite{Sirunyan:2017mzd,Sirunyan:2018pse}, and LHCb~\cite{Aaij:2013zxa,Aaij:2014mza,Aaij:2016eyl} collaborations.
In the charmonium sector, a significant suppression of J/$\psi$ yields is observed at forward rapidity $y$, i.e. in the p-going direction, at low transverse momentum $p_{\rm T}$, with the effect vanishing with increasing $p_{\rm T}$. 
The suppression at midrapidity is compatible with the one at forward $y$, while at backward $y$, i.e.\ in the Pb-going direction, no suppression of the J/$\psi$ yields is observed~\cite{Abelev:2013yxa,Adam:2015iga,Acharya:2018yud}.
Interestingly, the $\psi$(2S) appears to be more suppressed than the J/$\psi$ at both forward and backward rapidity~\cite{Abelev:2014zpa}.
This observation cannot be explained by the first group of CNM effects discussed above and seems to indicate the need to consider additional final-state effects.
The centrality dependence of the J/$\psi$ and $\psi$(2S) suppression was also measured in p--Pb collisions at $\sqrt{s_{\rm NN}} = 5.02$~TeV~\cite{Adam:2015jsa,Adam:2016ohd}. 
The difference between the $\psi$(2S) and J/$\psi$ suppression increases with increasing centrality, especially at backward rapidity, indicating, once again, that shadowing or coherent parton energy-loss mechanisms are not enough to explain the $\psi$(2S) suppression~\cite{Adam:2016ohd}.
Complementarily, the ALICE Collaboration also studied the J/$\psi$ production at forward, mid, and backward rapidity as a function of the multiplicity of charged particles measured at midrapidity~\cite{Adamova:2017uhu}. 
Such study does not require the interpretation of the centrality classes in terms of collision geometry and allows for the investigation of rare events with the highest charged particle multiplicities.
An increase of the relative J/$\psi$ yields with the relative charged-particle multiplicity is observed. 
At forward rapidity the increase saturates towards the highest multiplicities, while at backward rapidity a hint of a faster-than-linear increase with multiplicity is seen.

Also, in high-multiplicity p--Pb events, long-range angular correlations between the J/$\psi$ at large rapidity and charged particles at midrapidity are observed~\cite{Acharya:2017tfn}.
These correlations are reminiscent of those observed in Pb--Pb collisions, which are often interpreted as signatures of the collective motion of the particles during the hydrodynamic evolution of the hot and dense medium.

More recently, the J/$\psi$ and $\psi$(2S) production cross sections were also measured in p--Pb collisions at $\sqrt{s_{\rm NN}} = 8.16$~TeV as a function of transverse momentum and rapidity~\cite{Acharya:2018kxc,Aaij:2017ypPb,Acharya:2020wwy} confirming, with better statistical precision, the earlier findings. 
Namely, a significant suppression of the J/$\psi$ is observed at forward rapidity but not at backward rapidity, and a stronger suppression of the $\psi$(2S) is seen, especially at backward rapidity.
The J/$\psi$ production at forward and backward rapidity as a function of the multiplicity of charged particles measured at midrapidity was also studied at $\sqrt{s_{\rm NN}} = 8.16$~TeV~\cite{Acharya:2020giw} confirming the earlier observations.

In the bottomonium sector, a significant suppression of the $\Upsilon$(1S) yield is observed at mid and forward rapidity, vanishing from low to high transverse momentum, while at backward rapidity the yields are consistent with the expectations from pp collisions~\cite{Acharya:2019lqc,Abelev:2014oea,Aaboud:2017cif,Aaij:2014mza,Aaij:2018scz}.
Interestingly, the excited $\Upsilon$(2S) state at midrapidity~\cite{Chatrchyan:2013nza,Aaboud:2017cif} and $\Upsilon$(3S) state at backward rapidity~\cite{Aaij:2018scz} appear to be more suppressed than the fundamental $\Upsilon$(1S) state, which is similar to the comparison of the $\psi$(2S) and J/$\psi$ discussed above.

This paper presents the centrality dependence of the production of inclusive J/$\psi$ and $\psi$(2S) in p--Pb collisions at $\sqrt{s_{\rm NN}} = 8.16$~TeV. 
The inclusive $\psi$(nS) production contains contributions from direct $\psi$(nS), from decays of higher-mass excited states in the case of the J/$\psi$  (mainly $\psi$(2S) and $\chi_{\rm c}$), as well as from non-prompt $\psi$(nS), from weak decays of beauty hadrons.
Section~\ref{sec:exp} briefly presents the experimental setup and event selection, Section~\ref{DataAnalysis} describes the data analysis procedure, while the results are presented and discussed in Section~\ref{Results}. A summary is given in Section~\ref{Conclusions}.

\section{Experimental apparatus and event selection}
\label{sec:exp}
A detailed description of the ALICE apparatus and its performance can be found in Refs.~\cite{Aamodt:2008zz,Abelev:2014ffa}.
The main detectors used in this analysis are briefly discussed below.

The ALICE muon spectrometer is used to detect muons in the pseudorapidity region $-4<\eta_{\mathrm{lab}}<-2.5$. It includes five tracking stations each having two planes of cathode pad chambers, with the third station being placed inside a dipole magnet with a field integral of 3 $\mathrm{T \cdot m}$. 
Two trigger stations, each composed of two planes of resistive plate chambers, provide the trigger for single muon as well as dimuon events with a programmable single-muon \pt threshold. The setup is completed by a set of absorbers. 
A front absorber made of carbon, concrete, and steel is placed between the nominal interaction point (IP) and the first tracking station, to remove hadrons coming from the interaction vertex. 
An iron filter is positioned between the tracking and trigger stations and absorbs the remaining hadrons escaping the front absorber and the low \pt muons originating from the decay of pions and kaons. 
Finally, a conical absorber surrounding the beam pipe protects the muon spectrometer against secondary particles produced by the primary particles emerging at large pseudorapidities and interacting with the beam pipe. 

The Silicon Pixel Detector (SPD), corresponding to the two innermost layers of the Inner Tracking System~\cite{Aamodt:2010aa} and covering the pseudorapidity ranges $|\eta_{\rm lab}| < 2$ (first layer) and $|\eta_{\rm lab}| < 1.4$ (second layer), is used to reconstruct the primary vertex of the collision.
The two V0 hodoscopes~\cite{Abbas:2013taa} have 32 scintillator tiles each, are placed on each side of the IP, and cover the pseudorapidity ranges $2.8 < \eta_{\rm lab} < 5.1$ and $-3.7 < \eta_{\rm lab} < -1.7$.
The coincidence of signals from the two hodoscopes defines the minimum bias (MB) trigger condition and a first luminosity signal during van der Meer scans~\cite{ALICE-PUBLIC-2018-002}.
The V0s are also used to remove beam-induced background.
A second luminosity signal in van der Meer scans is defined by the coincidence of signals from the two T0 arrays, which are located on opposite sides of the IP ($4.6 < \eta_{\rm lab} < 4.9$ and $-3.3 < \eta_{\rm lab} < -3.0$).
Each array consists of 12 quartz Cherenkov counters read out by photomultiplier tubes~\cite{Bondila:2005xy}.
Finally, two Zero Degree Calorimeters (ZDC)~\cite{ALICE:2012aa} are placed along the beam axis at $\pm 112.5$ m from the IP.
Each ZDC is composed of a neutron calorimeter (ZN), positioned between the two beam pipes downstream of the first machine dipole that separates the beams, and a proton calorimeter (ZP), installed externally to the outgoing beam pipe. 
The ZN are used to estimate the centrality of the collision (described in Section~\ref{DataAnalysis}) and to remove beam-induced background events.

The data were collected in 2016 with two beam configurations obtained by reverting the direction of the proton and lead ion beams. The corresponding acceptance ranges of the muon spectrometer, in terms of dimuon centre-of-mass rapidity, are $-4.46<y_{\mathrm{cms}}<-2.96$ and $2.03<y_{\mathrm{cms}}<3.53$. The backward and forward rapidity intervals correspond to the muon spectrometer being located in the Pb-going and p-going direction, and are denoted as Pb--p and p--Pb, respectively.

The non-symmetric rapidity ranges arise from the energy-per-nucleon asymmetry of the p and Pb beams, which shifts the rapidity of the nucleon--nucleon centre-of-mass system with respect to the laboratory system by 0.465 units of rapidity in the direction of the proton beam.
The events were collected using an opposite-sign dimuon trigger, which requires the coincidence of the MB trigger condition and two opposite-sign track segments in the muon trigger chambers.
For the data samples used here, the programmable online \pt threshold for each muon track was set to 0.5~GeV/$c$.
This threshold is not sharp in \pt and the single-muon trigger efficiency is about 50\% at $p^{\mu}_{\rm T} = 0.5$~GeV/$c$ and reaches a plateau value of about 96\% at $p^{\mu}_{\rm T} \simeq 1.5$ GeV/$c$.
Beam-induced background was removed using the timing information provided by the V0 and the ZDC. 
The events are classified in classes of centrality according to the energy deposited in the ZN located in the direction of the Pb beam, as will be discussed in Section~\ref{DataAnalysis}.
Events in which two or more interactions occur in the same colliding bunch (in-bunch pile-up) or during the readout time of the SPD (out-of-bunch pile-up) are removed using the information from the SPD and V0.
The integrated luminosity for the two beam configurations is $\textit{L}_{\mathrm{int}}=12.8\pm0.3\;\mathrm{nb^{-1}}$ for Pb--p and $\textit{L}_{\mathrm{int}}=8.4\pm0.2\;\mathrm{nb^{-1}}$ for p--Pb collisions.

\section{Data analysis}
\label{DataAnalysis}

In this section the various elements involved in the cross section and the nuclear modification factor measurements are discussed.

In p--Pb collisions, a centrality determination based on the charged-particle multiplicity can be biased by fluctuations related to the variation of the event topology, which are unrelated to the collision geometry.
In contrast, an event selection depending on the energy deposited in the ZDC by nucleons emitted in the nuclear de-excitation process after the collision or knocked out by the nucleons participating in the collision (participant or wounded nucleons) should not be affected by this kind of bias. In this analysis the centrality estimation is based on a hybrid method, as described in detail in Refs.~\cite{Adam:2014qja,ALICE-PUBLIC-2018-011}. In this approach, the centrality classes are determined using the ZN detector, while the average number of binary nucleon--nucleon collisions $\langle \Ncoll \rangle$ and the average nuclear overlap function $\langle T_{\mathrm{pPb}} \rangle$ for each centrality class are obtained assuming that the charged-particle multiplicity measured at midrapidity scales with the number of participant nucleons $\Npart=\Ncoll+1$. 
The centrality classes used in this analysis and the corresponding $\langle \Ncoll \rangle$ and $\langle T_{\mathrm{pPb}} \rangle$ as well as their uncertainties, which reflect possible remaining biases (as discussed in Refs.~\cite{Adam:2014qja,ALICE-PUBLIC-2018-011}), are shown in Table~\ref{tabZNncall}. Monte Carlo (MC) simulations reproducing the LHC running conditions indicate that a residual pile-up may be present in the 2\% most central collisions. The 0--2\% centrality interval is therefore excluded and a 2\% systematic uncertainty is conservatively assigned to the results in the other centrality classes.
Furthermore, the 90--100\% centrality interval is also excluded as the dimuon trigger may suffer from residual background contamination. It is worth noting that the previous analysis at $\sqrt{s_{\rm NN}} = 5.02$ TeV was performed in the wider 80--100\% centrality class where such possible contamination was not apparent. 

\begin{table}[!htb]
	\centering
	\caption{\label{tabZNncall} The average number of binary nucleon--nucleon collisions $\langle N_{\rm coll} \rangle$ and average nuclear overlap function $\langle T_{\rm pPb} \rangle$, along with their systematic uncertainty, for the used centrality classes.}
	\begin{tabular}{lccccc}
		\hline
		\hline
		ZN class &  $\langle N_{\rm coll} \rangle$ &  Total syst on $\langle N_{\rm coll} \rangle$ (\%) & $\langle T_{\rm pPb} \rangle$ & Total syst on $\langle T_{\rm pPb} \rangle$ (\%) \\ 
		\hline
		2--10\%   & 12.7 & 4.8 & 0.175 & 4.8\\ 
		10--20\%  & 11.5 & 3.1 & 0.159 & 3.3\\ 
		20--40\%  & 9.81 & 1.7 & 0.135 & 2.1\\ 
		40--60\%  & 7.09 & 4.1 & 0.0978& 4.2\\  
		60--80\%  & 4.28 & 4.6 & 0.0590& 4.8\\ 
		\hline
		20--30\%  & 10.4 & 1.8 & 0.143 & 2.2\\ 
		30--40\%  & 9.21 & 2.0 & 0.127 & 2.4\\ 
		40--50\%  & 7.82 & 3.4 & 0.108 & 3.7\\ 
		50--60\%  & 6.37 & 4.6 & 0.0879& 4.8\\ 
		60--70\%  & 4.93 & 5.1 & 0.0680& 5.3\\  
		70--80\%  & 3.63 & 4.4 & 0.0501& 4.6\\ 
		80--90\%  & 2.53 & 1.7 & 0.0349& 2.1\\ 
		\hline
		\hline
	\end{tabular}
\end{table}

Charmonium candidates are built by forming pairs of opposite-sign charged tracks that were reconstructed by the tracking chambers of the muon spectrometer satisfying the following criteria.
Each muon track candidate should be within $-4<\eta_{\rm lab}^{\mu}<-2.5$ to avoid the edges of the acceptance. 
The tracks crossing the thicker part of the absorber are removed with the condition that the radial transverse position of the muon track at the end of the front absorber must be in the range $17.6<R_{\mathrm{abs}}<89.5\;\mathrm{cm}$. 
The tracks must match a track segment in the muon trigger chambers above the aforementioned \pt threshold of 0.5 GeV/$c$. 
The rapidity of the muon pair should be within the fiducial acceptance of the muon spectrometer, namely $2.03 < y_{\mathrm{cms}} < 3.53$ and $-4.46 < y_{\mathrm{cms}} < -2.96$, for the p--Pb and Pb--p data samples, respectively. 

\begin{figure}[!b]
	\includegraphics[width=0.5\columnwidth]{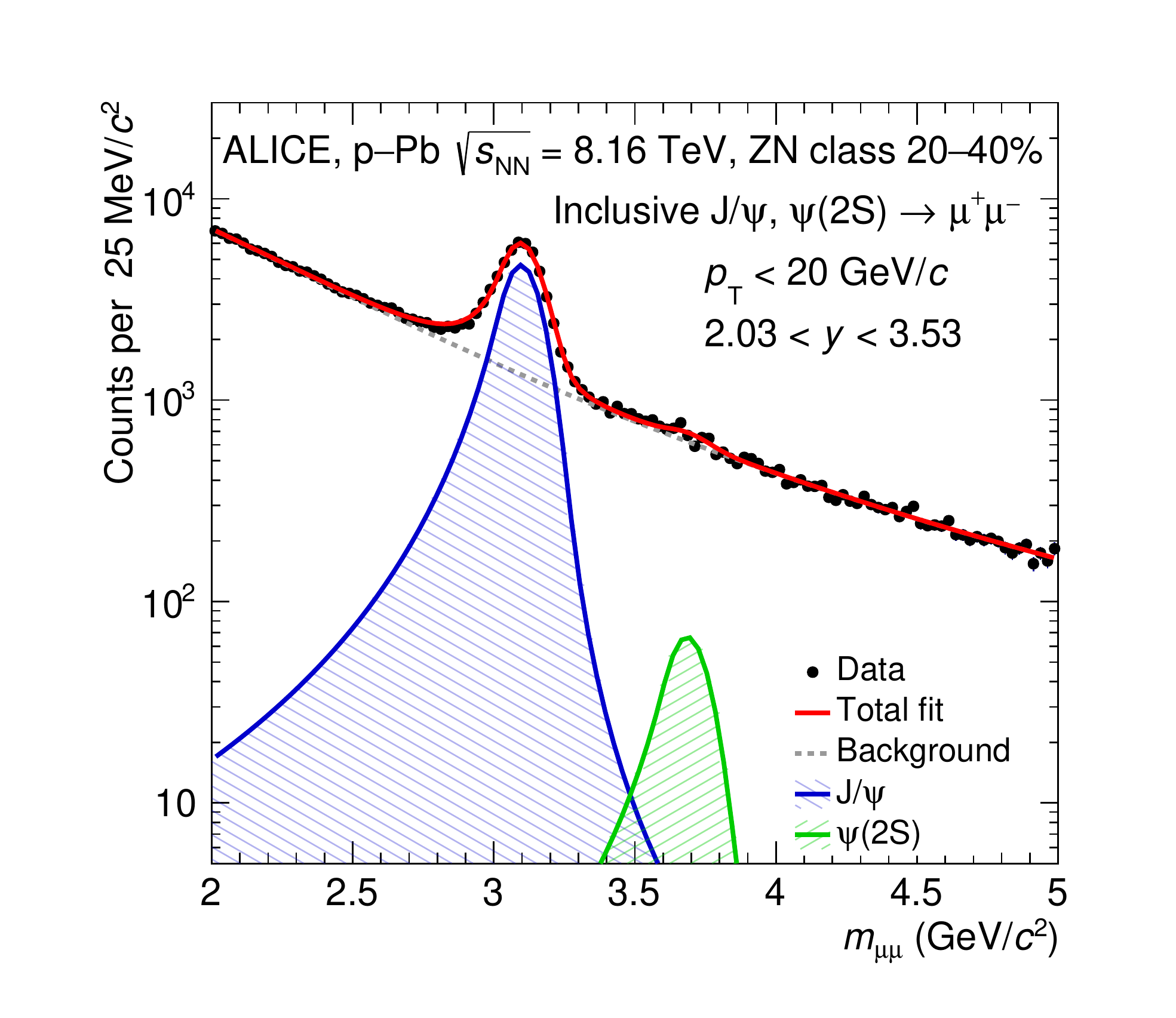}
	\includegraphics[width=0.5\columnwidth]{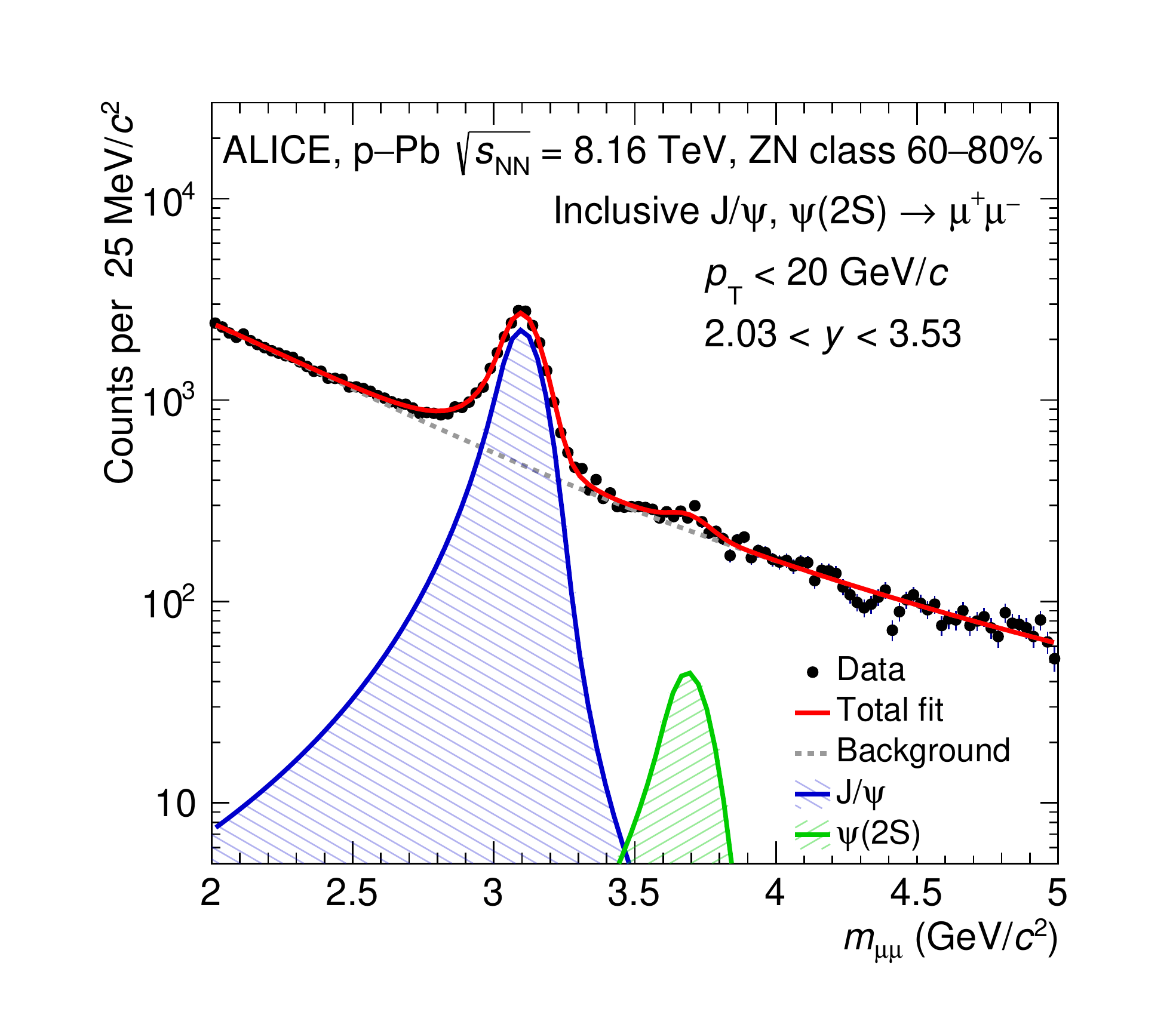}
	\includegraphics[width=0.5\columnwidth]{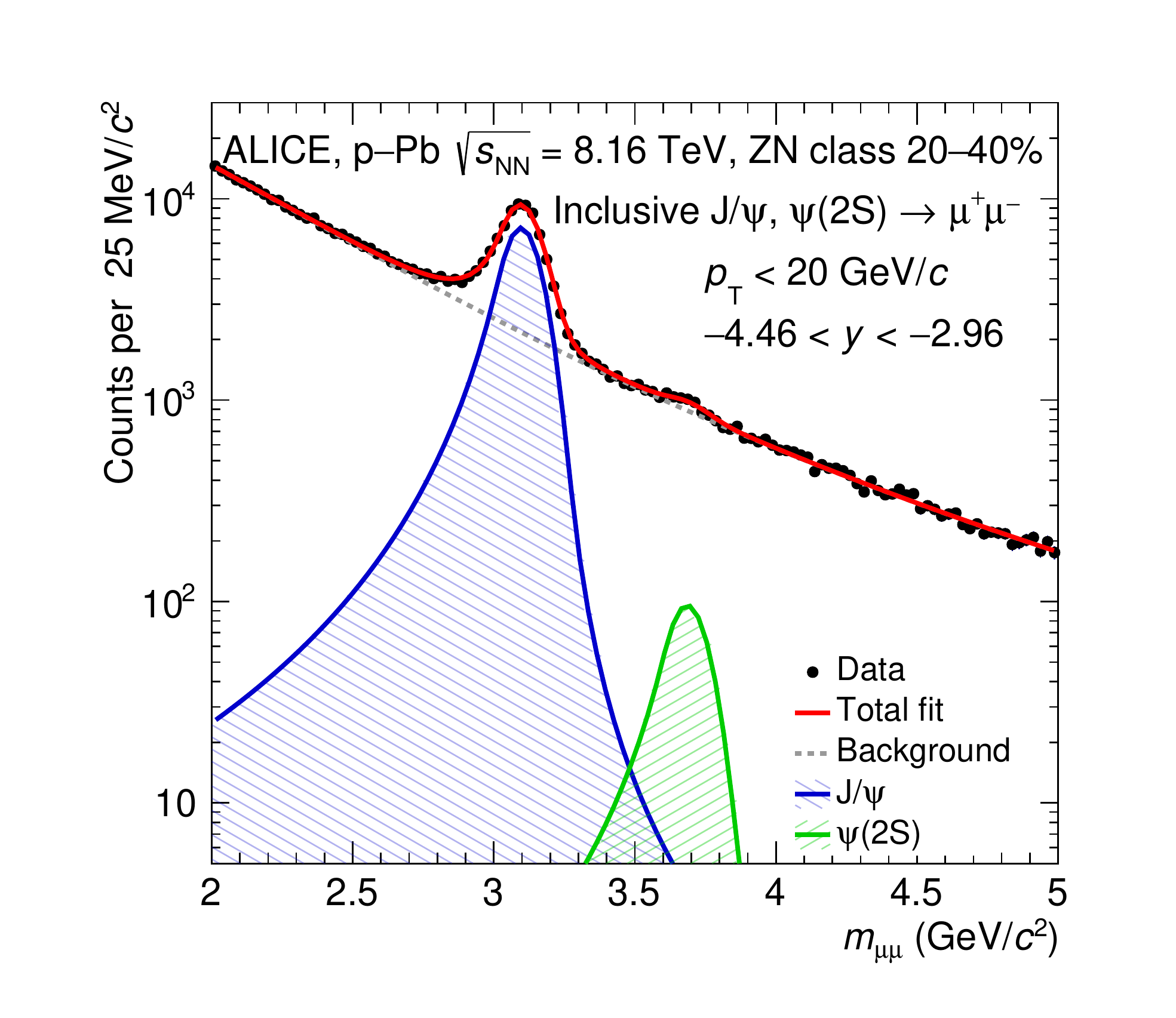}
	\includegraphics[width=0.5\columnwidth]{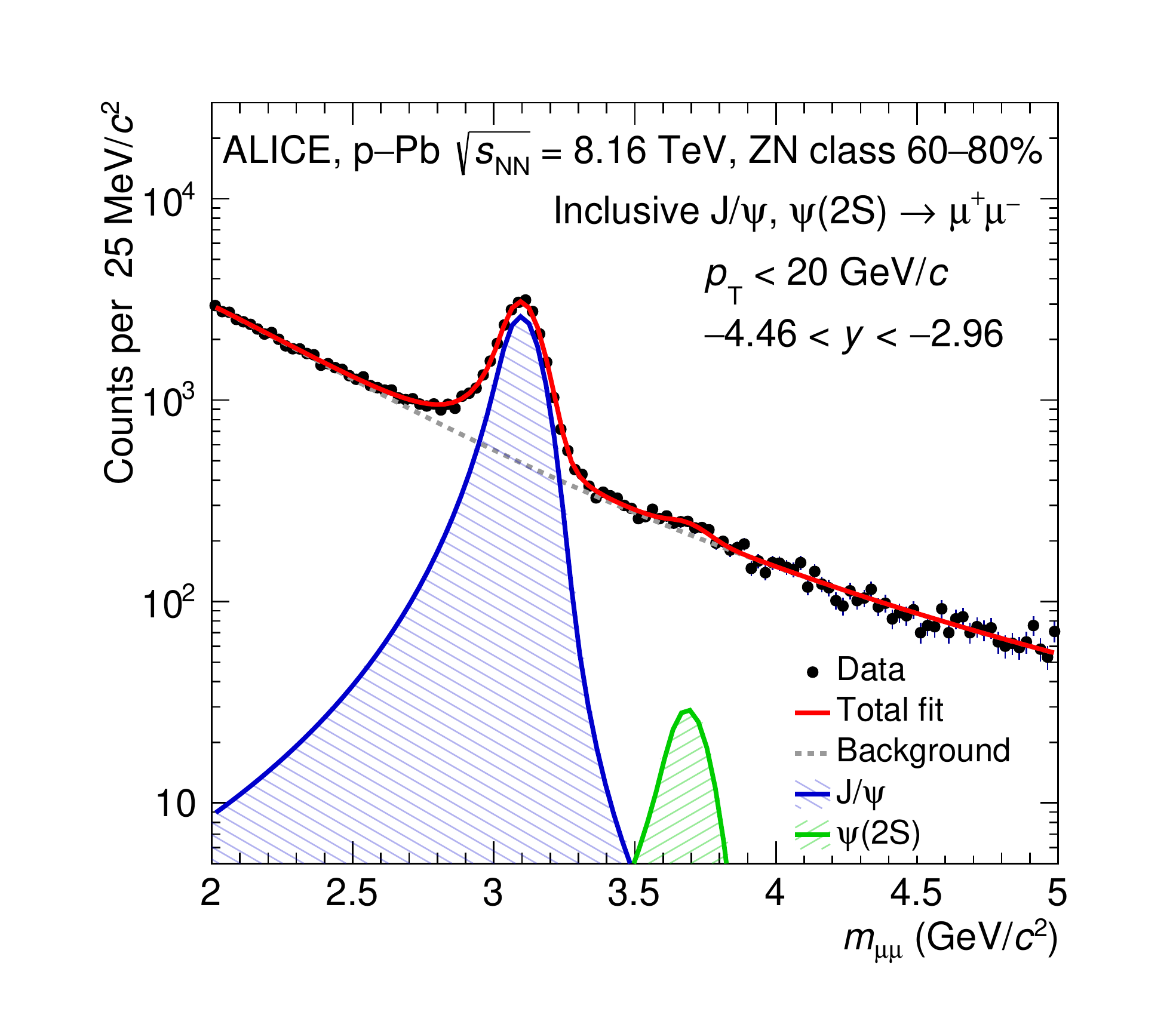}
	\caption{\label{figFit_Minv} Fit to the dimuon invariant mass distribution for the p--Pb (top panels) and Pb--p (bottom panels) data sets, for the 20--40\% (left panels) and 60--80\% (right panels) ZN centrality classes. The extended Crystal Ball function is used to describe the $\jpsi$ and $\psip$ signals, while a Variable Width Gaussian function is used for the background. The red line represents the total fit.}
\end{figure}

The charmonium signal is estimated with a binned maximum likelihood fit to the dimuon invariant mass distribution. The $\jpsi$ and $\psip$ mass shapes are described with a Crystal Ball function with asymmetric tails on both sides of the peak (denoted as extended Crystal Ball) or a pseudo-Gaussian function~\cite{ALICE-PUBLIC-2015-006}. The $\jpsi$ mass and width are free parameters of the fit, while the other parameters, which correspond to the non-Gaussian tails of the signal shape, are fixed to those extracted from MC simulations. In addition, other sets of tails obtained from fits to the centrality-integrated invariant mass distribution in p--Pb at $\sqrt{s_{\mathrm{NN}}}=8.16$ TeV and in pp collisions at $\sqrt{s}=8$ TeV are used to test the stability of the fit and are included in the evaluation of the charmonium signal and its systematic uncertainty. 
The $\psip$ fit parameters, apart from the amplitude, are constrained to those of the $\jpsi$, since its signal-to-background ratio is rather small. 
For the position of the mass peak, the following relation is used $m_{\psip}=m_{\jpsi} + m_{\psip}^{\rm PDG} - m_{\jpsi}^{\rm PDG}$, where the value obtained from the $\jpsi$ fit is shifted by the difference between the two mass poles reported by the PDG~\cite{Tanabashi:2018oca}. 
The $\psip$ width is fixed to the $\jpsi$ one, applying a correction factor given by the ratio of the widths obtained in MC simulations ($\sigma_{\psip}=\sigma_{\jpsi} \times \sigma_{\psip}^{\rm MC} \big/ \sigma_{\jpsi}^{\rm MC}$). 
The background continuum is parameterised by either a Gaussian having a mass-dependent width or the product of a fourth degree polynomial function and an exponential. 
Finally, to test the background description, the signal is extracted using different fit ranges ($2 < m_{\mu\mu} < 5$ GeV/$c^2$ and $2.2 < m_{\mu\mu} < 4.5$ GeV/$c^2$).
The number of $\jpsi$ and $\psip$ and their statistical uncertainties are evaluated as the averages of the results of each test, i.e.\ the aforementioned signal extraction variations, and of their statistical uncertainty, respectively. The systematic uncertainty is given by the root-mean-square of the distribution of the results.
For the $\psi$(2S), an additional uncertainty of 5\% is added in quadrature. It corresponds to the uncertainty on the $\psi$(2S) width obtained from the large pp data sample used to validate the assumption on the relative widths for J/$\psi$ and $\psi$(2S) from the MC~\cite{Acharya:2017hjh}.
 In Fig.~\ref{figFit_Minv} the fits to the dimuon invariant mass distribution for the forward and the backward rapidity ranges are shown for two centrality classes. 
 
The product of the detector acceptance and the reconstruction efficiency ($A\times\varepsilon$) is evaluated with a MC simulation in which $\jpsi$ and $\psip$ are generated unpolarised according to the results obtained in pp collisions by the ALICE~\cite{Abelev:2011md,Acharya:2018uww}, CMS~\cite{Chatrchyan:2013cla}, and LHCb~\cite{Aaij:2013nlm,Aaij:2014qea} collaborations. 
In order to realistically describe the $\jpsi$ and $\psip$ spectra, the MC input $\pt$ and $\y$ shapes are tuned directly on data performing an iterative procedure~\cite{Acharya:2018kxc}. 
The decay products of the generated charmonia are then propagated inside a realistic description of the ALICE detector, based on GEANT 3.21~\cite{Brun:1994aa}.
The $\pt$- and $\y$-integrated ($A\times\varepsilon$) values are $0.264 \pm 0.001$ ($0.235 \pm 0.001$) in the p--Pb (Pb--p) data sample for the $\jpsi$  and $0.280 \pm 0.008$ ($0.250 \pm 0.004$) for the $\psip$.
The larger ($A\times\varepsilon$) in the p--Pb than Pb--p data taking period is due to different running conditions.
The quoted uncertainties are the systematic uncertainties on the input $\pt$ and $\y$ shapes used for the MC generation, which are evaluated comparing the ($A \times \varepsilon$) values obtained using different input MC distributions. For the J/$\psi$, these were obtained by adjusting the input MC distributions to the data in various $p_{\rm T}$ and $y$ intervals. For the $\psi(2S)$, due to the larger statistical uncertainties of the data, the input $\pt$ and $\y$ shapes used for the J/$\psi$ were considered in addition to the ones tuned directly on the $\psi(2S)$ data.
For the $\jpsi$ the uncertainty on the $\pt$-integrated ($A\times\varepsilon$) is 0.5\% for both p--Pb and Pb--p, varying with $\pt$ from 1\% to 3\%, while for the $\psip$ the uncertainty amounts to 3\% and 1.5\% in p--Pb and Pb--p collisions, respectively.
The same values of ($A \times \varepsilon$) are used for all centrality classes since no dependence on the detector occupancy  is observed within the multiplicities reached in p--Pb collisions. 
Possible changes of ($A\times\varepsilon$) due to shape variations of the $p_{\rm T}$- and $y$-differential cross sections with centrality are accounted for in the systematic uncertainties by using different $p_{\rm T}$ shapes extracted from different centrality intervals as inputs to the MC simulations. 
The corresponding systematic uncertainty on the $\pt$-integrated $\jpsi$ ($A\times\varepsilon$) varies from 1.6\% (2.5\%) to 1.7\% (2.7\%) as a function of centrality, while as a function of $p_{\rm T}$ in different centrality classes it varies from 1.2\% (1.4\%) to 4.4\% (2.2\%) in Pb--p (p--Pb) collisions. 

The normalisation of the $\jpsi$ and $\psip$ yields is obtained following the prescription described in Ref.~\cite{Adam:2014qja}. It is based on the evaluation of the number of minimum bias events $N_{\mathrm{MB}}^{i}$ for each centrality class $i$ as $N_{\mathrm{MB}}^{i}=F_{2\mu/\mathrm{MB}}^{i} \times N_{2\mu}^{i}$, where $F_{2\mu/\mathrm{MB}}^{i}$ is the inverse of the probability of having a dimuon-triggered event in a MB-triggered one, and $N_{2\mu}^{i}$ is the number of analyzed dimuon-triggered events. The value of $F_{2\mu/\mathrm{MB}}^{i}$ depends on the centrality class and increases from central to peripheral events, passing from $384\pm3$ to $1855\pm18$ in p--Pb and from $161\pm1$ to $2036\pm16$ in Pb--p collisions, for the 2--10\% and 80--90\% centrality classes, respectively. 
The quoted systematic uncertainties, which vary between 1\% and 1.4\%, contain two contributions.
The first one, which is correlated in $p_{\rm T}$ and centrality and amounts to 1\%, is estimated by comparing the centrality-integrated $F_{2\mu/\mathrm{MB}}$ obtained with the method described above with the one obtained using the information of the online trigger counters, as described in Ref.~\cite{Abelev:2013yxa}.
The second one, which is not correlated in centrality, is obtained comparing $F_{2\mu/\mathrm{MB}}^{i}$ evaluated with two different methods, as detailed in Ref.~\cite{Adam:2015jsa}. Namely, $F_{2\mu/\mathrm{MB}}^{i}$ can be evaluated directly in each centrality class, or derived from the centrality integrated $F_{2\mu/\mathrm{MB}}$ factor normalised by the ratio of $N_{\mathrm{MB}}^{i}/N_{\mathrm{MB}}$ to $N_{2\mu}^{i}/N_{2\mu}$. The resulting systematic uncertainty varies from 0.1\% (0.1\%) to 0.8\% (1\%) in Pb--p (p--Pb) collisions.

The inclusive cross section for $\jpsi$ and $\psip$ for centrality class $i$ is calculated using the expression
\begin{equation}
\sigma_{\mathrm{pPb}}^{i,\ensuremath{\psi\mathrm{(nS)}}}= \frac{N_{\ensuremath{\psi\mathrm{(nS)}}\rightarrow\mu^{+}\mu^{-}}^{i}}{(A\times\varepsilon)_{\ensuremath{\psi\mathrm{(nS)}}\rightarrow\mu^{+}\mu^{-}} \times N_{\mathrm{MB}}^{i} \times {\rm B.R.}_{\psi\mathrm{(nS)}\rightarrow\mu^{+}\mu^{-}}} \times \sigma_{\mathrm{MB}},
\label{CS}
\end{equation}
where $N_{\ensuremath{\psi\mathrm{(nS)}}\rightarrow\mu^{+}\mu^{-}}^{i}$ is the raw yield for the given resonance, $(A \times \epsilon)_{\ensuremath{\psi\mathrm{(nS)}}\rightarrow\mu^{+}\mu^{-}}$ is the corresponding product of the detector acceptance and reconstruction efficiency, and $ {\rm B.R.}_{\psi\mathrm{(nS)}\rightarrow\mu^{+}\mu^{-}}$ is the branching ratio of the corresponding dimuon decay channel as reported in Ref.~\cite{Tanabashi:2018oca}.
The integrated luminosity $L_{\rm int}$ of the analyzed data sample is given by the ratio of the equivalent number of minimum bias events $N_{\mathrm{MB}}$ to the cross section for events satisfying the minimum bias trigger condition $\sigma_{\mathrm{MB}}$. The latter is evaluated through a van der Meer scan and results in a value of $2.09 \pm 0.04$ b for p--Pb collisions and $2.10 \pm 0.04$ b for Pb--p~\cite{ALICE-PUBLIC-2018-002}, where the quoted uncertainties are the systematic uncertainties. 
The integrated luminosity can be independently calculated using the luminosity signal provided by the T0 detector. 
The difference between the integrated luminosity obtained with the V0 and T0 detectors amounts to 1.1\% (0.6\%)~\cite{ALICE-PUBLIC-2018-002} in the p--Pb (Pb--p) data sample  and is assigned as a further systematic uncertainty of $\sigma_{\mathrm{MB}}$. The correlated uncertainty on $\sigma_{\mathrm{MB}}$ for the p--Pb and Pb--p data samples are 0.5\% and 0.7\%, respectively~\cite{ALICE-PUBLIC-2018-002}.

The relative modification between the two charmonium states in proton--nucleus collisions can be firstly observed through the evaluation of the ratio ${\rm B.R.}_{\psip\rightarrow\mu^{+}\mu^{-}}\sigma_{\psip} / {\rm B.R.}_{\jpsi\rightarrow\mu^{+}\mu^{-}}\sigma_{\jpsi}$, where the systematic uncertainties on trigger, tracking, and matching efficiencies, as well as on the luminosity, which are common for the $\jpsi$ and $\psip$, cancel out.
The only remaining systematic uncertainties are those related to the signal extraction and to the shape of the input $p_{\rm T}$ and $y$ distribution used for the MC simulations.
In turn this ratio can be normalised to the same quantity evaluated in pp collisions, providing a direct access to the relative $\psip$ production modification with respect to $\jpsi$ moving from a pp to a p--Pb collision system. Since there is no measurement available at $\sqrtS = 8.16$ TeV in pp collisions, the ratio $\psip/\jpsi$ is evaluated through an interpolation procedure using ALICE data at $\sqrtS = 5$, 7, 8, and 13 TeV in the interval $2.5 < y < 4$~\cite{Acharya:2017hjh,Abelev:2014qha,Adam:2015rta}.  
The uncertainty associated to the interpolated value contains a contribution of 6\% due to the energy-interpolation procedure and a further 1\% contribution due to the rapidity-extrapolation procedure~\cite{Acharya:2020wwy}. In addition an extra 1\% is included due to the assumption of non-flat dependence of the ratio as a function of $\sqrt{s}$, according to the NRQCD+CGC calculations \cite{Ma:2010yw,Ma:2014mri}. 
The results of the interpolation procedure are reported in Ref.~\cite{ALICE-PUBLIC-2020-007}.

The nuclear modification factor as a function of centrality is calculated using the following expression
\begin{equation}
Q_{\mathrm{pPb}}^{i,\ensuremath{\psi\mathrm{(nS)}}\rightarrow\mu^{+}\mu^{-}}=\frac{N_{\ensuremath{\psi\mathrm{(nS)}}\rightarrow\mu^{+}\mu^{-}}^{i}}{\langle T_{\mathrm{pPb}}^{i} \rangle \times N_{\mathrm{MB}}^{i} \times (A\times\varepsilon)_{\ensuremath{\psi\mathrm{(nS)}}\rightarrow\mu^{+}\mu^{-}} \times B.R._{\psi\mathrm{(nS)}\rightarrow\mu^{+}\mu^{-}} \times \sigma_{\ensuremath{\psi\mathrm{(nS)}}}^{\mathrm{pp}}},
\label{QpPb}
\end{equation}
where $\langle T_{\mathrm{pPb}}^{i} \rangle$ is the nuclear overlap function for the centrality class $i$, while $\sigma_{\ensuremath{\psi\mathrm{(nS)}}}^{\rm pp}$ is the $\psi(n{\rm S})$  production cross section in proton--proton collisions. 
The notation $Q_{\mathrm{pPb}}$ is used instead of the usual $R_{\mathrm{pPb}}$ in order to point out the possible bias in the centrality determination, which depends on the loose correlation between the centrality estimator and the collision geometry~\cite{Adam:2014qja}.
The J/$\psi$ cross section in pp collisions at $\sqrtS~=~8.16$~TeV is obtained from the available results in the interval $2.5 < y < 4$ for inclusive J/$\psi$ production at $\sqrtS = 8$~TeV from ALICE~\cite{Adam:2015rta} and LHCb~\cite{Aaij:2013yaa} using the energy and rapidity extrapolation procedure described in Ref.~\cite{Acharya:2018kxc}. 
A resulting first contribution of 7.1\% to the systematic uncertainty of the extrapolation procedure is correlated in $p_{\rm T}$, $y$, and centrality.
A second contribution of 1.8\% (1.5\%) for the $p_{\rm T}$-integrated cross section and ranging from 3.0\% to 4.6\% (2.9\% to 4.7\%) for the $p_{\rm T}$-differential cross section at backward (forward) rapidity, correlated with centrality, arises from the energy and rapidity interpolation procedures (see Ref.~\cite{Acharya:2018kxc} for details). 
The $\psi$(2S) cross section in pp collisions at $\sqrtS = 8.16$ TeV is obtained from the extrapolated J/$\psi$ cross section and the interpolated $\psip/\jpsi$ ratio. 
The related total systematic uncertainty is 9.4\% and is correlated in $p_{\rm T}$, $y$, and centrality.
The resulting extrapolated cross sections are reported in Ref.~\cite{ALICE-PUBLIC-2020-007}.

In addition to the various contributions to the systematic uncertainty discussed above, the following sources, which are common for the J/$\psi$ and $\psi$(2S) states, are also taken into account. 
The systematic uncertainty of the trigger efficiency includes two contributions, one related to the intrinsic efficiency of each trigger chamber and one related to the muon trigger response function. 
The former is calculated from the uncertainties on the trigger chamber efficiencies measured from data and applied to simulations and it amounts to 1\%. 
The latter is obtained from the difference between the ($A\times\varepsilon$) obtained using the response function in data or in MC simulations and for the $p_{\rm T}$-integrated case this uncertainty is 2.9\% for Pb--p and 2.4\% for p--Pb, and it varies between 1\% and 4\% as a function of $\pt$.  
The total systematic uncertainty of the trigger efficiency, obtained by adding in quadrature the aforementioned contributions, is 3.1\% for Pb--p and 2.6\% for p--Pb, varying as a function of $\pt$ from 1.4\% up to 4.1\%.
The evaluation of the systematic uncertainty on the tracking efficiency follows a similar approach as reported in Ref.~\cite{Abelev:2014zpa}. 
The discrepancy between the efficiencies in data and MC corresponds to 2\% in Pb--p and 1\% in p--Pb, without any appreciable dependence on the dimuon kinematics and event centrality. 
Finally, the choice of the $\chi^{2}$ selection applied for the definition of the matching between tracks in the trigger and tracking chambers leads to a 1\% systematic uncertainty. 

In Table~\ref{tabsyst}, a summary of all the sources of systematic uncertainty which contribute to the cross section and nuclear modification factor measurements is reported.

\begin{table}[!htb]
	\centering
	\caption{\label{tabsyst} Summary of the systematic uncertainties (in percentage) of the quantities associated to the measurements of the differential J/$\psi$ cross section and $Q_{\rm pPb}$ of J/$\psi$ and $\psi$(2S). The uncertainties for the $p_{\rm T}$-differential case are indicated in parentheses if the values are different from the $p_{\rm T}$-integrated case. When appropriate, a range of variation (for centrality, rapidity, or $p_{\rm T}$ intervals) of the uncertainty is given. Type I, II, and III stands for uncertainties correlated over centrality, rapidity, or $p_{\rm T}$, respectively.}
	\resizebox{\columnwidth}{!}{
		\begin{tabular}{lccccc}
			\hline
			\hline
			&    \multicolumn{2}{c} {J/$\psi$}                                                                           & \multicolumn{2}{c} {$\psi$(2S)}   \\
			\hline
			Sources of uncertainty  & $-$ 4.46 $<$ $y_{\rm cms}$ $<$ $-$ 2.96 &  2.03 $<$ $y_{\rm cms}$ $<$ 3.53  & $-$ 4.46 $<$ $y_{\rm cms}$ $<$ $-$ 2.96  &  2.03 $<$ $y_{\rm cms}$ $<$ 3.53   \\ 
			& cent. (cent. and $p_{\rm T}$)     	    &  cent. (cent. and $p_{\rm T}$)	     & cent.		                            &  cent.                    	      \\
			\hline
			Signal extraction       & 3.0--3.3 (2.2--6.8) 			    & 2.8--3.1 (2.6--4.2)		             & 7.1--15.9			            & 7.6--12.8		              \\
			Trigger efficiency (I)  & 3.1 (1.4--4.1)  	 		    & 2.6 (1.4--4.1)			     & 3.1				            & 2.6 	        		      \\
			Tracking efficiency (I) & 2  					    & 1 				     & 2					    & 1 				      \\
			Matching efficiency (I) & 1  					    & 1 				     & 1					    & 1 				      \\
			MC input (I)            & 0.5 (1--2) 		                    & 0.5 (1--3)	                     & 1.5		                            &  3 	                                      \\ 
			MC input            & 1.6--1.7 (1.2--4.4) 		            & 2.5--2.7 (1.4--2.2)	             & 1.6--1.7		                            &  2.5--2.7 	                                      \\ 

			$F_{\rm norm}$ (I,III)  & 1  					    & 1 				     & 1					    & 1 				      \\
			$F_{\rm norm}$ (III)    & 0.1--0.8   			            & 0.1--1.0			     & 0.1--0.8                                   & 0.1--1.0	\\
			Pile-up (III)                 & 2					    & 2                                      & 2                                            & 2                                       \\  
			\hline
			\multicolumn{5}{c} {Uncertainties related to cross section only}  \\
			\hline
			$\sigma_{\rm MB}$ (I,III)         & 2.2        & 2.1	    & --	 & --   	  \\
			$\sigma_{\rm MB}$ (I,II,III)            & 0.7        & 0.5	    & --	 & --   	  \\
			BR (I,II,III)                              & 0.6        & 0.6	    & --	 & --   	  \\ 
			\hline
			\multicolumn{5}{c}  {Uncertainties related to $Q_{\rm pPb}$ only} \\
			\hline
			$\langle T_{\rm pPb} \rangle$ (II,III)                       & 2.1--4.8          & 2.1--4.8			    & 2.1--4.8		        & 2.1--4.8			 \\
			pp reference (I)                             & 1.8 (3.0--4.6)    & 1.5 (2.9--4.7)                   & --	                & --                             \\
			pp reference (I,II,III)                      & 7.1  		 & 7.1  			    & 9.4			& 9.4 	         \\
			\hline
			\hline
		\end{tabular}
	}
\end{table}

\section{Results}
\label{Results}
\subsection{$p_{\rm T}$-differential cross section of inclusive $\rm J/\psi$ for various centrality classes}
\begin{figure}[!tb]
\includegraphics[width=0.5\columnwidth]{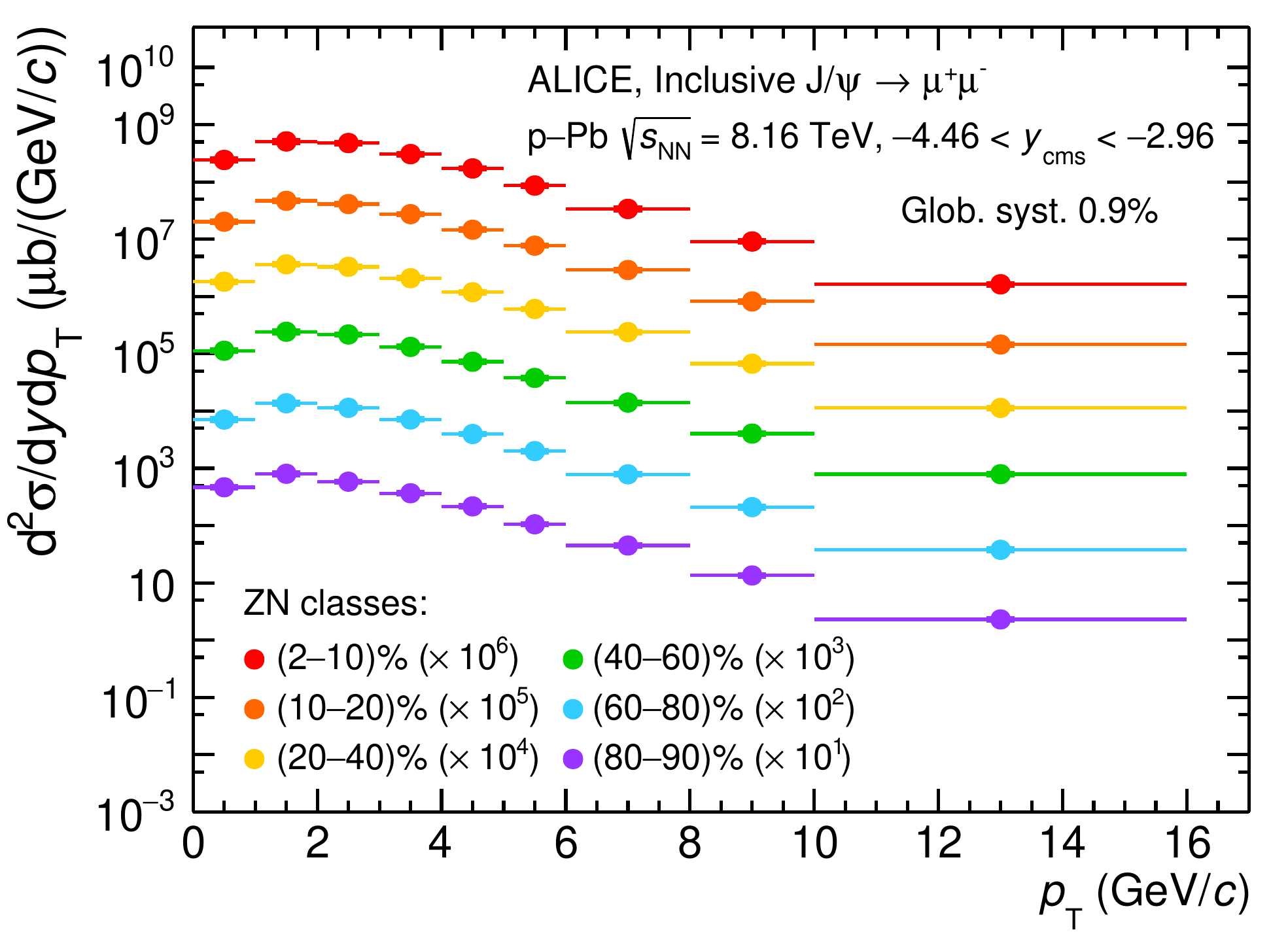}
\includegraphics[width=0.5\columnwidth]{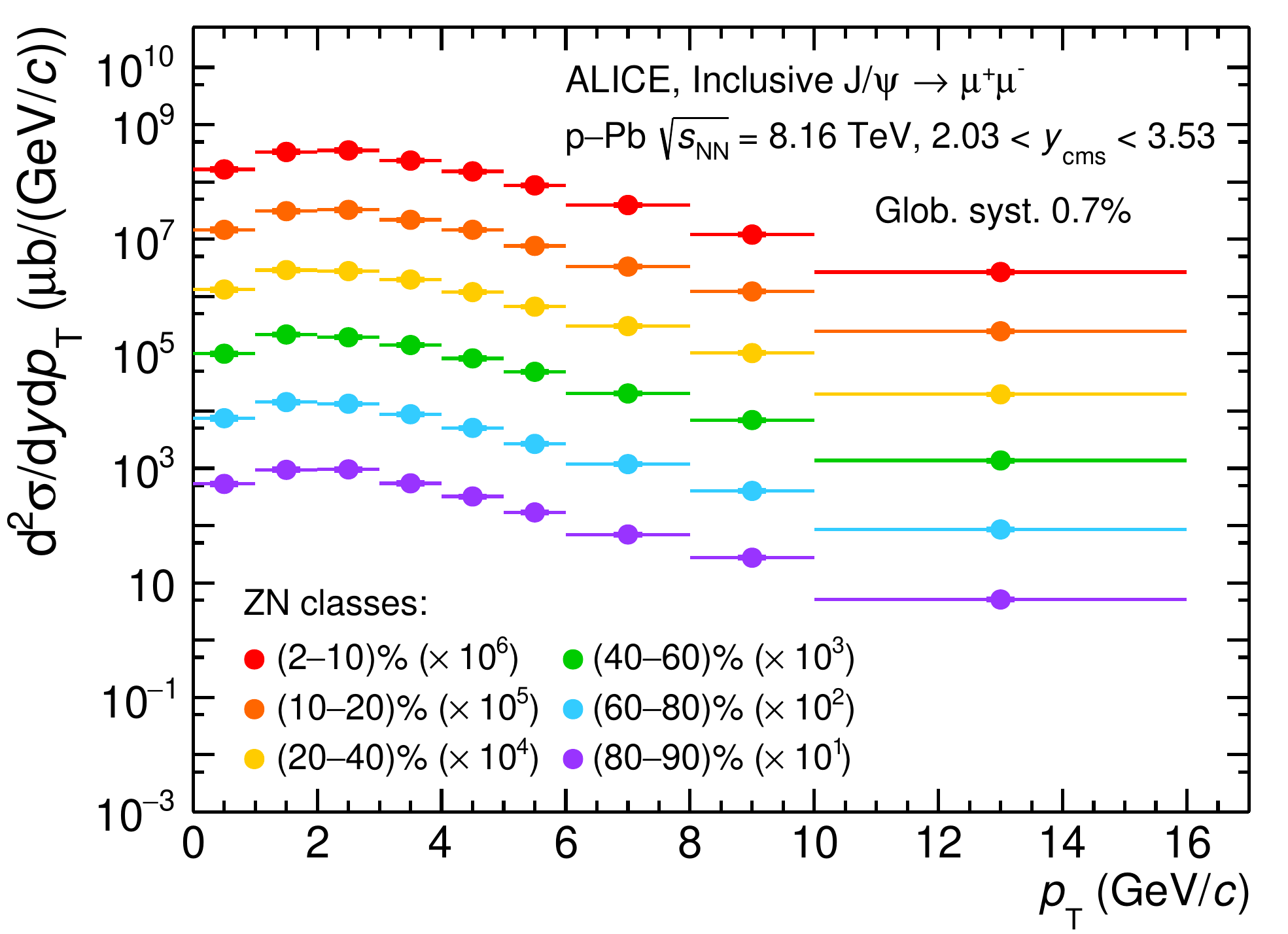}
\caption{\label{CS_JPsi}Inclusive J/$\psi$ $p_{\rm T}$-differential cross section for different centrality classes at backward (left) and forward (right) rapidity in p--Pb collisions at $\sqrt{s_{\rm NN}} = 8.16$ TeV. The vertical error bars, representing the statistical uncertainties, and the boxes around the points, representing the uncorrelated systematic uncertainties, are smaller than the marker. The global systematic uncertainty, which is correlated over centrality, rapidity, and $p_{\rm T}$ and is obtained as the quadratic sum of the systematic uncertainty of the branching ratio and the correlated systematic uncertainty of $\sigma_{\rm MB}$, amounts to 0.9\% (0.7\%) at backward (forward) rapidity and is shown as text.}
\end{figure}

Figure~\ref{CS_JPsi} shows the $p_{\rm T}$-differential cross section of inclusive J/$\psi$ at backward (left) and forward (right) rapidity measured in six centrality classes: 2--10\%, 10--20\%, 20--40\%, 40--60\%, 60--80\%, and 80--90\%. The vertical error bars represent the statistical uncertainties and the open boxes the uncorrelated systematic uncertainties. A global systematic uncertainty, which is correlated over centrality, rapidity, and $p_{\rm T}$ and is obtained as the quadratic sum of the systematic uncertainty of the branching ratio and the correlated systematic uncertainty of $\sigma_{\rm MB}$ amounts to 0.9\% (0.7\%) at backward (forward) rapidity. 

\subsection{Inclusive \jpsi\ average transverse momentum and \pt\ broadening}
A first insight into the modification of \jpsi production in p--Pb collisions can be obtained by studying the average transverse momentum $\langle p_{\rm T}\rangle$ and the average squared  transverse momentum $\langle p^{2}_{\rm T}\rangle$ as a function of the collision centrality. 
The $\langle p_{\rm T}\rangle$ and $\langle p^{2}_{\rm T}\rangle$ are extracted for each centrality class by performing a fit of the $p_{\rm T}$-differential cross section with a widely used function proposed in Ref.~\cite{Yoh:1978id} and defined as 
\begin{eqnarray}
  f(p_{\rm T}) = C \frac{p_{\rm T}}{(1+(p_{\rm T}/p_{\rm 0})^{2})^{n}},
\label{eq:JPsiPsigma}
\end{eqnarray}
where $C$, $p_{\rm 0}$, and $n$ are free parameters of the fit. 
The central values of $\langle p_{\rm T}\rangle$ and $\langle p^{2}_{\rm T}\rangle$ are obtained from the fit using the quadratic sum of statistical and uncorrelated systematic uncertainties of the data points. 
The uncertainties on the free parameters obtained from the fit are propagated to the values of $\langle p_{\rm T}\rangle$ and $\langle p^{2}_{\rm T}\rangle$. 
The statistical and systematic uncertainties on $\langle p_{\rm T}\rangle$ and $\langle p^{2}_{\rm T}\rangle$ are obtained by performing the fit using, respectively, only the statistical or the uncorrelated systematic uncertainties on the data points. 
The range of integration on \pt\ for this calculation is limited to the $p_{\rm T}$ interval 0 $<$ $p_{\rm T}$ $<$ 16 GeV/$c$. Extending the integration range to infinity has a negligible effect with respect to the quoted uncertainties. 
Table~\ref{tabmeanpt} shows the values of $\langle p_{\rm T}\rangle$ and $\langle p^{2}_{\rm T}\rangle$ of inclusive $\rm J/\psi$ for each centrality class. Both $\langle p_{\rm T}\rangle$ and $\langle p^{2}_{\rm T}\rangle$ increase with increasing centrality, which indicates a hardening of the \jpsi \pt\ distribution from peripheral to central collisions in both rapidity intervals.  

\begin{table}[b]
\centering
\caption{Values of $\langle p_{\rm T}\rangle$ and $\langle p^{2}_{\rm T}\rangle$ of inclusive $\rm J/\psi$ in the range 0 $<$ $p_{\rm T}$ $<$ 16 GeV/$c$. The first uncertainty is statistical while the second one is systematic. The values along with the systematic uncertainty obtained from the pp cross section interpolated to $\sqrt{s} = 8.16$ TeV are also indicated.}
\resizebox{\columnwidth}{!}{
\begin{tabular}{lccccc}
\hline
\hline 
    &  \multicolumn{2}{c}{$-$ 4.46 $<$ $y_{\rm cms}$ $<$ $-$ 2.96} & \multicolumn{2}{c}{2.03 $<$ $y_{\rm cms}$ $<$ 3.53} \\
    & $\langle p_{\rm T}\rangle$ (GeV/$c$) &  $\langle p^{2}_{\rm T}\rangle$ (GeV$^{2}$/$c^{2}$) & $\langle p_{\rm T}\rangle$ (GeV/$c$) & $\langle p^{2}_{\rm T}\rangle$ (GeV$^{2}$/$c^{2}$)\\
\hline
centrality class   & \multicolumn{4}{c}{p--Pb} \\
\hline
2--10\%   & 2.753 $\pm$ 0.016 $\pm$ 0.027 & 10.919 $\pm$ 0.118 $\pm$ 0.186 & 3.094 $\pm$ 0.022 $\pm$ 0.029  & 14.016 $\pm$ 0.119 $\pm$ 0.223   \\  
10--20\%  & 2.760 $\pm$ 0.014 $\pm$ 0.027 & 10.959 $\pm$ 0.108 $\pm$ 0.189 & 3.094 $\pm$ 0.020 $\pm$ 0.029  & 14.051 $\pm$ 0.188 $\pm$ 0.219   \\
20--40\%  & 2.740 $\pm$ 0.011 $\pm$ 0.028 & 10.846 $\pm$ 0.095 $\pm$ 0.192 & 3.059 $\pm$ 0.015 $\pm$ 0.029  & 13.747 $\pm$ 0.135 $\pm$ 0.224   \\
40--60\%  & 2.700 $\pm$ 0.013 $\pm$ 0.027 & 10.549 $\pm$ 0.106 $\pm$ 0.184 & 3.007 $\pm$ 0.017 $\pm$ 0.029  & 13.303 $\pm$ 0.155 $\pm$ 0.226   \\
60--80\%  & 2.658 $\pm$ 0.016 $\pm$ 0.028 & 10.334 $\pm$ 0.130 $\pm$ 0.190 & 2.875 $\pm$ 0.020 $\pm$ 0.028  & 12.339 $\pm$ 0.177 $\pm$ 0.208   \\
80--90\%  & 2.594 $\pm$ 0.030 $\pm$ 0.032 & 10.037 $\pm$ 0.231 $\pm$ 0.208 & 2.811 $\pm$ 0.033 $\pm$ 0.029  & 11.836 $\pm$ 0.284 $\pm$ 0.211    \\
\hline
    & \multicolumn{4}{c}{pp} \\
\hline
      & 2.557 $\pm$ 0.035 & 9.678 $\pm$ 0.225  & 2.738 $\pm$ 0.037  & 11.242 $\pm$ 0.252  \\
\hline
\hline
\end{tabular}
}
\label{tabmeanpt}
\end{table}

\begin{figure}[tb]
\centering
\includegraphics[width=0.5\columnwidth]{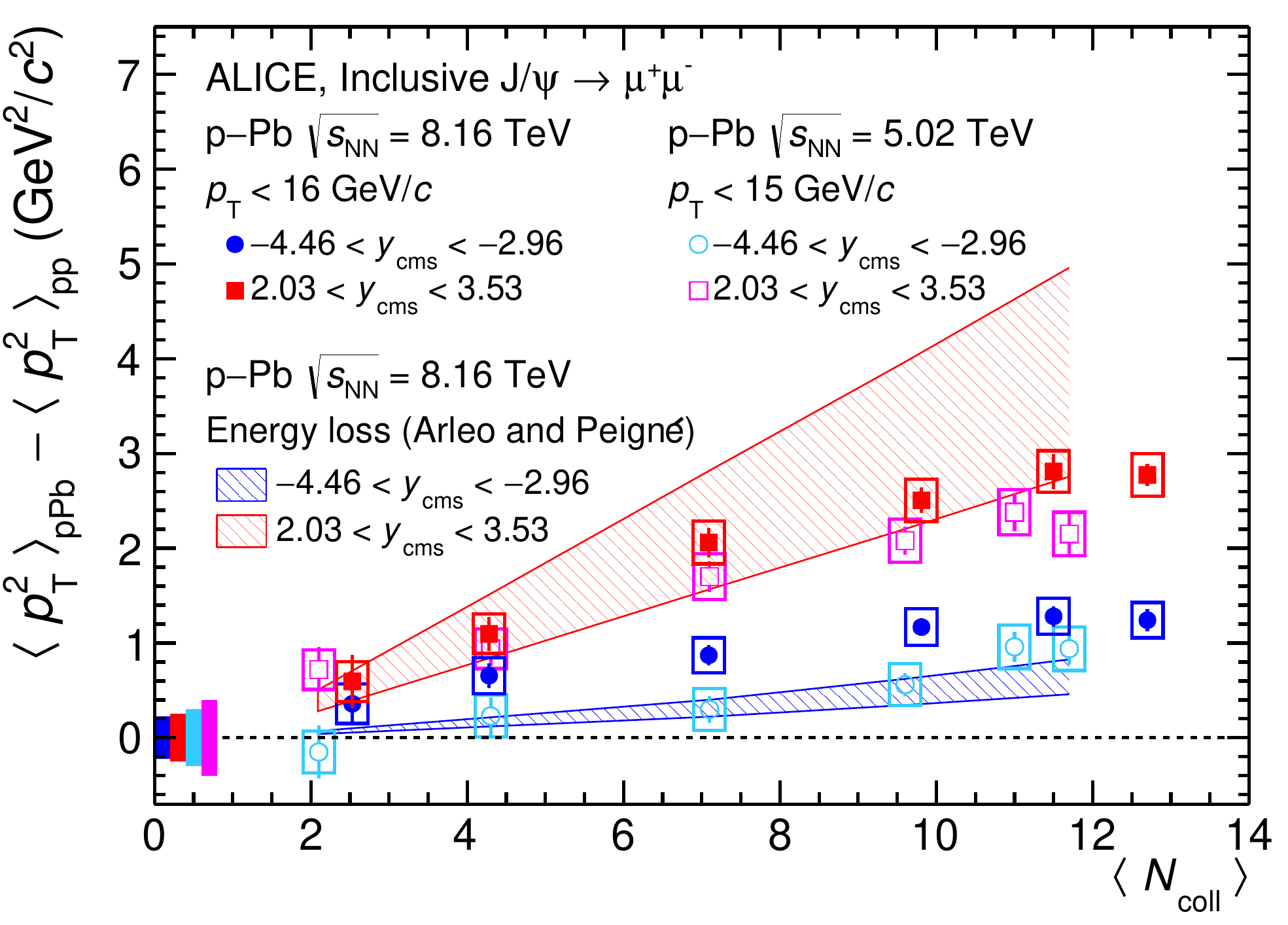}
\caption{\label{delta_avrg_pT2} \pt\ broadening of J/$\psi$, $\Delta \langle p^{2}_{\rm T}\rangle$, as a function of $\langle N_{\rm coll}\rangle$ at backward (blue circles) and forward (red squares) rapidity in p--Pb collisions at $\sqrt{s_{\rm NN}} = 8.16$ TeV compared to the results at $\sqrt{s_{\rm NN}} = 5.02$ TeV~\cite{Adam:2015jsa} and to energy loss model calculations~\cite{Arleo:2014oha}. The vertical error bars represent the statistical uncertainties and the boxes around the data points the systematic uncertainties.}
\end{figure}

The $p_{\rm T}$ broadening defined as the difference between the average squared transverse momentum in p--Pb and pp collisions ($\Delta \langle p^{2}_{\rm T}\rangle = {\langle p^{2}_{\rm T}\rangle}_{\rm pPb} - {\langle p^{2}_{\rm T}\rangle}_{\rm pp}$)
can be used to quantify the nuclear effects on the \jpsi production~\cite{Kang:2008us,Kang:2012am,Arleo:2013zua}. 
The value of ${\langle p^{2}_{\rm T}\rangle}_{\rm pp}$ is evaluated from the $p_{\rm T}$-differential cross section in pp collisions at $\sqrt{s} = 8.16$ TeV obtained with the interpolation procedure described in Ref.~\cite{Acharya:2018kxc}, and using the same $p_{\rm T}$ integration range as for p--Pb collisions. 
The resulting values are reported in Ref.~\cite{ALICE-PUBLIC-2020-007}.
Figure~\ref{delta_avrg_pT2} shows $\Delta \langle p^{2}_{\rm T}\rangle$ as a function of the number of binary collisions at backward and forward rapidity. 
In all cases, $\Delta \langle p^{2}_{\rm T}\rangle$ is larger than zero, indicating a broadening of the $p_{\rm T}$ distribution in p--Pb collisions compared to pp collisions.
For the most peripheral collisions, corresponding to $\langle N_{\rm coll}\rangle \sim 2.5$, the $\Delta \langle p^{2}_{\rm T}\rangle$ measured at backward $y$ is compatible, within uncertainties, with that at forward $y$.
In both backward and forward rapidity ranges the $p_{\rm T}$ broadening increases with increasing centrality.
However, the increase of $\Delta \langle p^{2}_{\rm T}\rangle$ is stronger in the p-going direction than in the Pb-going direction. 
Thus, nuclear effects appear to increase with the centrality of the collision and to be stronger in the p-going than in the Pb-going direction.
Here, it is worth noting that under the naive assumption of a $2 \rightarrow 1$ production process (${\rm gg} \rightarrow \psi{\rm (nS)}$),
the sampled $x$ ranges of the lead nuclei correspond to the shadowing and anti-shadowing regions for the p-going and lead-going direction measurements, respectively.
Also shown in Fig.~\ref{delta_avrg_pT2} are the results at $\sqrt{s_{\rm NN}} = 5.02$ TeV~\cite{Adam:2015jsa}.
The same trend of $\Delta \langle p^{2}_{\rm T}\rangle$ as a function of $\langle N_{\rm coll}\rangle$ is seen at both collision energies in the two rapidity ranges.
Overall, $\Delta \langle p^{2}_{\rm T}\rangle$ slightly increases with the collision energy.
The  $\Delta \langle p^{2}_{\rm T}\rangle$ as function of $\langle N_{\rm coll}\rangle$ is also compared in Fig.~\ref{delta_avrg_pT2} to the results of an energy loss model, which is based on a parameterisation of the prompt J/$\psi$ pp cross section and includes coherent energy loss effects from the incoming and outgoing partons~\cite{Arleo:2014oha}. The band in this model represents the uncertainty on the parton transport coefficient and the parameterisation used for the pp reference cross section. The model describes the centrality dependence of $\Delta \langle p^{2}_{\rm T}\rangle$ at forward rapidity reasonably well, but it underestimates the data at backward rapidity.

\subsection{Centrality dependence of the inclusive \jpsi nuclear modification factor}
\begin{figure}[b]
\includegraphics[width=0.5\columnwidth]{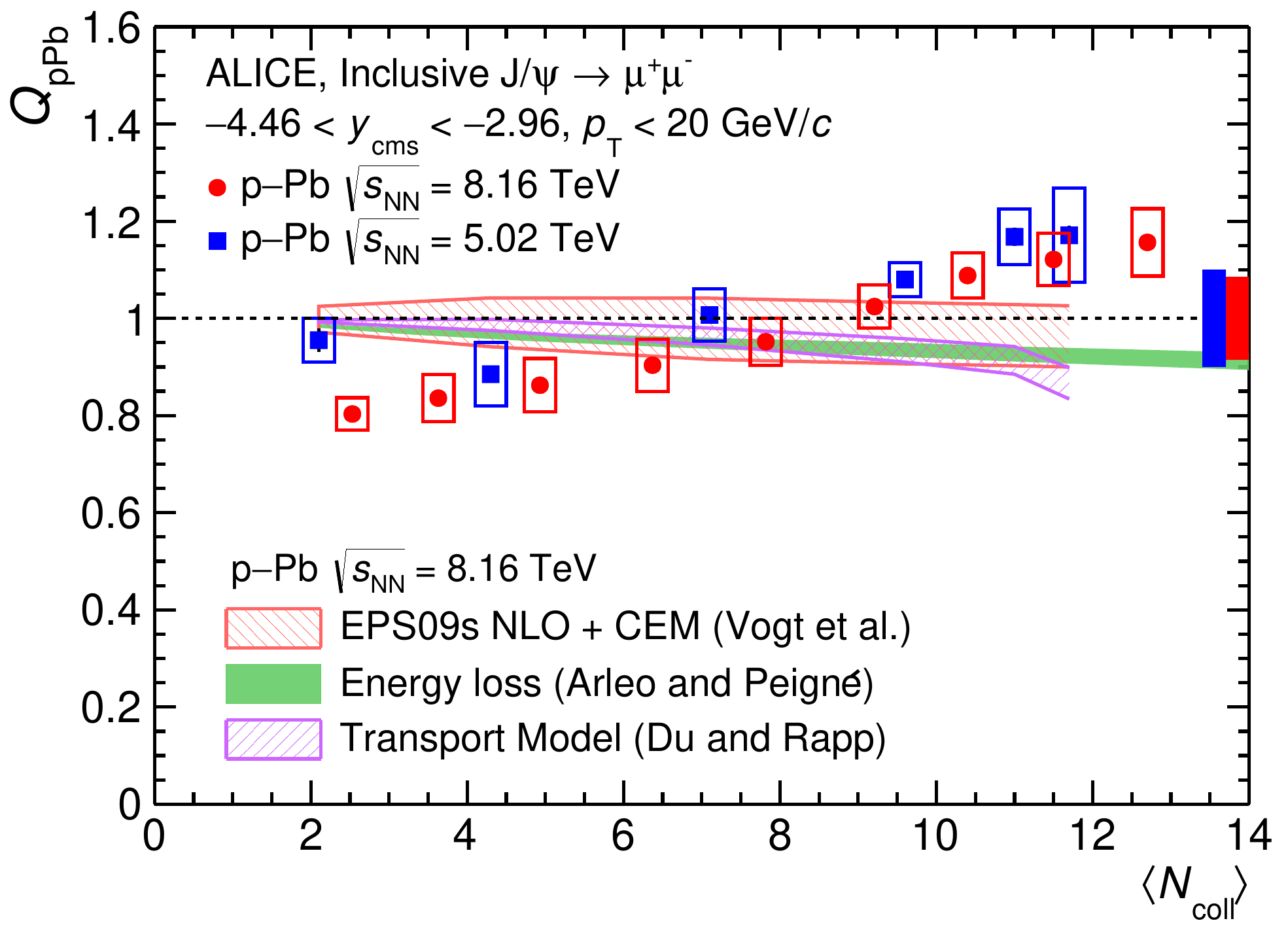}
\includegraphics[width=0.5\columnwidth]{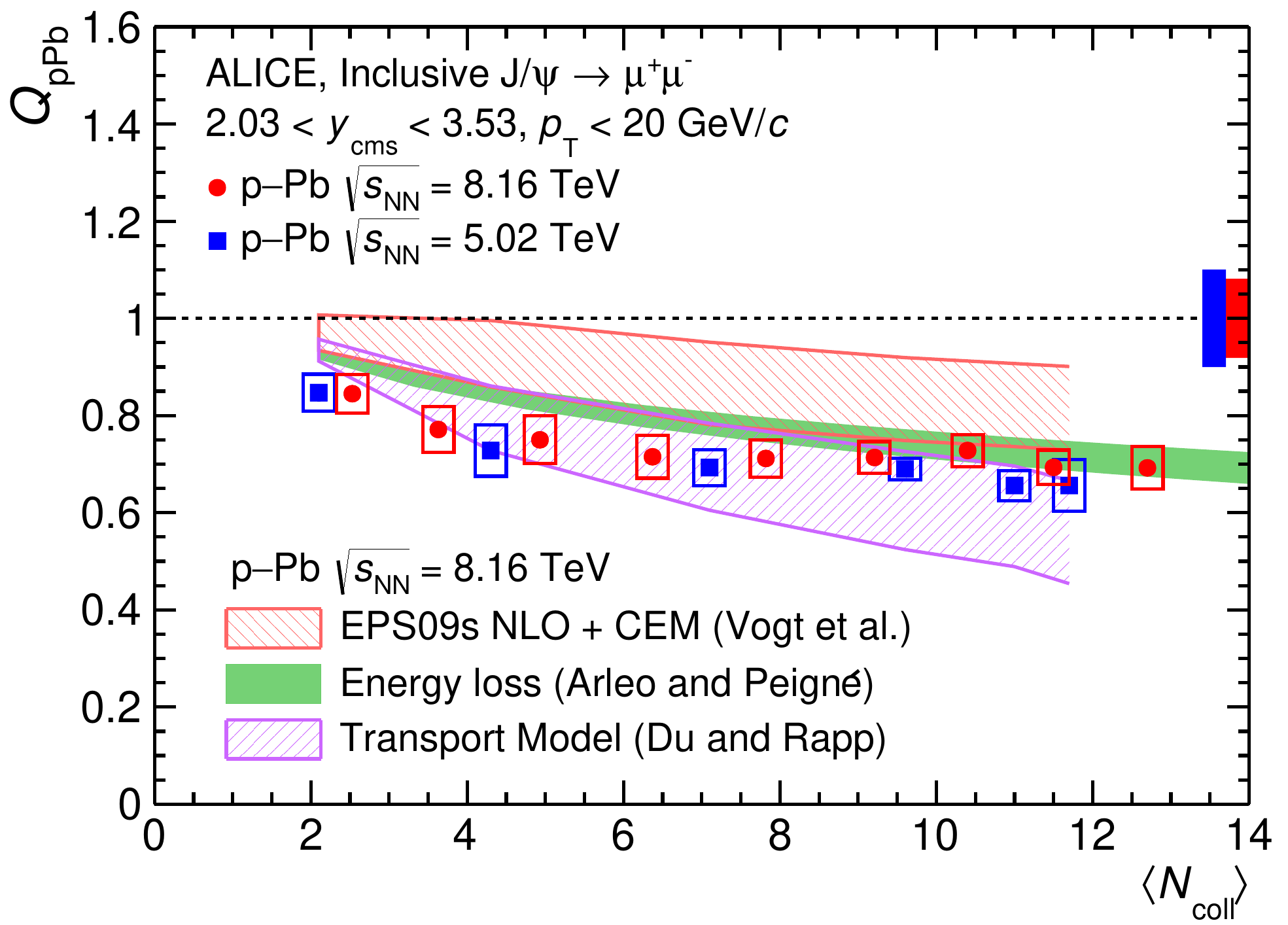}
\caption{\label{figQpPb2}Inclusive J/$\psi$ $Q_{\rm pPb}$ as a function of $\langle N_{\rm coll} \rangle$ at backward (left) and forward (right) rapidity in p--Pb collisions at $\sqrt{s_{\rm NN}} = 8.16$ TeV compared with the  results at $\sqrt{s_{\rm NN}} = 5.02$ TeV~\cite{Adam:2015jsa} and theoretical models~\cite{McGlinchey:2012bp, Arleo:2014oha,Du:2015wha}. The vertical error bars represent the statistical uncertainties and the boxes around the data points the uncorrelated systematic uncertainties. The boxes centered at $Q_{\rm pPb} = 1$ represent the systematic uncertainties correlated over centrality.}
\end{figure}

Figure~\ref{figQpPb2} shows the \pt-integrated $Q_{\rm pPb}$ of J/$\psi$ as a function of $\langle N_{\rm coll} \rangle$ in p--Pb collisions at $\sqrt{s_{\rm NN}} = 8.16$ TeV at backward and forward rapidity.   
At forward $y$, the production of inclusive J/$\psi$ in p--Pb collisions is suppressed with respect to expectations from pp collisions for all centrality classes.
Furthermore, $Q_{\rm pPb}$ decreases with increasing collision centrality from a value of $0.85 \pm 0.02 {\rm (stat.)} \pm 0.03 {\rm (syst.)}$ for the 80--90\% centrality class to $0.69 \pm 0.01 {\rm (stat.)} \pm 0.04 {\rm (syst.)}$ for the 2--10\% centrality class.
At backward $y$, on the contrary, a significant suppression is seen for the most peripheral collisions ($Q_{\rm pPb}^{80-90\%}  = 0.80 \pm 0.02 {\rm (stat.)} \pm 0.03 {\rm (syst.)}$) with $Q_{\rm pPb}$ increasing with increasing centrality and reaching values above unity for the most central collisions ($Q_{\rm pPb}^{2-10\%} = 1.16 \pm 0.01 {\rm (stat.)} \pm 0.07 {\rm (syst.)}$).
The $Q_{\rm pPb}$ as a function of $\langle N_{\rm coll} \rangle$ is compared with the  results at $\sqrt{s_{\rm NN}} = 5.02$ TeV~\cite{Adam:2015jsa}.
No strong dependence with the energy of the collision is observed in the two rapidity intervals.

Three model calculations are also shown in Fig.~\ref{figQpPb2} for comparison.
First, a next-to-leading order (NLO) Colour Evaporation Model (CEM)~\cite{McGlinchey:2012bp} using the EPS09 parameterisation of the nuclear modification of the gluon PDF at NLO is shown and denoted as ``EPS09s NLO + CEM". 
The band represents the systematic uncertainty of the calculation, which is dominated by the uncertainty of the EPS09 parameterisation. The second one is the energy loss model that was described in the Section 4.2.
Finally, the third one is a transport model~\cite{Du:2015wha} based on a thermal-rate equation framework, which implements the dissociation of charmonia in a hadron resonance gas. 
The fireball evolution implemented in this model includes the transition from a short QGP phase into the hadron resonance gas, through a mixed phase. 
The model uses a c$\overline{\rm c}$ production cross section ${\rm d}\sigma_{\rm c\overline{c}}/{\rm d}y = 0.57$~mb and a prompt J/$\psi$ production cross section in pp collisions of ${\rm d}\sigma^{\rm pp}_{{\rm J}/\psi}/{\rm d}y = 3.35$~$\mu$b.
Shadowing effects are included through the EPS09 parameterisation. 
In this case, the upper (lower) limit of this calculation corresponds to a 10\% (25\%) contribution of nuclear shadowing. 
The three models provide a satisfactory description of the centrality dependence of the inclusive J/$\psi$  $Q_{\rm pPb}$ at forward rapidity. 
However, at backward rapidity, all three calculations show a slightly decreasing trend of  $Q_{\rm pPb}$ with increasing centrality that appears opposite to the one indicated by the data.

It is worth noting that the model calculations discussed above are for prompt J/$\psi$ while the inclusive measurements contain a contribution from non-prompt J/$\psi$ too.
The $Q_{\rm pPb}^{\rm prompt}$ can be extracted from $Q_{\rm pPb}^{\rm incl}$ using the relation $Q_{\rm pPb}^{\rm prompt} = Q_{\rm pPb}^{\rm incl} + f_{\rm B} \cdot (Q_{\rm pPb}^{\rm incl} - Q_{\rm pPb}^{\text{non-prompt}}$), where $ f_{\rm B}$ is the ratio of non-prompt to prompt J/$\psi$ production cross sections in pp collisions and $Q_{\rm pPb}^{\text{non-prompt}}$ is the nuclear modification factor of the non-prompt J/$\psi$ mesons. The value of $f_{\rm B}$ is about 0.12 and was calculated from the LHCb measurements for $2 < y < 4.5$ and $p_{\rm T} < 14$ GeV/$c$ in pp collisions at $\sqrt{s} = 8$ TeV~\cite{Aaij:2013yaa}. 
The nuclear modification factor of non-prompt J/$\psi$ with $p_{\rm T} < 14$ GeV/$c$ measured by LHCb varies between $0.97 \pm 0.11$ and $1.10 \pm 0.13$ ($0.80 \pm 0.07$ and $0.89 \pm 0.09$) in the backward  (forward) rapidity interval of interest in p--Pb collisions at $\sqrt{s_{\rm NN}} = 8.16$ TeV~\cite{Aaij:2017ypPb}.
However, the centrality dependence of $Q_{\rm pPb}^{\text{non-prompt}}$ has not been measured yet, therefore $Q_{\rm pPb}^{\rm prompt}$ is estimated for each centrality class under the two extreme hypotheses of $Q_{\rm pPb}^{\text{non-prompt}} = 0.75$ (0.85) and $Q_{\rm pPb}^{\text{non-prompt}} = 0.95$ (1.25) at forward (backward) rapidity.
These hypotheses correspond to the same relative variation of  $Q_{\rm pPb}^{\text{non-prompt}}$ with centrality as observed for $Q_{\rm pPb}^{\rm incl}$.
The differences between $Q_{\rm pPb}^{\rm prompt}$ and $Q_{\rm pPb}^{\rm incl}$ are found to be below 9\% and 5\% at backward and forward rapidity, respectively. 
Thus, the conclusions outlined above, and also in the following, are expected to remain valid also for prompt J/$\psi$.

\subsection{Centrality-differential inclusive \jpsi $Q_{\rm pPb}$ as a function of \pt}
\begin{figure}[t]
\includegraphics[width=0.5\columnwidth]{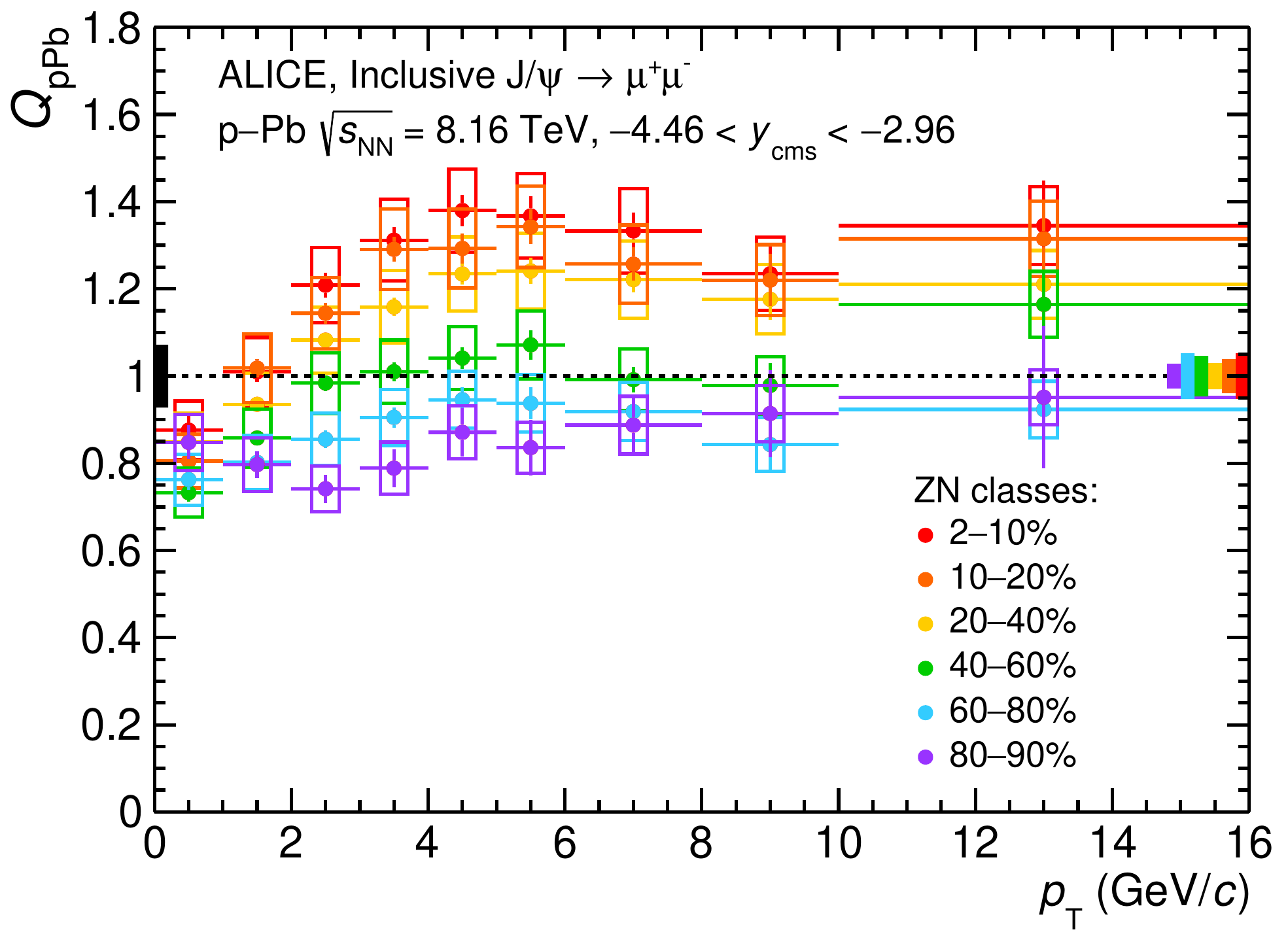}
\includegraphics[width=0.5\columnwidth]{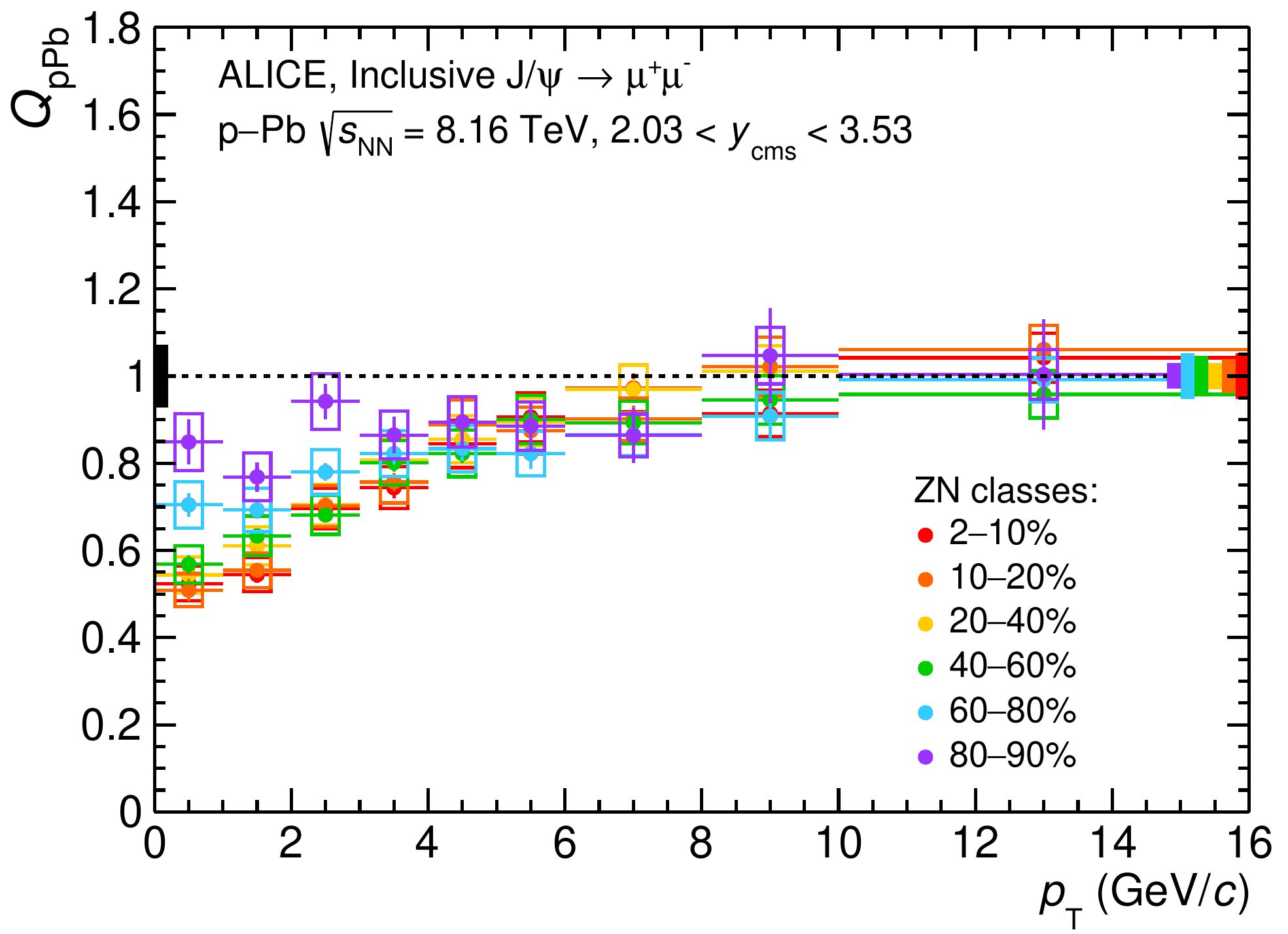}
\caption{\label{figQpPbvsptall}Inclusive J/$\psi$ $Q_{\rm pPb}$ as a function of $p_{\rm T}$ for various centrality classes at backward (left) and forward (right) rapidity. The vertical error bars represent the statistical uncertainties and the open boxes around the data points the uncorrelated systematic uncertainties. The full coloured boxes centered at $Q_{\rm pPb} = 1$ on the right are the systematic uncertainties due to pile-up, $\langle T_{\rm pPb} \rangle$, and $F_{\rm norm}$, while the full black box on the left of each panel shows the global systematic uncertainties.}
\end{figure}

Figure~\ref{figQpPbvsptall} shows the inclusive J/$\psi$ $Q_{\rm pPb}$ as a function of $p_{\rm T}$ at backward and forward rapidity for all centrality classes considered in this analysis. 
At backward rapidity, a slight suppression is seen at low $p_{\rm T}$ for all centralities.
However, while almost no $p_{\rm T}$ dependence is observed for the most peripheral collisions, for all other centralities $Q_{\rm pPb}$ increases with $p_{\rm T}$ reaching a plateau for $p_{\rm T} \gtrsim 5$ GeV/$c$, with the value of the plateau being largest for more central collisions.
For the three most central classes, $Q_{\rm pPb}$ is above unity for $p_{\rm T} \gtrsim 2$ GeV/$c$.
A similar behavior is also observed for prompt D mesons at midrapidity ($-0.96 < y_{\rm cms} < 0.04$) measured in p--Pb collisions at  $\sqrt{s_{\rm NN}} = 5.02$~TeV~\cite{Acharya:2019mno}.
In contrast, at forward rapidity, $Q_{\rm pPb}$ is below or consistent with unity for all $p_{\rm T}$ in all centrality classes. 
At low $p_{\rm T}$, a centrality dependent hierarchy of $Q_{\rm pPb}$ is observed, showing a stronger suppression in central collisions compared to peripheral ones. 
For all centralities, $Q_{\rm pPb}$ smoothly increases towards unity at high $p_{\rm T}$.

\begin{figure}[b]
\includegraphics[width=0.5\columnwidth]{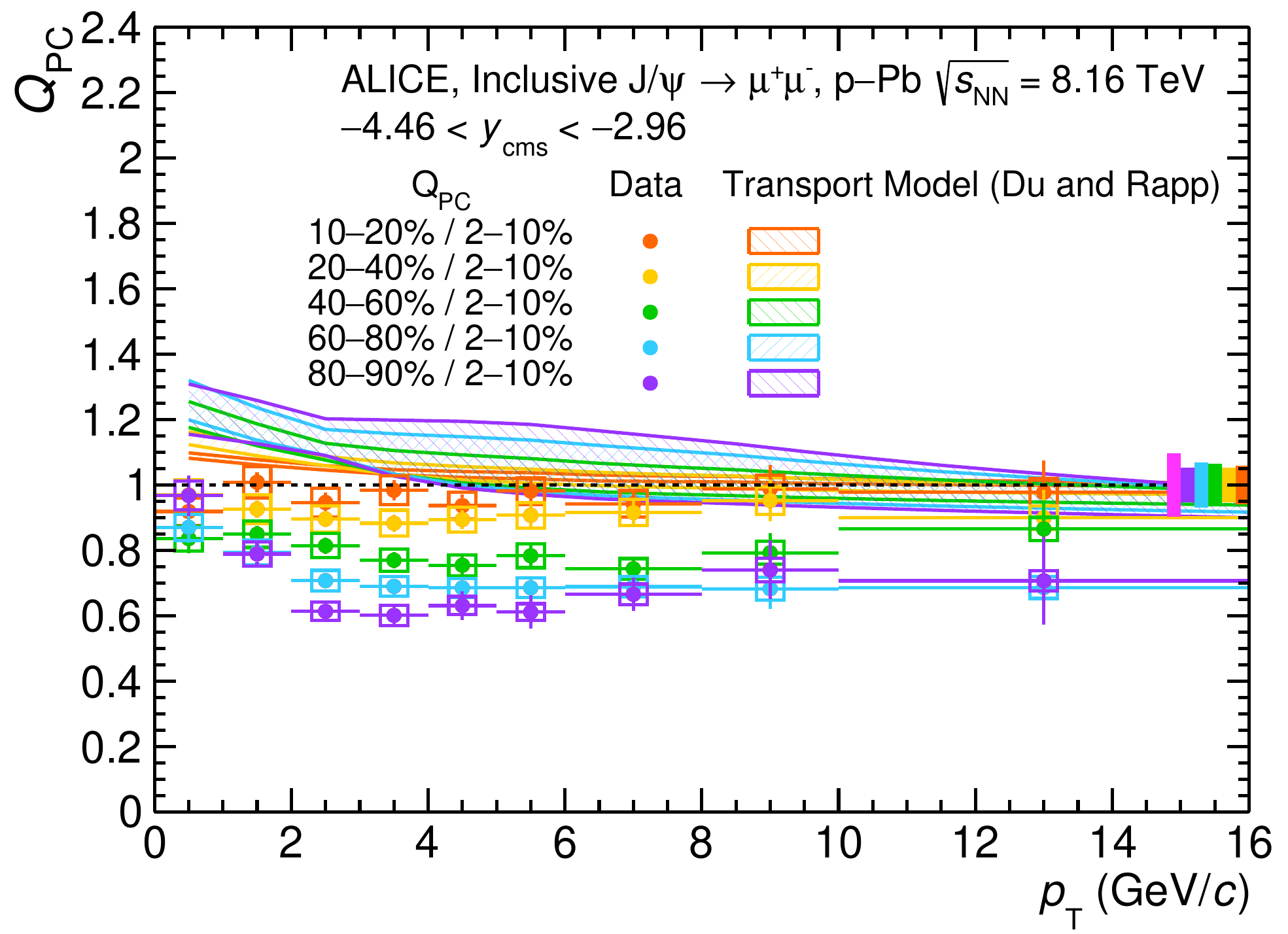}
\includegraphics[width=0.5\columnwidth]{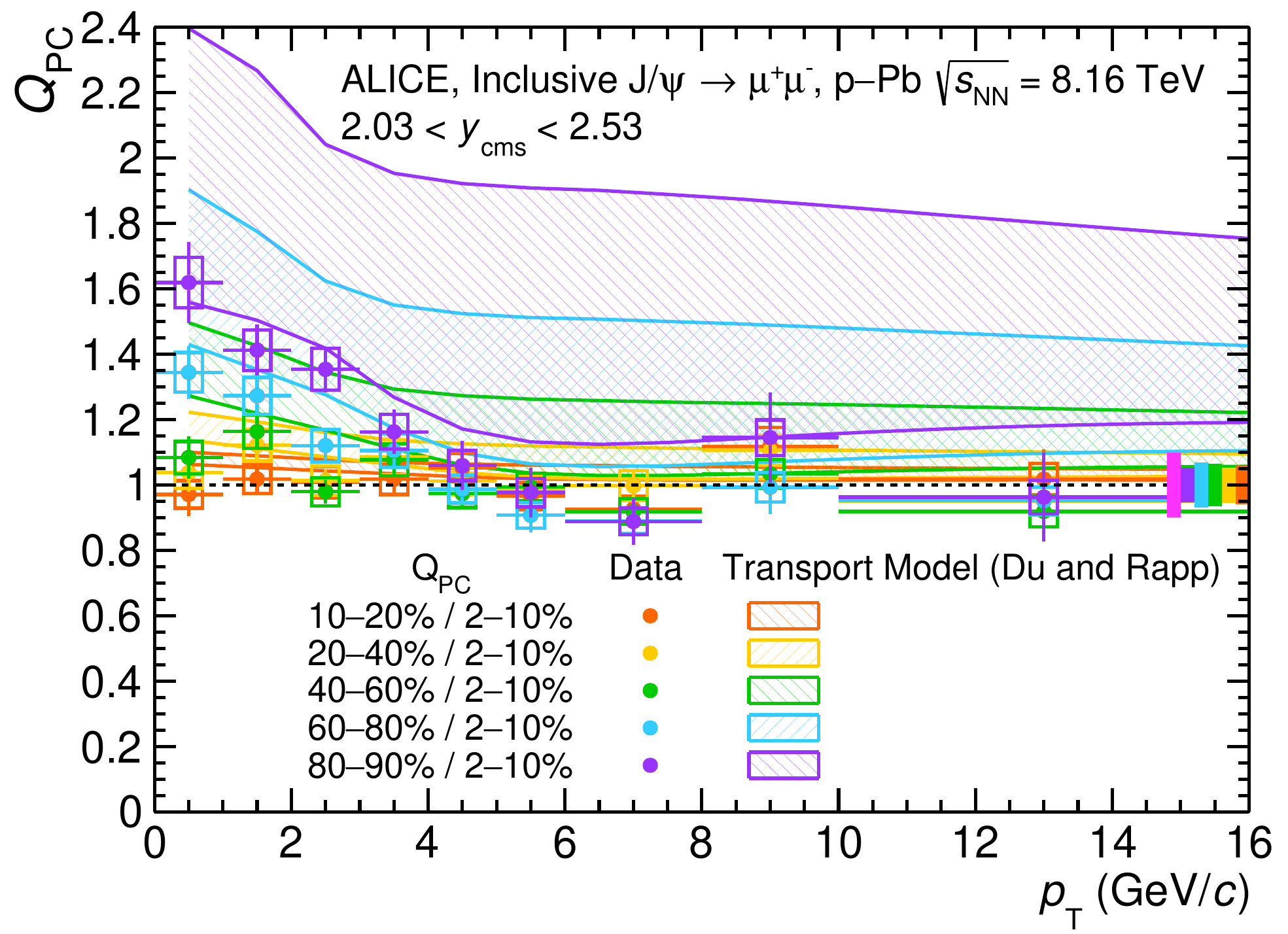}
\caption{\label{figQPCvsptall}Inclusive J/$\psi$ $Q_{\rm PC}$ as a function of $p_{\rm T}$ for various centrality classes at backward (left) and forward (right) rapidity compared to the theoretical calculations~\cite{Du:2015wha}. The vertical error bars represent the statistical uncertainties and the boxes around the data points the uncorrelated systematic uncertainties. The boxes centered at $Q_{\rm PC} = 1$ are the systematic uncertainties due to pile-up, $\langle T_{\rm pPb} \rangle$, and $F_{\rm norm}$.}
\end{figure}

The different shapes of the evolution of $Q_{\rm pPb}$ with $p_{\rm T}$ for the various centralities can be better appreciated by forming the ratio $Q_{\rm PC}$ of the $Q_{\rm pPb}$ in peripheral to that in central collisions. 
Figure~\ref{figQPCvsptall} shows the inclusive J/$\psi$ $Q_{\rm PC}$ as a function of $p_{\rm T}$ at backward and forward rapidity. 
The centrality-correlated systematic uncertainties cancel when calculating the ratio. 
The  $Q_{\rm PC}$ could, therefore, provide stronger constraints to the theoretical calculations.
Transport model calculations by Du et al.~\cite{Du:2015wha} are also shown in Fig.~\ref{figQPCvsptall} for comparison.
At backward rapidity, the model calculations tend to overestimate the measured $Q_{\rm PC}$ for all centrality classes. The centrality dependent hierarchy of the measured $Q_{\rm PC}$ is also not reproduced by the model calculation.
At forward rapidity, the transport model calculations qualitatively describe the $p_{\rm T}$ and centrality dependence of the inclusive J/$\psi$ $Q_{\rm PC}$, but do systematically overestimate the measurements.

\begin{figure}[H]
\includegraphics[width=0.5\columnwidth]{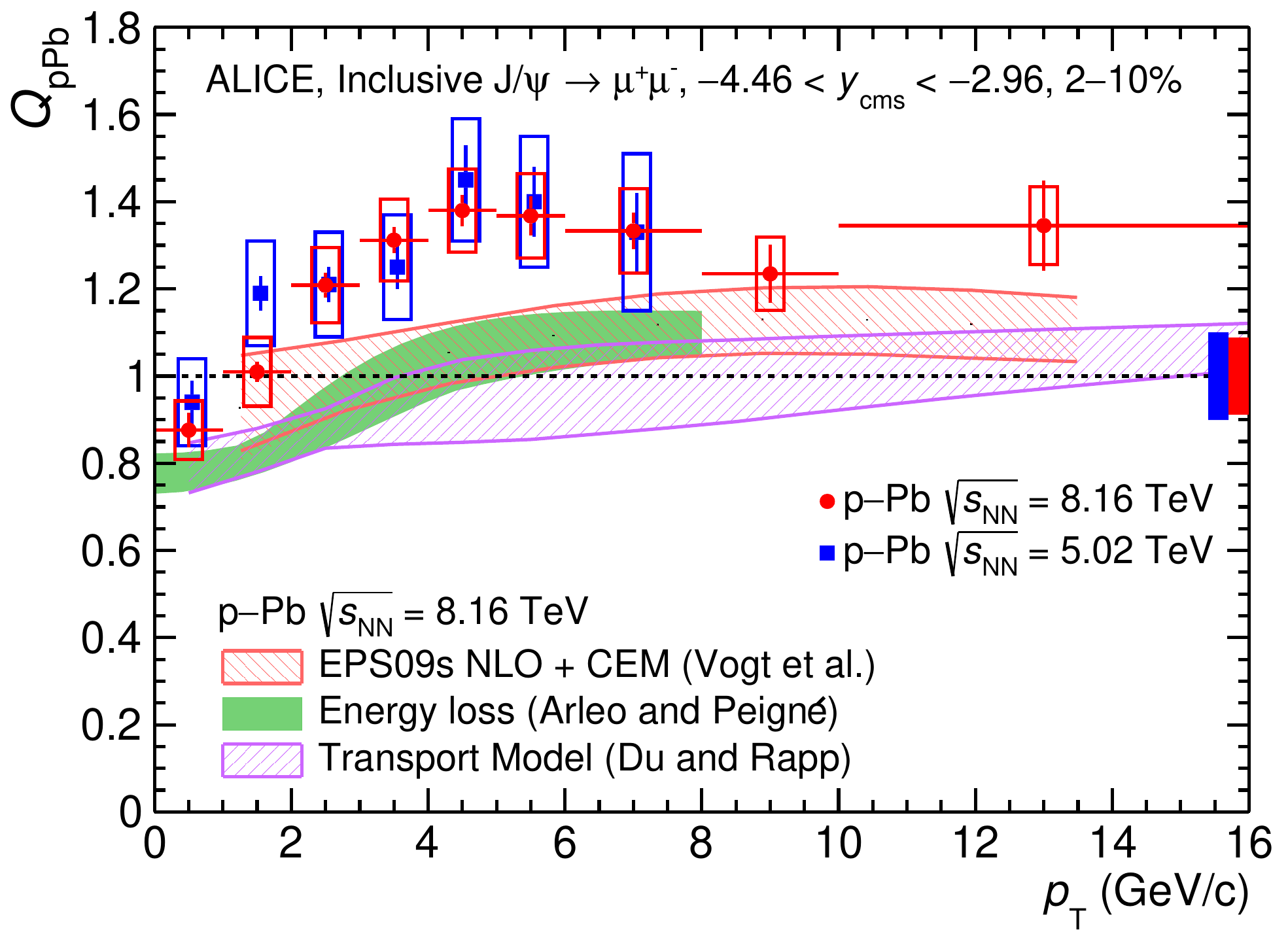}
\includegraphics[width=0.5\columnwidth]{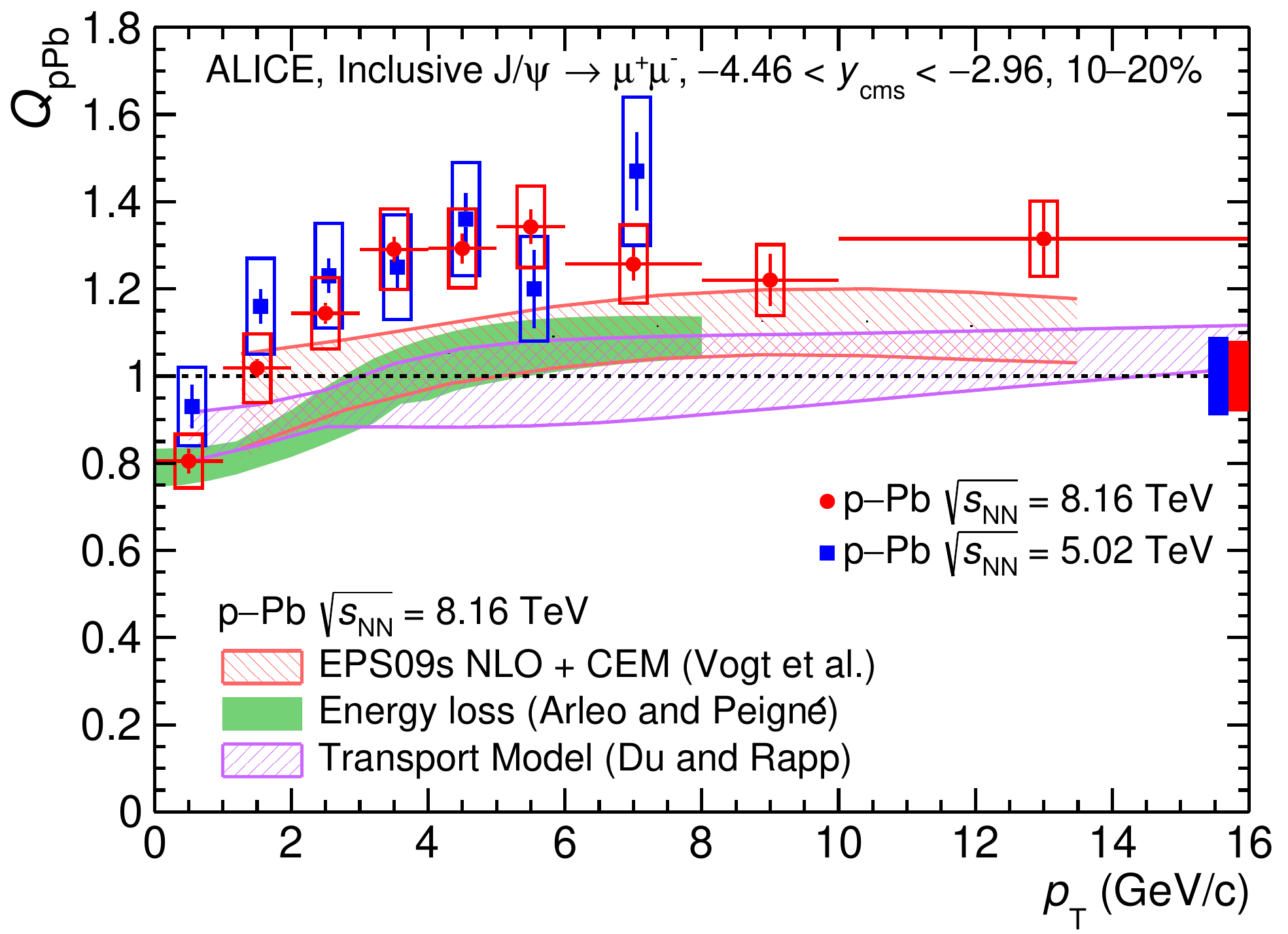}\\
\includegraphics[width=0.5\columnwidth]{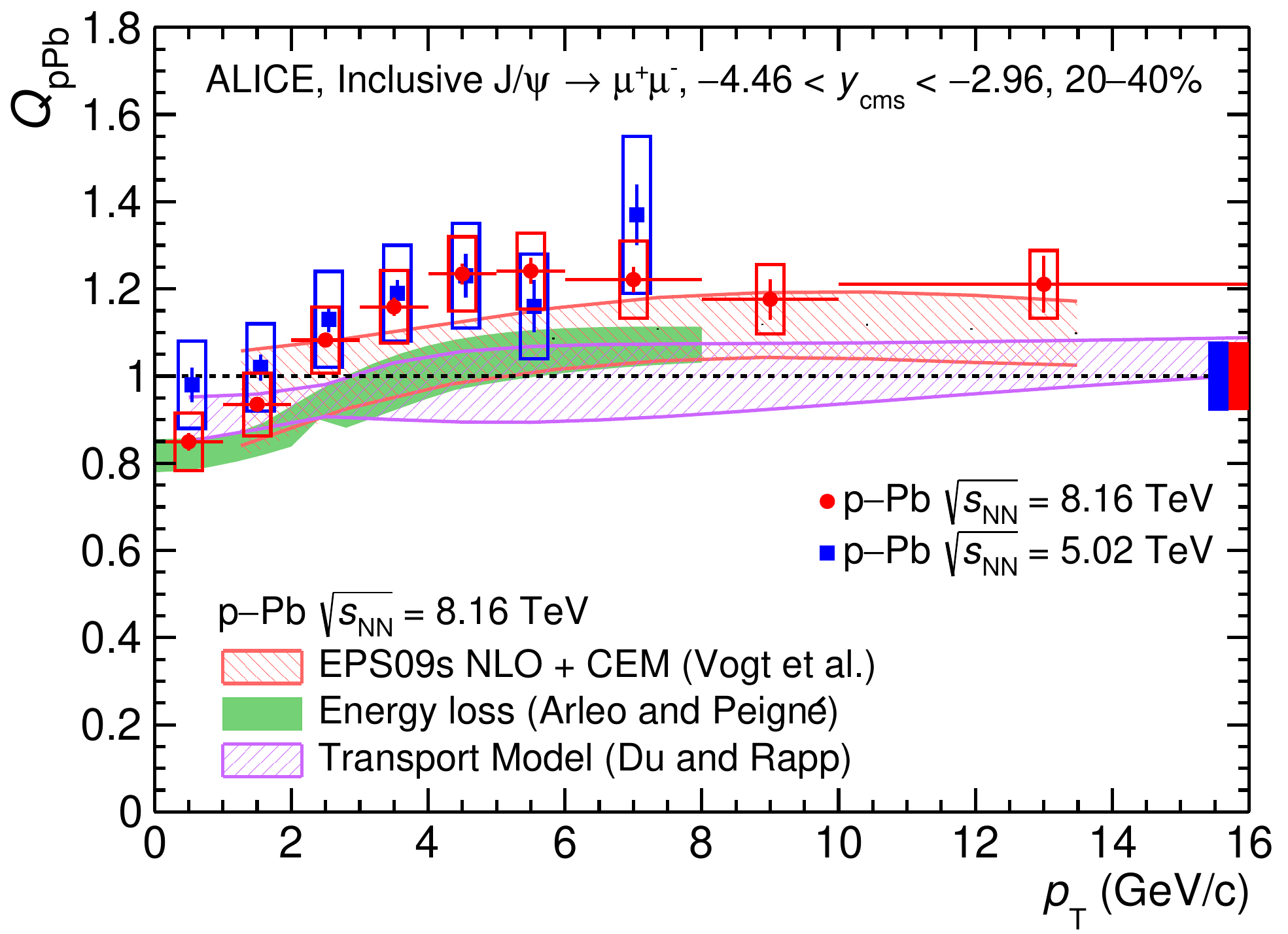}
\includegraphics[width=0.5\columnwidth]{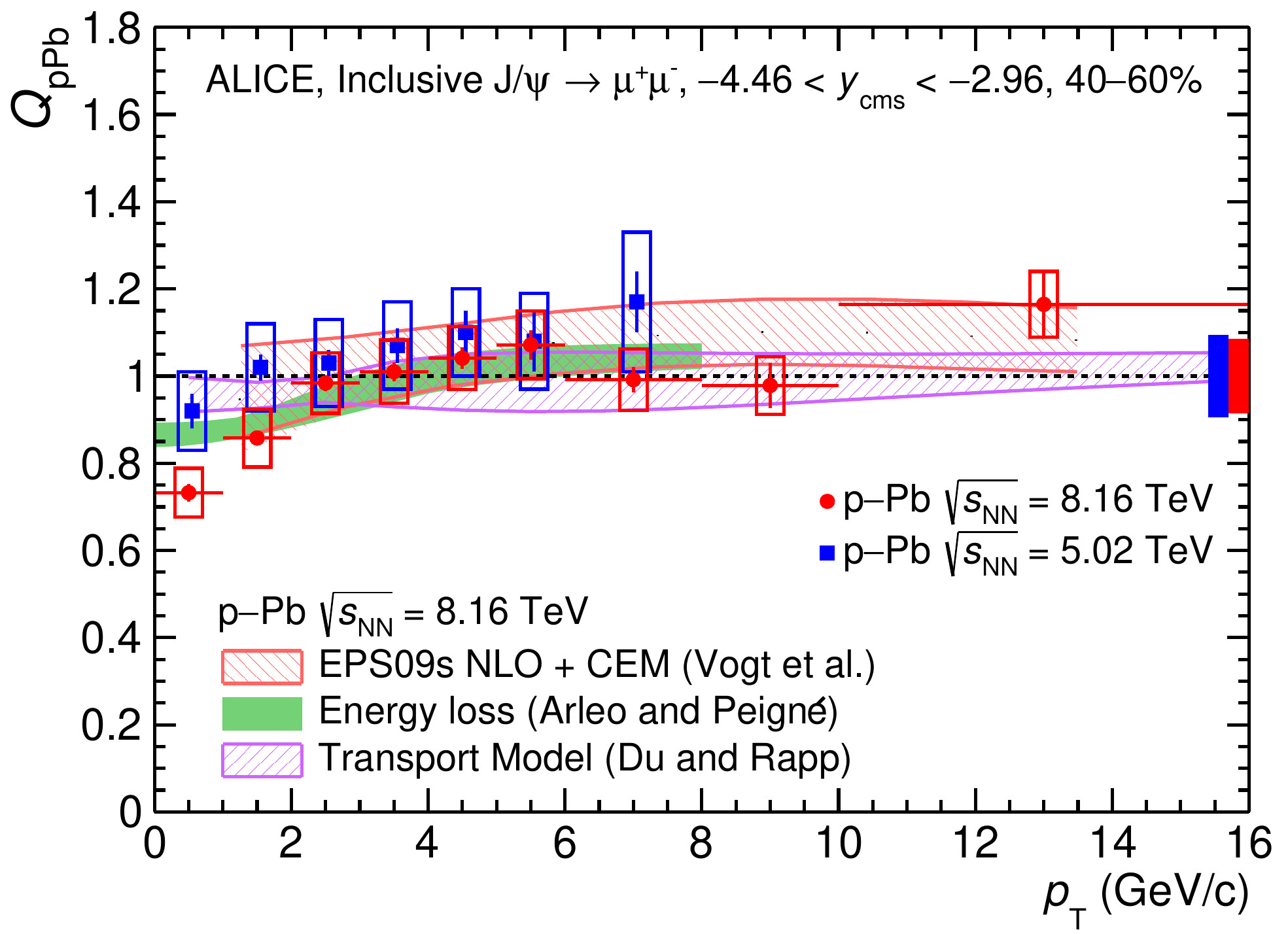}\\
\includegraphics[width=0.5\columnwidth]{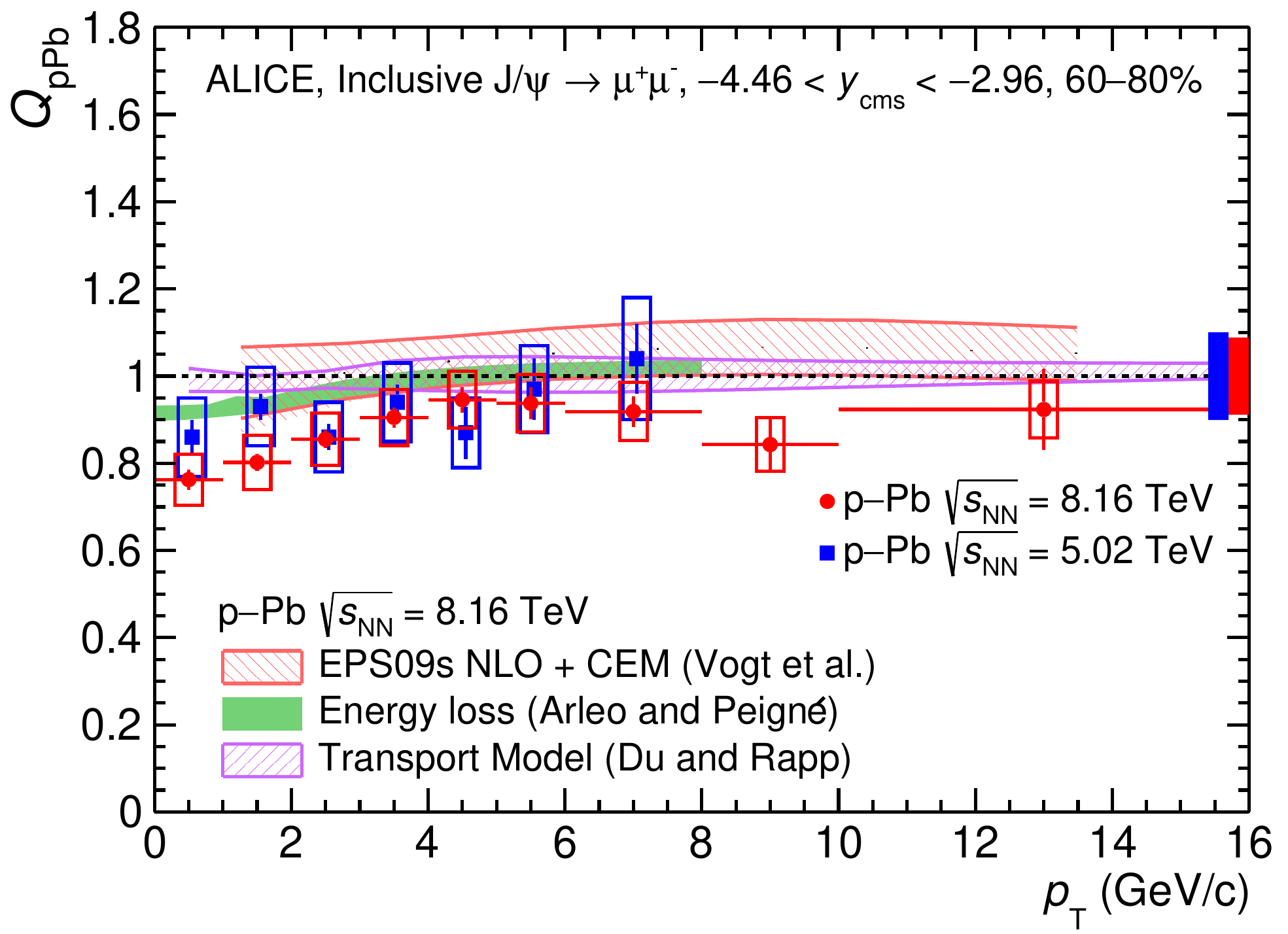}
\includegraphics[width=0.5\columnwidth]{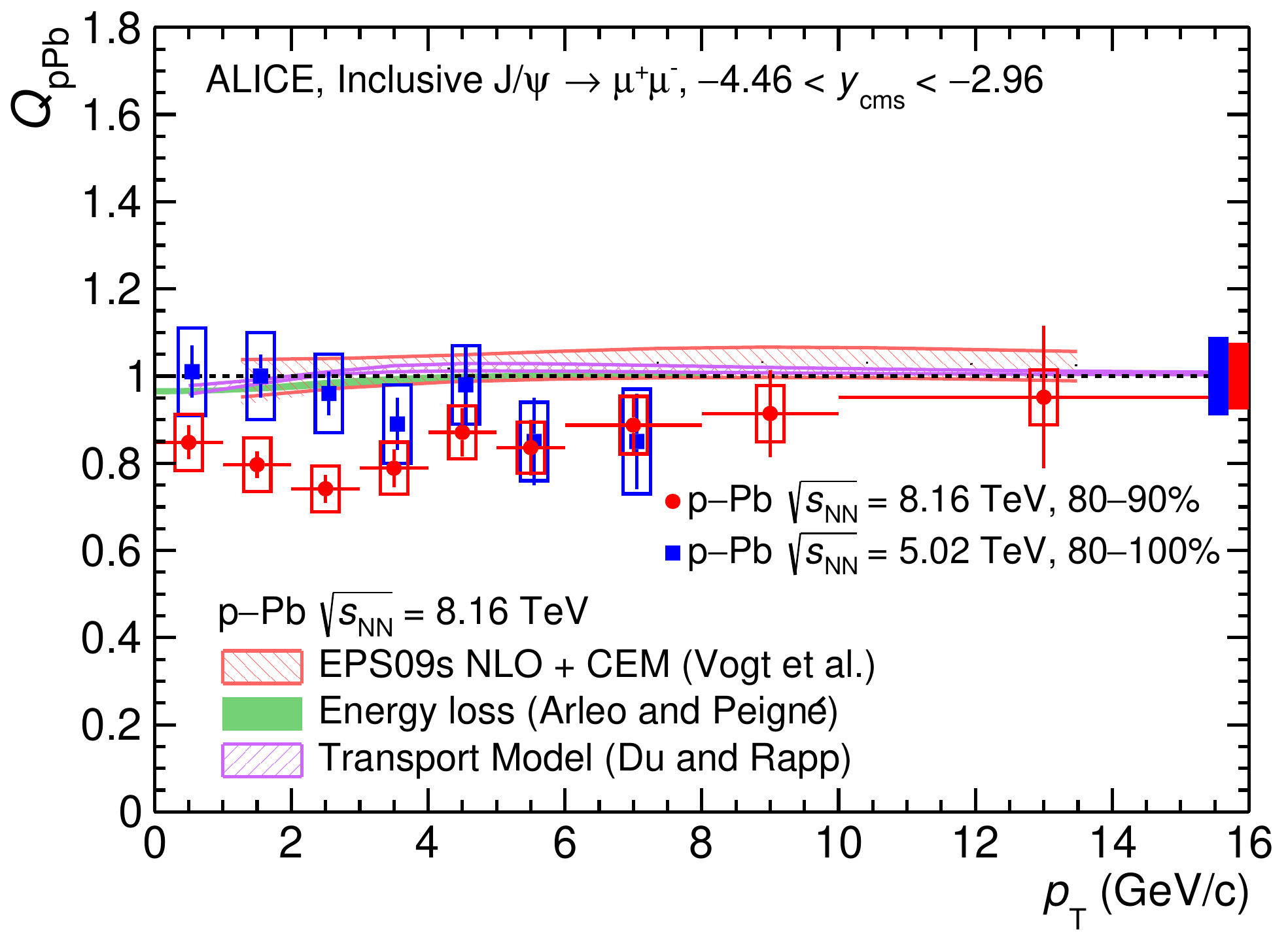}\\
\caption{\label{figQpPbvsptfwd}Inclusive J/$\psi$ $Q_{\rm pPb}$ as a function of $p_{\rm T}$ for 2--10\%, 10--20\%, 20--40\%, 40--60\%, 60--80\%, and 80--90\% ZN centrality classes at backward rapidity in p--Pb collisions at $\sqrt{s_{\rm NN}} = 8.16$ TeV compared with the  results at $\sqrt{s_{\rm NN}} = 5.02$ TeV~\cite{Adam:2015jsa} and with the theoretical calculations~\cite{McGlinchey:2012bp, Arleo:2014oha,Du:2015wha}. The vertical error bars show the statistical uncertainties, the open boxes the uncorrelated systematic uncertainties, and the full boxes centered at $Q_{\rm pPb} = 1$ the correlated systematic uncertainties.}
\end{figure}

\begin{figure}[H]
\includegraphics[width=0.5\columnwidth]{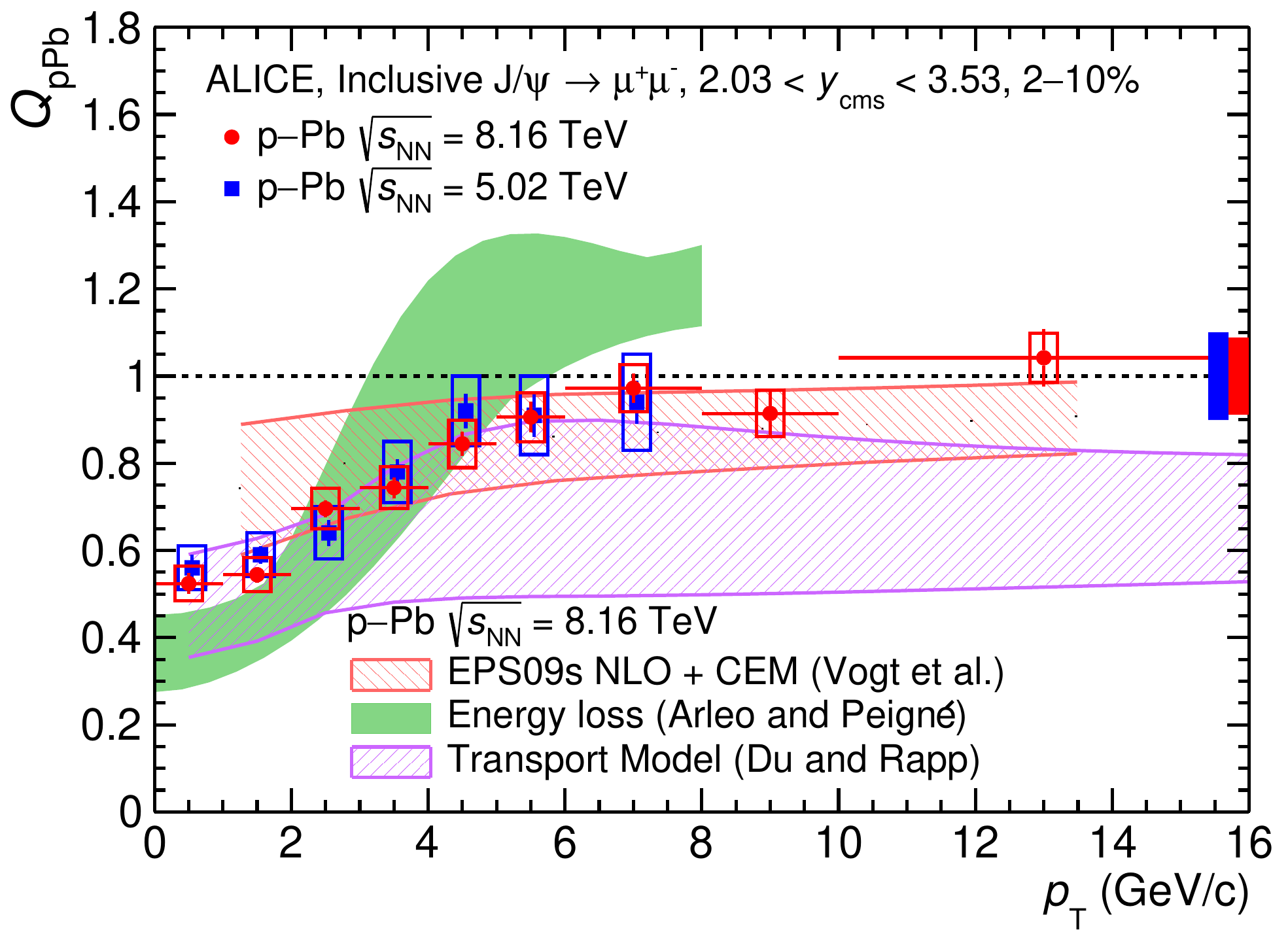}
\includegraphics[width=0.5\columnwidth]{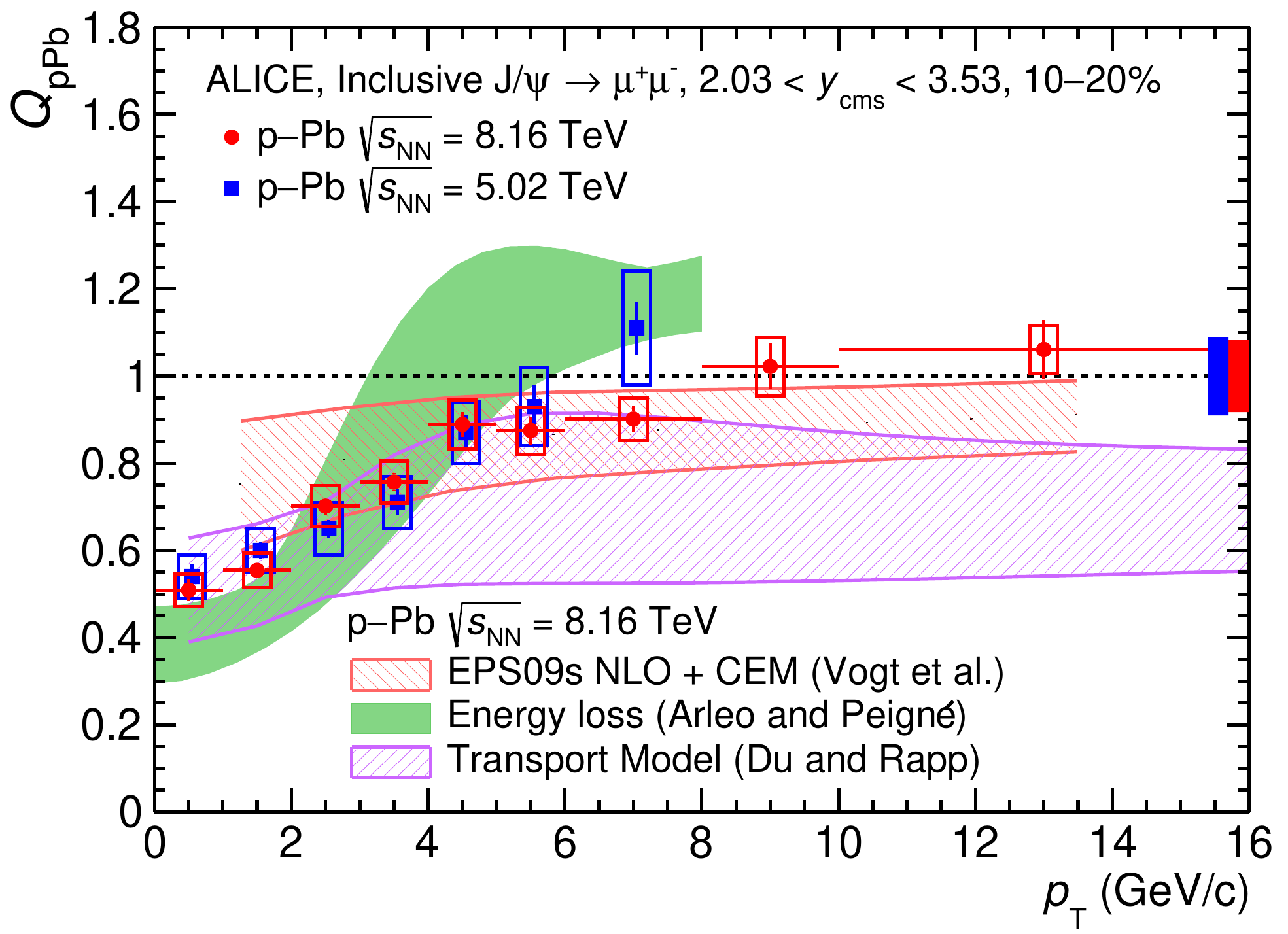}\\
\includegraphics[width=0.5\columnwidth]{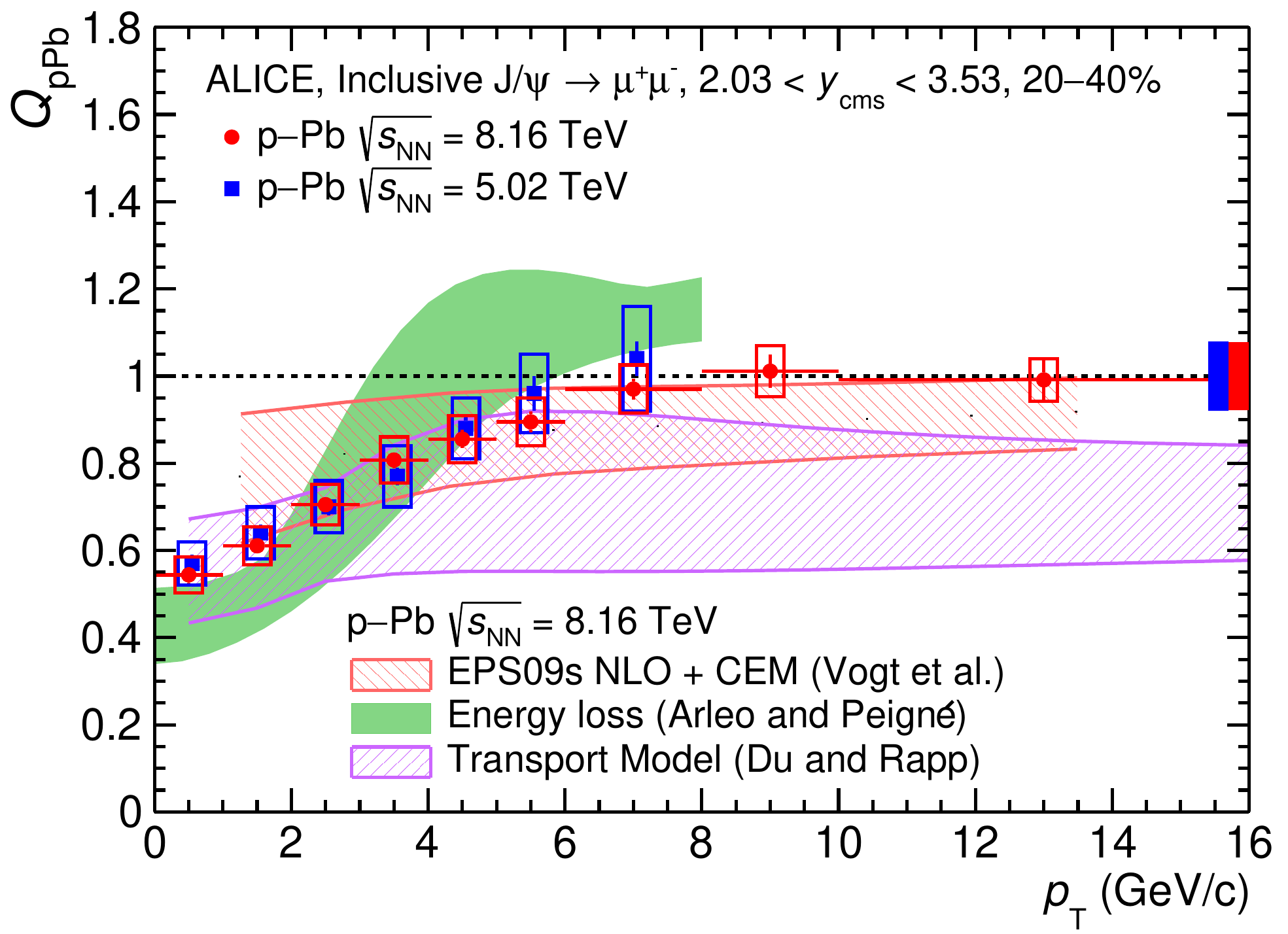}
\includegraphics[width=0.5\columnwidth]{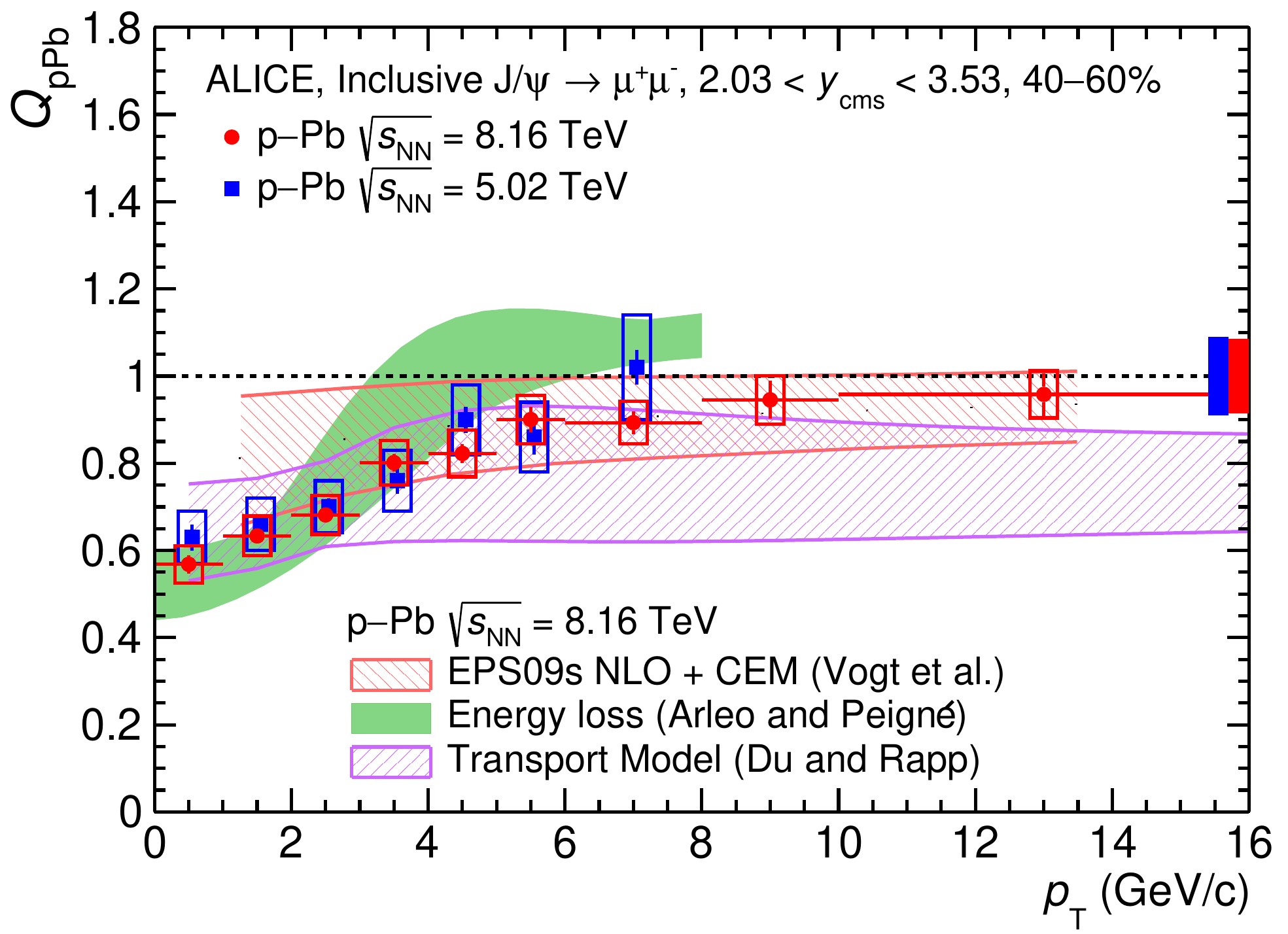}\\
\includegraphics[width=0.5\columnwidth]{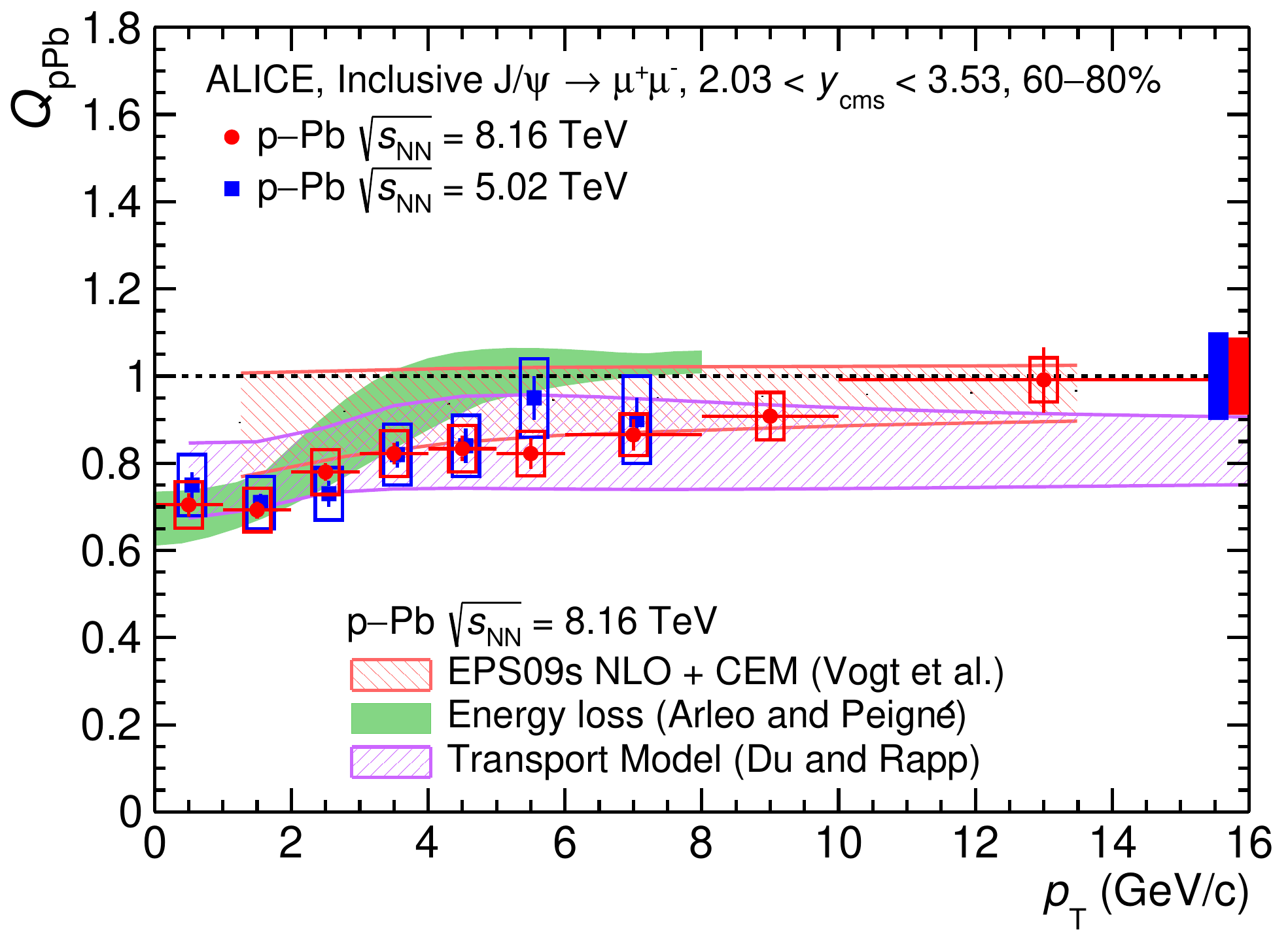}
\includegraphics[width=0.5\columnwidth]{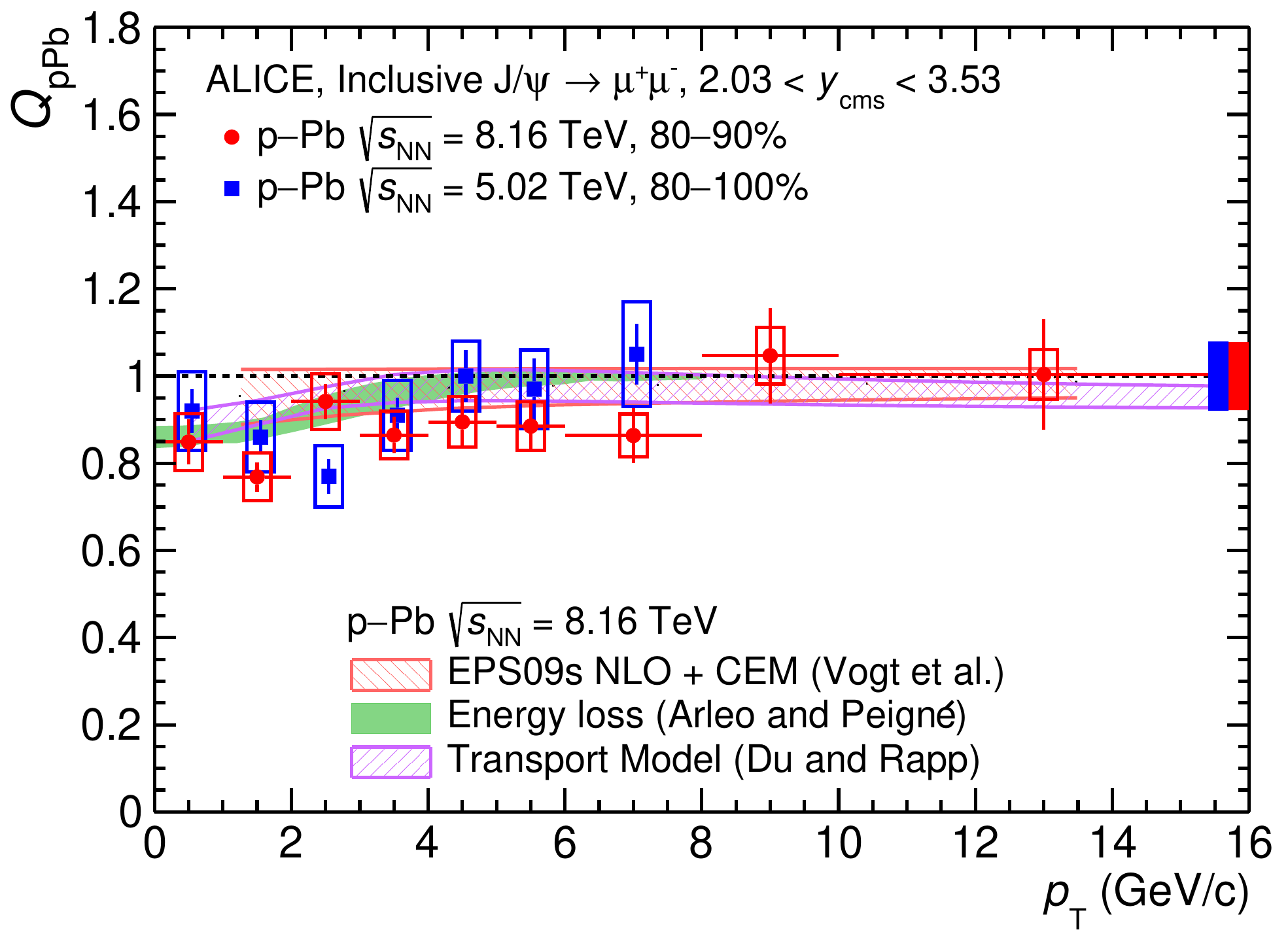}\\
\caption{\label{figQpPbvsptbck}Inclusive J/$\psi$ $Q_{\rm pPb}$ as a function of $p_{\rm T}$ for 2--10\%, 10--20\%, 20--40\%, 40--60\%, 60--80\%, and 80--90\% ZN centrality classes at forward rapidity in p--Pb collisions at $\sqrt{s_{\rm NN}} = 8.16$ TeV compared with the  results at $\sqrt{s_{\rm NN}} = 5.02$ TeV~\cite{Adam:2015jsa} and with theoretical calculations~\cite{McGlinchey:2012bp, Arleo:2014oha,Du:2015wha}. The vertical error bars show the statistical uncertainties, the open boxes the uncorrelated systematic uncertainties, and the full boxes centered at $Q_{\rm pPb} = 1$ the correlated systematic uncertainties.}
\end{figure}

The J/$\psi$ $Q_{\rm pPb}$ as a function of $p_{\rm T}$ is shown separately for the six centrality classes in Figs.~\ref{figQpPbvsptfwd} and~\ref{figQpPbvsptbck} for the backward and forward rapidity regions and is compared with the  results at $\sqrt{s_{\rm NN}} = 5.02$ TeV~\cite{Adam:2015jsa} and the same model calculations discussed previously. 
The results are similar at both collision energies in the two rapidity ranges, indicating that the mechanisms behind the modification of the J/$\psi$ production in p--Pb collisions do not depend strongly on the collision energy. 
It is worth noting that the $p_{\rm T}$ range is extended up to 16 GeV/$c$ at $\sqrt{s_{\rm NN}} = 8.16$ TeV and that the most peripheral centrality is 80--90\% at the highest energy while it was 80--100\% at the lowest one.

At backward rapidity, the EPS09s NLO + CEM~\cite{McGlinchey:2012bp} calculations show a mild increase of $Q_{\rm pPb}$ with $p_{\rm T}$ for all centralities, but more pronounced towards more central collisions. 
The EPS09s NLO + CEM $Q_{\rm pPb}$ is above unity for all centralities but the strength of the anti-shadowing effect is stronger the more central the collisions are.
The description of the data by the EPS09s NLO + CEM calculations is rather poor, except for the 40--60\% centrality class.
For more central collisions the calculations underestimate the data, but overestimate them for more peripheral collisions.
Similar observations can be drawn from the energy loss~\cite{Arleo:2014oha} calculations, which in the common $p_{\rm T}$ region are compatible with the EPS09s NLO + CEM calculations. 
Only for the more central collisions the $p_{\rm T}$ dependence appears steeper for the energy loss model and closer to the data, but the overall magnitude is lower than the measured $Q_{\rm pPb}$.
The transport model~\cite{Du:2015wha} calculations, which are in general terms lower than the EPS09 + CEM and quite similar to the energy loss ones, only describe the inclusive J/$\psi$ $Q_{\rm pPb}$ in the 40--60\% centrality class, while underestimating it for more central collisions and overestimating it for more peripheral ones.

At forward rapidity, the differences between the EPS09 + CEM and the energy loss calculations are more pronounced.
On the contrary, the transport model calculations are rather similar to the EPS09 + CEM ones, though on the lower edge. 
The uncertainties of the model calculations are also larger at forward than at backward rapidity, especially for the most central collisions.
The description of the data by the EPS09 + CEM calculations is fair for all centralities, especially for $p_{\rm T} \gtrsim 4$ GeV/$c$. 
Below 4~GeV/$c$, the model tends to overestimate the measured $Q_{\rm pPb}$.
The $p_{\rm T}$ dependence of the energy loss calculation appears steeper than that in data, except for the most peripheral class. 
The model tends to underestimate the measured $Q_{\rm pPb}$ at low $p_{\rm T}$ and to overestimate it at high $p_{\rm T}$ in all the other centrality classes.
The transport model describes the data fairly well in all centrality classes for $p_{\rm T} \lesssim 8$~GeV/$c$ but tends to overestimate the $Q_{\rm pPb}$ at higher $p_{\rm T}$.

\subsection{Inclusive $\psi$(2S) to J/$\psi$ ratio and double ratio}

\begin{figure}[b]
\includegraphics[width=0.5\columnwidth]{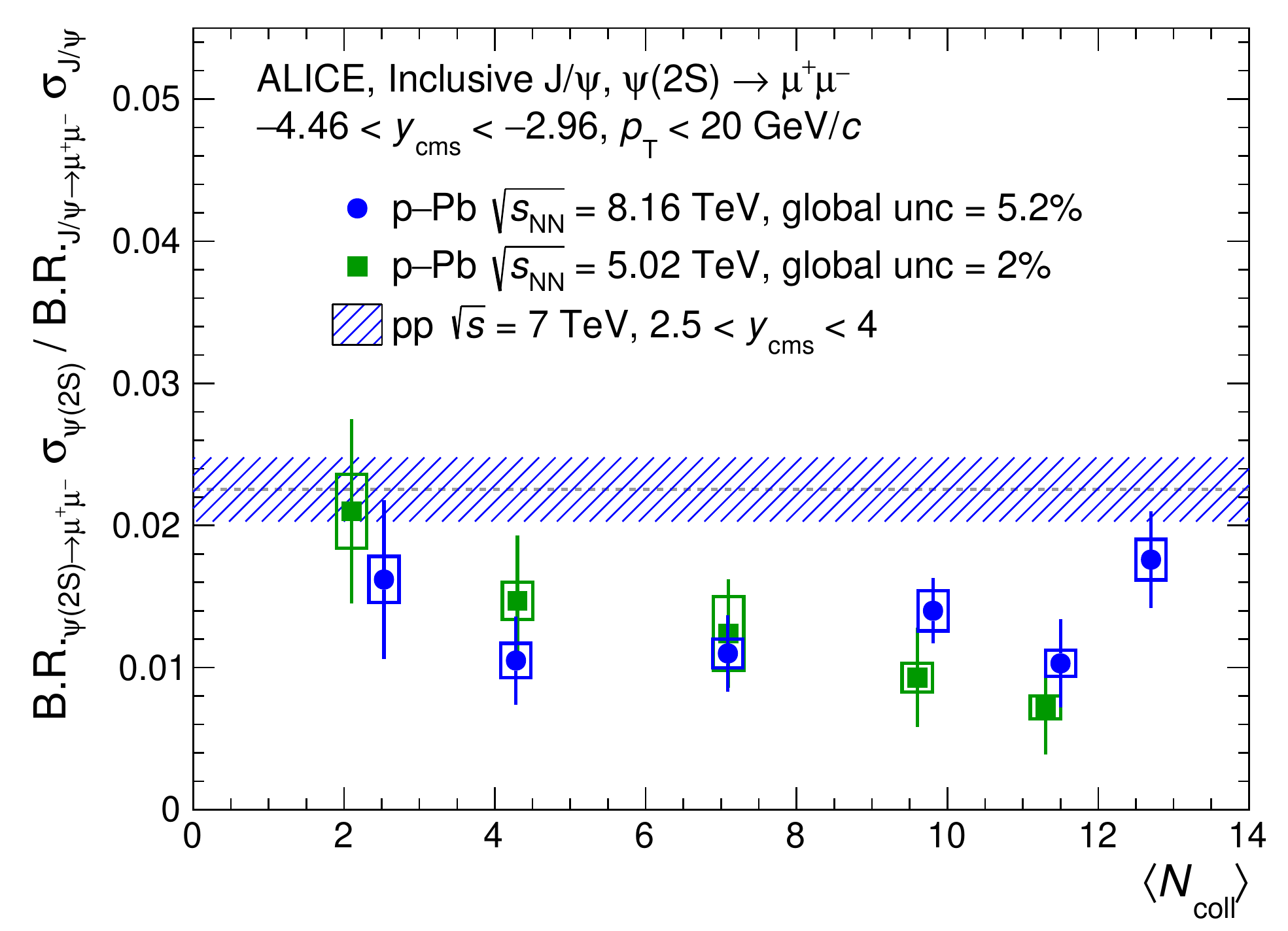}
\includegraphics[width=0.5\columnwidth]{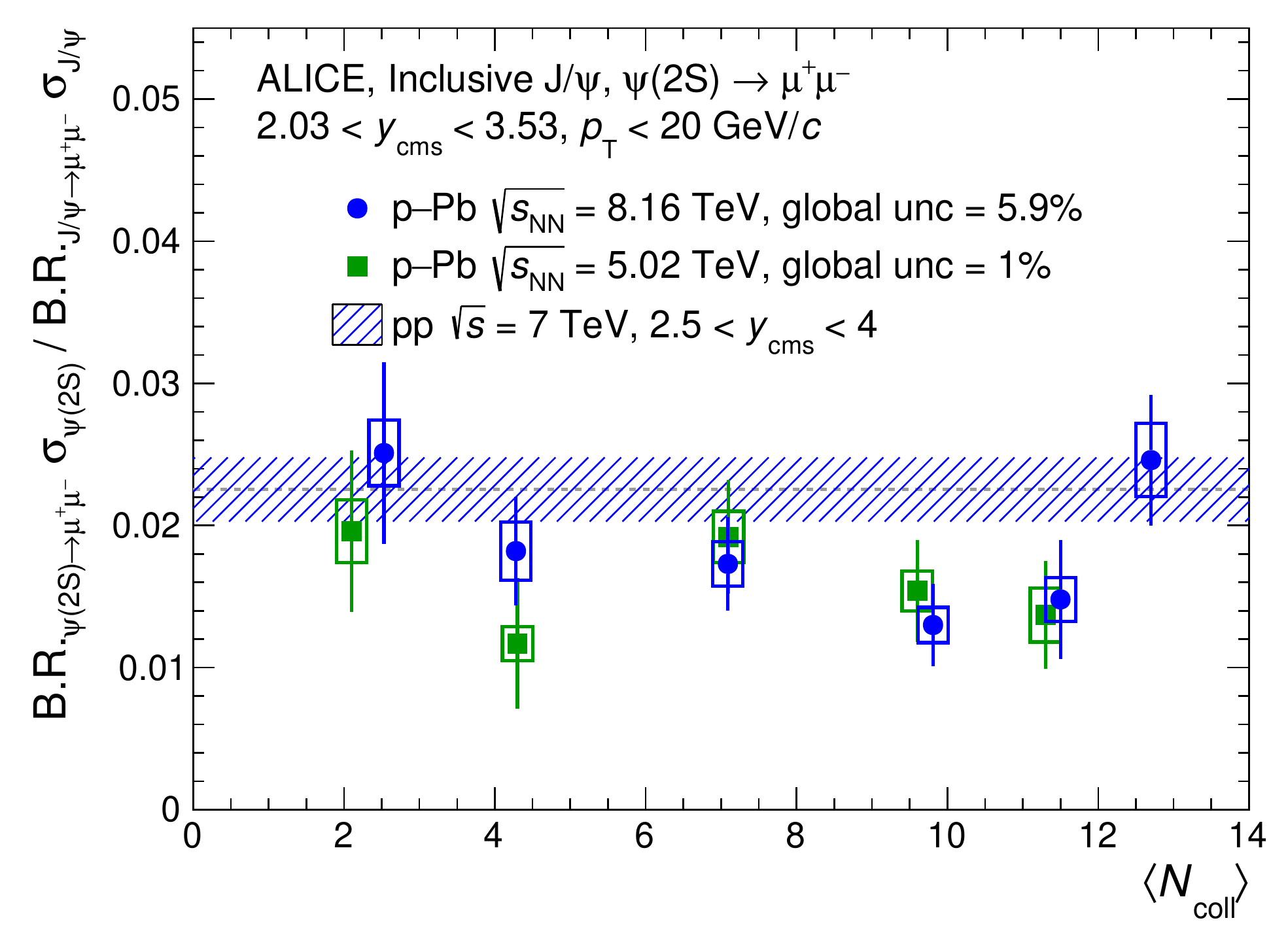}
\caption{\label{PsiPtoJPsi}B.R.$_{\psi(\rm 2S)\rightarrow\mu^{+}\mu^{-}}{\sigma_{\psi(\rm 2S)}}$/B.R.$_{J/\psi\rightarrow\mu^{+}\mu^{-}}{\sigma_{\rm J/\psi}}$ as a function of $\langle N_{\rm coll} \rangle$ at backward (left) and forward (right) rapidity compared with the measurement in pp collisions at $\sqrt{s} = 7$~TeV~\cite{Abelev:2014qha} (line with the band representing the total uncertainty), and to the  results at $\sqrt{s_{\rm NN}} = 5.02$ TeV~\cite{Adam:2016ohd}. Vertical error bars represent the statistical uncertainties, while the open boxes correspond to the systematic uncertainties.}
\end{figure}

\sloppy The relative production of the excited $\psi$(2S) state compared to that of the J/$\psi$ state can be quantified by the $\psi$(2S)/J/$\psi$ ratio, which is defined here as B.R.$_{\psi(\rm 2S)\rightarrow\mu^{+}\mu^{-}}{\sigma_{\psi(\rm 2S)}}$/B.R.$_{J/\psi\rightarrow\mu^{+}\mu^{-}}{\sigma_{\rm J/\psi}}$. 
The relative modification of the production of the two states in p--Pb collisions with respect to pp collisions is then obtained by comparing the $\psi$(2S)/J/$\psi$ ratio in the two collision systems.
Several systematic uncertainties cancel in the $\psi$(2S)/J/$\psi$ ratio, and the remaining ones are due to the signal extraction and the MC input shapes. 
The centrality dependence of the $\psi$(2S)/J/$\psi$ ratio at backward and forward rapidity in p--Pb collisions at $\sqrt{s_{\rm NN}} = 8.16$ TeV are shown in Fig.~\ref{PsiPtoJPsi}. 
The results are compared with the same ratio in p--Pb collisions at $\sqrt{s_{\rm NN}} = 5.02$ TeV~\cite{Adam:2016ohd} as well as in pp collisions at $\sqrt{s} = 7$ TeV~\cite{Abelev:2014qha}. 
Firstly, the $\psi$(2S)/J/$\psi$ ratio does not exhibit any significant dependence on the collision energy.
Secondly, the ratio appears to be smaller in p--Pb than in pp collisions, in both explored rapidity regions and for all centralities, except the most peripheral, where the uncertainty is considerably large, and the most central ones.
Here, it is important to note that also no significant energy dependence is observed in the $\psi$(2S)/J/$\psi$ ratio in pp collisions~\cite{Acharya:2017hjh}.
Thus, the production of the $\psi$(2S) in p--Pb collisions appears to be suppressed compared to that of the J/$\psi$ with respect to the expectation from pp collisions. 
Thirdly, given the current experimental uncertainties, no clear trend of the ratio as a function of centrality can be drawn.
Finally, the suppression of the $\psi$(2S) relative to the J/$\psi$ in p--Pb compared to pp collisions appears to be stronger in the Pb-going (backward rapidity) than in the p-going direction (forward rapidity).

\begin{figure}[tb]
\includegraphics[width=0.5\columnwidth]{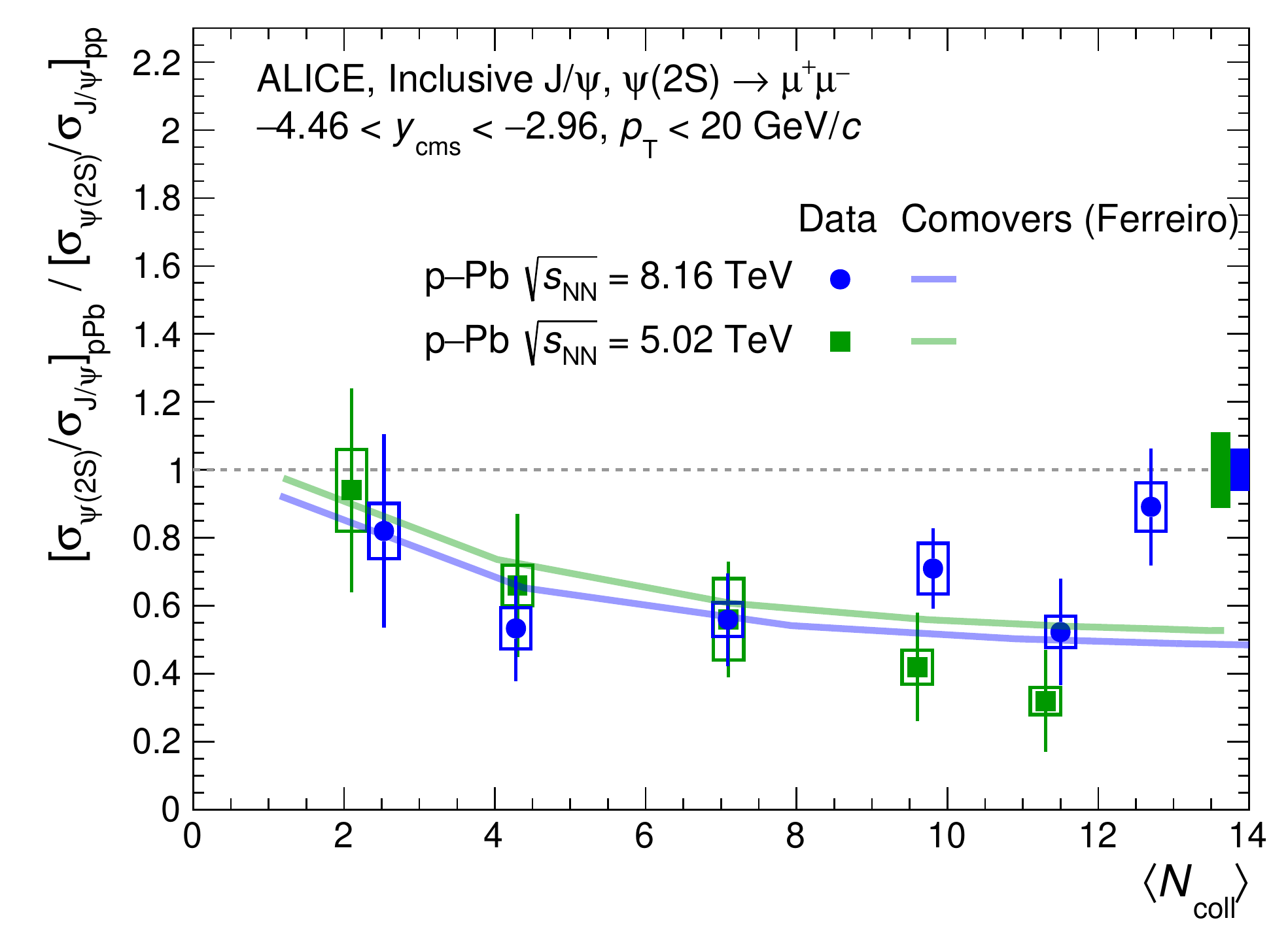}
\includegraphics[width=0.5\columnwidth]{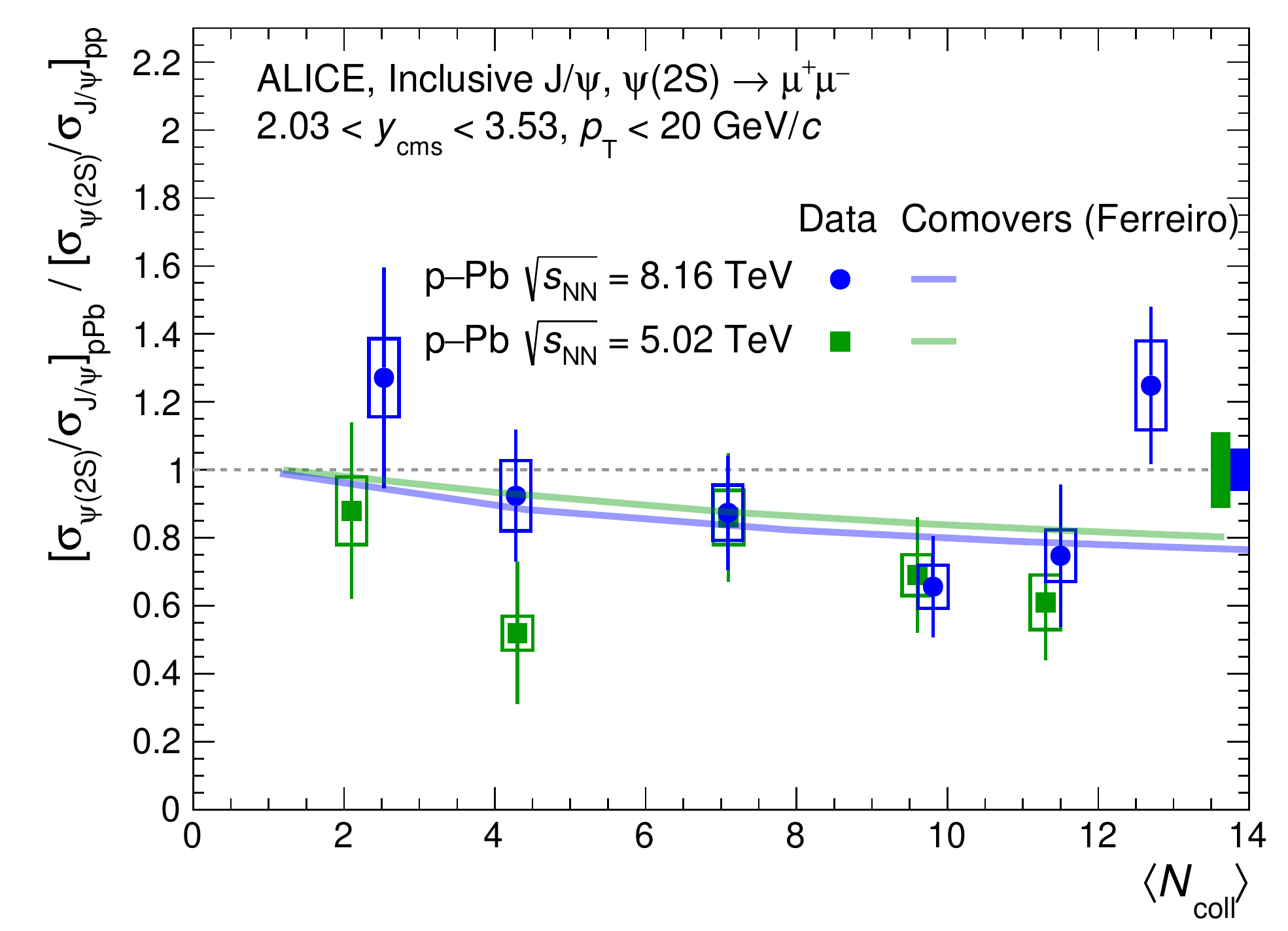}
\caption{\label{DPsiPtoJPsi} Double ratio $[\sigma_{\psi(\rm 2S)}/\sigma_{\rm J/\psi}]_{\rm pPb}/[\sigma_{\psi(\rm 2S)}/\sigma_{\rm J/\psi}]_{\rm pp}$ as a function of $\langle N_{\rm coll} \rangle$ at backward (left) and forward (right) rapidity compared with  the  one at $\sqrt{s_{\rm NN}} = 5.02$ TeV~\cite{Adam:2015jsa}. The vertical error bars represent the statistical uncertainties and the open boxes around the data points the uncorrelated systematic uncertainties. The boxes around unity represent the correlated systematic uncertainty and correspond to the uncertainty on the ratio $\psi(2S)/\jpsi$ in pp collisions. Experimental points are compared with the theoretical predictions of the comovers model at $\sqrt{s_{\rm NN}} = 5.02$ TeV (green line~\cite{Ferreiro:2014bia}) and $\sqrt{s_{\rm NN}} = 8.16$ TeV (blue line~\cite{Albacete:2016veq,Albacete:2017qng}).}
\end{figure}

\sloppy The same conclusions can be also drawn from the so-called double ratio, i.e.\ the ratio of the $\psi$(2S) to the J/$\psi$ cross section in p--Pb collisions divided by the same ratio in pp collisions, $[\sigma_{\psi(\rm 2S)}/\sigma_{\rm J/\psi}]_{\rm pPb}/[\sigma_{\psi(\rm 2S)}/\sigma_{\rm J/\psi}]_{\rm pp}$.
Figure~\ref{DPsiPtoJPsi} shows the double ratio $[\sigma_{\psi(\rm 2S)}/\sigma_{\rm J/\psi}]_{\rm pPb}/[\sigma_{\psi(\rm 2S)}/\sigma_{\rm J/\psi}]_{\rm pp}$ as a function of centrality in the backward and forward rapidity regions for p--Pb collisions at $\sqrt{s_{\rm NN}} = 8.16$ TeV. 
For the cross section ratio in pp collisions, the energy and rapidity interpolated value discussed in Section ~\ref{DataAnalysis} is used.
The double ratio is also compared with the one measured in p--Pb collisions at $\sqrt{s_{\rm NN}} = 5.02$ TeV~\cite{Adam:2016ohd}.
Calculations from the Comovers + EPS09LO model~\cite{Ferreiro:2014bia} are also shown in Fig.~\ref{DPsiPtoJPsi} for comparison. 
In the Comovers + EPS09LO model, resonances may be dissociated via interactions with ``comoving particles" (their nature, partonic or hadronic, not being defined in the model) produced in the same rapidity region. 
The dissociation is governed by the comover interaction cross sections, $\sigma^{{\rm co-J/}\psi} = 0.65$ mb and  $\sigma^{{\rm co-}\psi{\rm(2S)}} = 6$ mb, which are fixed from fits to low-energy experimental data. 
The main source of uncertainty in this model is the nPDF parameterisation, which is strongly correlated between the J/$\psi$ and the $\psi$(2S) and thus cancels out when calculating the cross section ratio.  
Overall, the agreement between the model calculations and the measurements is good at both collision energies. 
The decrease of the double ratio with increasing collision energy in the model is due to the increase of the comover density.
The measurement uncertainties do not allow for the experimental confirmation of such decrease of the double ratio.

\subsection{Centrality dependence of the inclusive $\psi$(2S) nuclear modification factor}
\begin{figure}[tb]
\includegraphics[width=0.5\columnwidth]{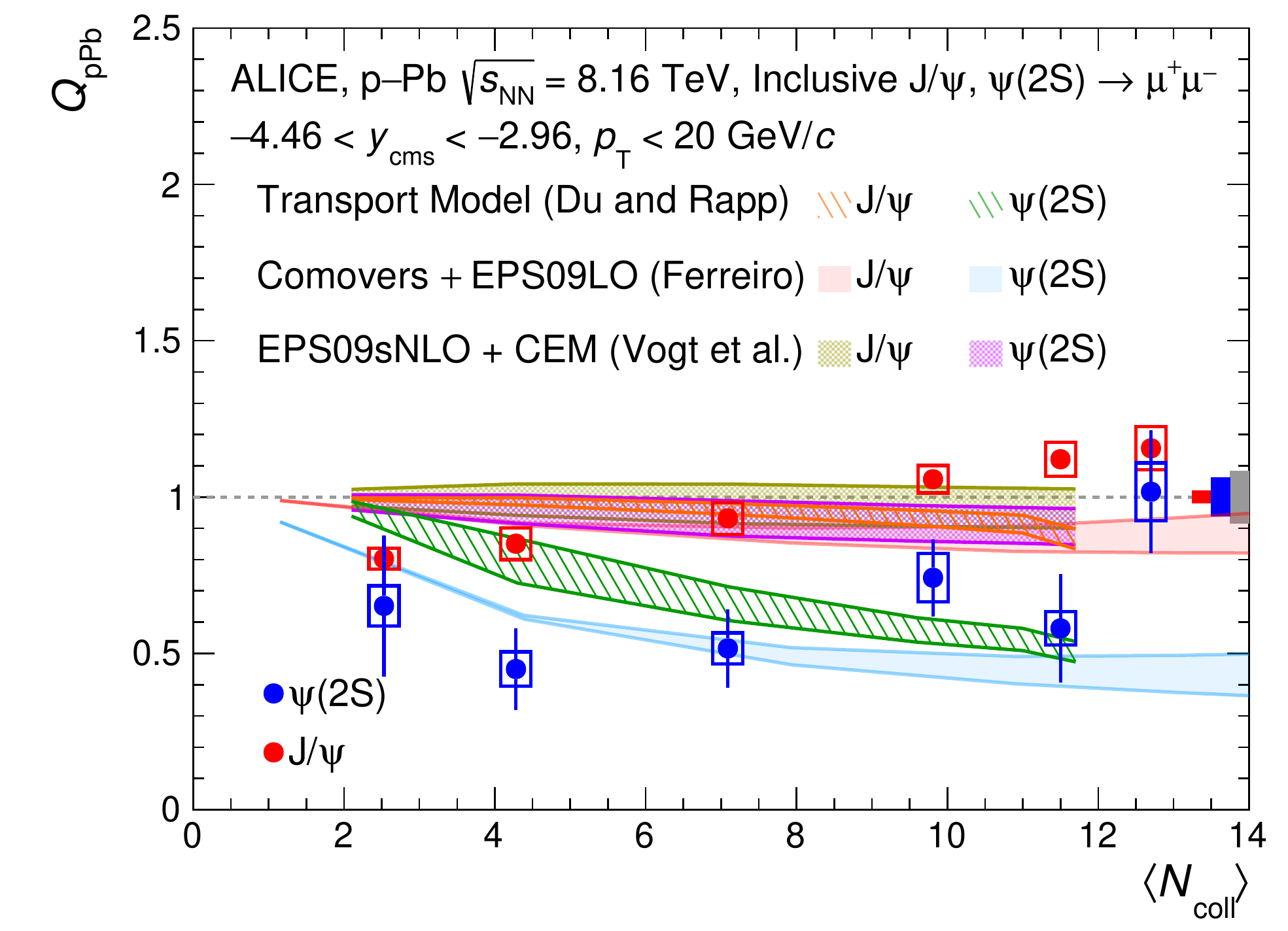}
\includegraphics[width=0.5\columnwidth]{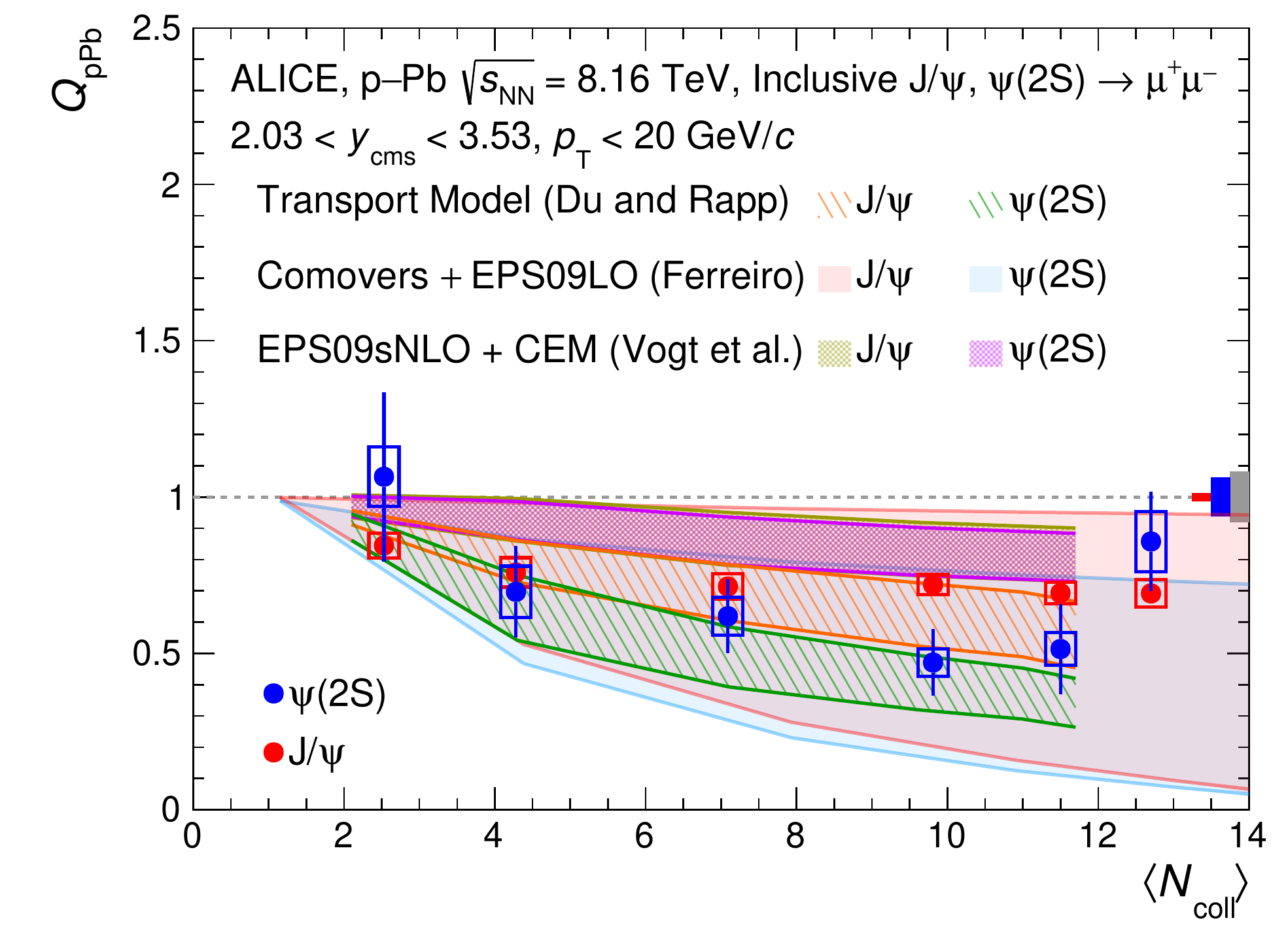}
\caption{\label{QpPbPsiP}Inclusive $\psi$(2S) $Q_{\rm pPb}$ as a function of $\langle N_{\rm coll} \rangle$ at backward (left) and forward (right) rapidity compared to J/$\psi$ $Q_{\rm pPb}$ and with the theoretical models. Vertical error bars represent the statistical uncertainties, while the open boxes around the data points correspond to the uncorrelated systematic uncertainties. The red and blue boxes around unity represent the correlated systematic uncertainty specific to the J/$\psi$ and $\psi$(2S), respectively. The grey box corresponds to the common systematic uncertainty correlated over $\langle N_{\rm coll}\rangle$.}
\end{figure}

The nuclear modification factor of the $\psi$(2S) is calculated using Eq.~\ref{QpPb}. 
Figure~\ref{QpPbPsiP} shows the inclusive $\psi$(2S) $Q_{\rm pPb}$ as a function of $\langle N_{\rm coll} \rangle$, for the backward and forward rapidity intervals, compared with the inclusive J/$\psi$ $Q_{\rm pPb}$. 
At forward rapidity, the suppression and its centrality dependence are similar for the $\psi$(2S) and the J/$\psi$.
At backward rapidity, on the contrary, a systematically stronger suppression of the $\psi$(2S) relative to the J/$\psi$ is observed, except for the most peripheral and most central collisions, where the large uncertainties prevent a firm conclusion.
The $\psi$(2S) $Q_{\rm pPb}$ at $\sqrt{s_{\rm NN}} = 8.16$ TeV shows the same dependence with the centrality of the collision than at $\sqrt{s_{\rm NN}} = 5.02$ TeV~\cite{Adam:2015jsa}.

Also shown in Fig.~\ref{QpPbPsiP} are the results of model calculations.
The EPS09s NLO + CEM calculations~\cite{McGlinchey:2012bp} of $Q_{\rm pPb}$ are very similar for both $\psi$(2S) and J/$\psi$. 
The model fails to describe $\psi$(2S) results at forward rapidity, while the J/$\psi$ results lie in the lower edge of the model calculation.
At backward rapidity, the model calculation is close to the J/$\psi$ data, although exhibiting different centrality trends, but fails at explaining the stronger $\psi$(2S) suppression.
The transport model~\cite{Du:2015wha} calculations yield significantly smaller $Q_{\rm pPb}$ for the $\psi$(2S) than for the J/$\psi$, with the difference being more pronounced in the Pb-going direction, where this difference increases with increasing centrality. 
The description of the forward rapidity results is fair for both charmonium states.
At backward rapidity, the model tends to overestimate the $\psi$(2S) measurement in the most peripheral centrality classes. 
In this model, the lower $Q_{\rm pPb}$ for the $\psi$(2S) than for the J/$\psi$ is caused by a larger suppression of the $\psi$(2S) in the short QGP and the hadron resonance gas phases. 
Finally, the Comovers + EPS09LO model~\cite{Ferreiro:2014bia} predicts a significantly lower $Q_{\rm pPb}$ for the $\psi$(2S) than for the J/$\psi$ in the backward rapidity region. 
In the forward rapidity region the model uncertainties are too large to draw any firm conclusion. 
It is worth noting that the model uncertainties are largely correlated between the J/$\psi$ and $\psi$(2S), as they are dominantly due to the nPDF parameterisation, and thus mostly cancel when calculating the double ratio as shown in Fig.~\ref{DPsiPtoJPsi}.
Nuclear shadowing is included using the EPS09 LO parameterisation~\cite{Eskola:2009uj} and the uncertainties of this parameterisation dominate the uncertainties of the model.
The effect of the comovers, responsible for the stronger suppression of the $\psi$(2S) compared to the J/$\psi$, is stronger at backward rapidity due to the larger density of comovers in the Pb-going direction~\cite{Ferreiro:2014bia}.
This model provides a fair description of $\psi$(2S) $Q_{\rm pPb}$ at backward rapidity.
However, the trend with centrality exhibited for the J/$\psi$ does not reproduce the one observed in the data.
Although not shown in the figure, the energy loss model~\cite{Arleo:2014oha} predicts sensibly the same $Q_{\rm pPb}$ for the two reported charmonium states. Only models including final-state interactions are able to describe, at least qualitatively, a stronger suppression of the less bound $\psi(2S)$ state than of the more tightly bound J/$\psi$ state.

As for the $\jpsi$, it is also possible to estimate the $Q_{\rm pPb}^{\rm prompt}$ of $\psip$. In this case, the value of $f_{B}$ is about 0.18 and it is calculated using the LHCb measurements in pp collisions at $\sqrt{s} = 7$ TeV for $\pt < 16$ GeV/$\textit{c}$ and $2 < y < 4.5$ \cite{Aaij:2012ag}. Since the non-prompt $\psip$ $Q_{\rm pPb}$ has not been measured yet as a function of centrality, it is conservatively assumed to vary between 0.4 and 1 in all centrality classes for both forward and backward rapidity. That variation range for non-prompt $\psip$ $Q_{\rm pPb}$ englobes the centrality-integrated non-prompt $\psip$  $R_{\rm pPb}$ measured by LHCb at backward and forward rapidity in p--Pb collisions at $\sqrt{s_{\rm NN}} = 5.02$~TeV~\cite{Aaij:2016eyl} as well as all the inclusive $\psip$ $Q_{\rm pPb}$ reported here. The $Q_{\rm pPb}^{\rm prompt}$ calculated under these assumptions is compatible within uncertainties with the inclusive one, showing a maximum difference of 25\% with respect to the latter.

\section{Summary}
\label{Conclusions}

The study of the centrality dependence of the J/$\psi$ and $\psi$(2S) production in p--Pb collisions at $\sqrt{s_{\rm NN}} = 8.16$ TeV using the energy deposited in the neutron ZDC located in the Pb-going direction as the centrality estimator is presented. 
The J/$\psi$ $\langle p_{\rm T}\rangle$ and $\langle p^{2}_{\rm T}\rangle$ are reported for different centrality classes in the forward and backward rapidity regions covered by the ALICE muon spectrometer. The $\Delta \langle p^{2}_{\rm T}\rangle$ measurement shows a $p_{\rm T}$ broadening, relative to pp collisions, that increases from peripheral to central collisions, with larger values at forward than at backward $y$, except for the most peripheral events where similar values are seen in both rapidity intervals. 

At forward rapidity, a clear suppression of J/$\psi$ in p--Pb collisions compared to pp collisions is observed, which increases from peripheral to central collisions.
At backward rapidity, the trend is opposite: the production of J/$\psi$ relative to expectations from pp collisions is suppressed in peripheral collisions but enhanced in central collisions. 
The $p_{\rm T}$- and centrality-differential measurements of the J/$\psi$ $Q_{\rm pPb}$ indicate a stronger suppression in central than in peripheral collisions at low $p_{\rm T}$ and forward rapidity, but with $Q_{\rm pPb}$ approaching unity at high $p_{\rm T}$ for all centrality classes. 
At backward rapidity, an enhancement is observed in central compared to peripheral collisions for $p_{\rm T} > 3$ GeV/$c$. 

The ratio B.R.$_{\psi(\rm 2S)\rightarrow\mu^{+}\mu^{-}}{\sigma_{\psi(\rm 2S)}}$/B.R.$_{J/\psi\rightarrow\mu^{+}\mu^{-}}{\sigma_{\rm J/\psi}}$ is compatible with the pp measurement in the most central and most peripheral collisions (within large uncertainties), whereas a decrease is observed in the semi-central and semi-peripheral events.
Thus, in those centrality classes, the $\psi$(2S) production relative to the J/$\psi$ is suppressed in p--Pb collisions compared to pp collisions. 
The nuclear modification factor of the $\psi$(2S) is compatible, within large uncertainties, with the one of the J/$\psi$ in the most central and most peripheral events, but a stronger suppression of the $\psi$(2S) is observed in semi-central and semi-peripheral events, especially at backward rapidity. 

The results presented here at $\sqrt{s_{\rm NN}} = 8.16$ TeV confirm with improved statistical precision the earlier observations at $\sqrt{s_{\rm NN}} = 5.02$ TeV and extend the $p_{\rm T}$ reach up to 16 GeV/$c$ for the J/$\psi$ analysis. 
No significant dependence with collision energy is observed.

Theoretical models employing nPDF or energy loss mechanisms describe the centrality dependence of the J/$\psi$ nuclear modification factor at forward rapidity but do not reproduce the shape at backward rapidity. 
The $p_{\rm T}$ dependence of the J/$\psi$ $Q_{\rm pPb}$ in central collisions is not well described by the nPDF or energy loss based models, while the agreement is fair in peripheral collisions.

Among the three models considered, the one based only on nPDF cannot reproduce the $\psi$(2S) suppression. 
The model including final-state comover interactions describes the stronger $\psi$(2S) suppression at backward and forward rapidity, although the large model uncertainty prevents a firm conclusion at forward rapidity. 
The transport model is in good agreement at forward rapidity, but overestimates the $\psi$(2S) results at backward rapidity, especially in peripheral collisions.  

The results presented here stress the need for a sound theoretical understanding of the production of quarkonia, including the excited states, in proton--nucleus collisions. 
Further experimental results expected from the future Run 3 and Run 4 of the LHC will push further our understanding of nuclear effects.


\newenvironment{acknowledgement}{\relax}{\relax}
\begin{acknowledgement}
\section*{Acknowledgements}

The ALICE Collaboration would like to thank all its engineers and technicians for their invaluable contributions to the construction of the experiment and the CERN accelerator teams for the outstanding performance of the LHC complex.
The ALICE Collaboration gratefully acknowledges the resources and support provided by all Grid centres and the Worldwide LHC Computing Grid (WLCG) collaboration.
The ALICE Collaboration acknowledges the following funding agencies for their support in building and running the ALICE detector:
A. I. Alikhanyan National Science Laboratory (Yerevan Physics Institute) Foundation (ANSL), State Committee of Science and World Federation of Scientists (WFS), Armenia;
Austrian Academy of Sciences, Austrian Science Fund (FWF): [M 2467-N36] and Nationalstiftung f\"{u}r Forschung, Technologie und Entwicklung, Austria;
Ministry of Communications and High Technologies, National Nuclear Research Center, Azerbaijan;
Conselho Nacional de Desenvolvimento Cient\'{\i}fico e Tecnol\'{o}gico (CNPq), Financiadora de Estudos e Projetos (Finep), Funda\c{c}\~{a}o de Amparo \`{a} Pesquisa do Estado de S\~{a}o Paulo (FAPESP) and Universidade Federal do Rio Grande do Sul (UFRGS), Brazil;
Ministry of Education of China (MOEC) , Ministry of Science \& Technology of China (MSTC) and National Natural Science Foundation of China (NSFC), China;
Ministry of Science and Education and Croatian Science Foundation, Croatia;
Centro de Aplicaciones Tecnol\'{o}gicas y Desarrollo Nuclear (CEADEN), Cubaenerg\'{\i}a, Cuba;
Ministry of Education, Youth and Sports of the Czech Republic, Czech Republic;
The Danish Council for Independent Research | Natural Sciences, the VILLUM FONDEN and Danish National Research Foundation (DNRF), Denmark;
Helsinki Institute of Physics (HIP), Finland;
Commissariat \`{a} l'Energie Atomique (CEA) and Institut National de Physique Nucl\'{e}aire et de Physique des Particules (IN2P3) and Centre National de la Recherche Scientifique (CNRS), France;
Bundesministerium f\"{u}r Bildung und Forschung (BMBF) and GSI Helmholtzzentrum f\"{u}r Schwerionenforschung GmbH, Germany;
General Secretariat for Research and Technology, Ministry of Education, Research and Religions, Greece;
National Research, Development and Innovation Office, Hungary;
Department of Atomic Energy Government of India (DAE), Department of Science and Technology, Government of India (DST), University Grants Commission, Government of India (UGC) and Council of Scientific and Industrial Research (CSIR), India;
Indonesian Institute of Science, Indonesia;
Centro Fermi - Museo Storico della Fisica e Centro Studi e Ricerche Enrico Fermi and Istituto Nazionale di Fisica Nucleare (INFN), Italy;
Institute for Innovative Science and Technology , Nagasaki Institute of Applied Science (IIST), Japanese Ministry of Education, Culture, Sports, Science and Technology (MEXT) and Japan Society for the Promotion of Science (JSPS) KAKENHI, Japan;
Consejo Nacional de Ciencia (CONACYT) y Tecnolog\'{i}a, through Fondo de Cooperaci\'{o}n Internacional en Ciencia y Tecnolog\'{i}a (FONCICYT) and Direcci\'{o}n General de Asuntos del Personal Academico (DGAPA), Mexico;
Nederlandse Organisatie voor Wetenschappelijk Onderzoek (NWO), Netherlands;
The Research Council of Norway, Norway;
Commission on Science and Technology for Sustainable Development in the South (COMSATS), Pakistan;
Pontificia Universidad Cat\'{o}lica del Per\'{u}, Peru;
Ministry of Science and Higher Education, National Science Centre and WUT ID-UB, Poland;
Korea Institute of Science and Technology Information and National Research Foundation of Korea (NRF), Republic of Korea;
Ministry of Education and Scientific Research, Institute of Atomic Physics and Ministry of Research and Innovation and Institute of Atomic Physics, Romania;
Joint Institute for Nuclear Research (JINR), Ministry of Education and Science of the Russian Federation, National Research Centre Kurchatov Institute, Russian Science Foundation and Russian Foundation for Basic Research, Russia;
Ministry of Education, Science, Research and Sport of the Slovak Republic, Slovakia;
National Research Foundation of South Africa, South Africa;
Swedish Research Council (VR) and Knut \& Alice Wallenberg Foundation (KAW), Sweden;
European Organization for Nuclear Research, Switzerland;
Suranaree University of Technology (SUT), National Science and Technology Development Agency (NSDTA) and Office of the Higher Education Commission under NRU project of Thailand, Thailand;
Turkish Atomic Energy Agency (TAEK), Turkey;
National Academy of  Sciences of Ukraine, Ukraine;
Science and Technology Facilities Council (STFC), United Kingdom;
National Science Foundation of the United States of America (NSF) and United States Department of Energy, Office of Nuclear Physics (DOE NP), United States of America.
\end{acknowledgement}

\bibliographystyle{utphys}   
\bibliography{Jpsi_Psi2S_pPb8Centrality}

\newpage
\appendix

%
%

\section{The ALICE Collaboration}
\label{app:collab}

\begingroup
\small
\begin{flushleft}
S.~Acharya\Irefn{org141}\And 
D.~Adamov\'{a}\Irefn{org95}\And 
A.~Adler\Irefn{org74}\And 
J.~Adolfsson\Irefn{org81}\And 
M.M.~Aggarwal\Irefn{org100}\And 
S.~Agha\Irefn{org14}\And 
G.~Aglieri Rinella\Irefn{org34}\And 
M.~Agnello\Irefn{org30}\And 
N.~Agrawal\Irefn{org10}\textsuperscript{,}\Irefn{org54}\And 
Z.~Ahammed\Irefn{org141}
% ...
\And
\renewcommand\labelenumi{\textsuperscript{\theenumi}~}

\section*{Affiliation notes}
\renewcommand\theenumi{\roman{enumi}}
\begin{Authlist}
\item \Adef{org*}Deceased
\item \Adef{orgI}Italian National Agency for New Technologies, Energy and Sustainable Economic Development (ENEA), Bologna, Italy
\item \Adef{orgII}Dipartimento DET del Politecnico di Torino, Turin, Italy
\item \Adef{orgIII}M.V. Lomonosov Moscow State University, D.V. Skobeltsyn Institute of Nuclear, Physics, Moscow, Russia
\item \Adef{orgIV}Department of Applied Physics, Aligarh Muslim University, Aligarh, India
\item \Adef{orgV}Institute of Theoretical Physics, University of Wroclaw, Poland
\end{Authlist}

\section*{Collaboration Institutes}
\renewcommand\theenumi{\arabic{enumi}~}
\begin{Authlist}
\item \Idef{org1}A.I. Alikhanyan National Science Laboratory (Yerevan Physics Institute) Foundation, Yerevan, Armenia
\item \Idef{org2}Bogolyubov Institute for Theoretical Physics, National Academy of Sciences of Ukraine, Kiev, Ukraine
\item \Idef{org3}Bose Institute, Department of Physics  and Centre for Astroparticle Physics and Space Science (CAPSS), Kolkata, India
\item \Idef{org4}Budker Institute for Nuclear Physics, Novosibirsk, Russia
\item \Idef{org5}California Polytechnic State University, San Luis Obispo, California, United States
\item \Idef{org6}Central China Normal University, Wuhan, China
\item \Idef{org7}Centre de Calcul de l'IN2P3, Villeurbanne, Lyon, France
\item \Idef{org8}Centro de Aplicaciones Tecnol\'{o}gicas y Desarrollo Nuclear (CEADEN), Havana, Cuba
\item \Idef{org9}Centro de Investigaci\'{o}n y de Estudios Avanzados (CINVESTAV), Mexico City and M\'{e}rida, Mexico
\item \Idef{org10}Centro Fermi - Museo Storico della Fisica e Centro Studi e Ricerche ``Enrico Fermi', Rome, Italy
\item \Idef{org11}Chicago State University, Chicago, Illinois, United States
\item \Idef{org12}China Institute of Atomic Energy, Beijing, China
\item \Idef{org13}Comenius University Bratislava, Faculty of Mathematics, Physics and Informatics, Bratislava, Slovakia
\item \Idef{org14}COMSATS University Islamabad, Islamabad, Pakistan
\item \Idef{org15}Creighton University, Omaha, Nebraska, United States
\item \Idef{org16}Department of Physics, Aligarh Muslim University, Aligarh, India
\item \Idef{org17}Department of Physics, Pusan National University, Pusan, Republic of Korea
\item \Idef{org18}Department of Physics, Sejong University, Seoul, Republic of Korea
\item \Idef{org19}Department of Physics, University of California, Berkeley, California, United States
\item \Idef{org20}Department of Physics, University of Oslo, Oslo, Norway
\item \Idef{org21}Department of Physics and Technology, University of Bergen, Bergen, Norway
\item \Idef{org22}Dipartimento di Fisica dell'Universit\`{a} 'La Sapienza' and Sezione INFN, Rome, Italy
\item \Idef{org23}Dipartimento di Fisica dell'Universit\`{a} and Sezione INFN, Cagliari, Italy
\item \Idef{org24}Dipartimento di Fisica dell'Universit\`{a} and Sezione INFN, Trieste, Italy
\item \Idef{org25}Dipartimento di Fisica dell'Universit\`{a} and Sezione INFN, Turin, Italy
\item \Idef{org26}Dipartimento di Fisica e Astronomia dell'Universit\`{a} and Sezione INFN, Bologna, Italy
\item \Idef{org27}Dipartimento di Fisica e Astronomia dell'Universit\`{a} and Sezione INFN, Catania, Italy
\item \Idef{org28}Dipartimento di Fisica e Astronomia dell'Universit\`{a} and Sezione INFN, Padova, Italy
\item \Idef{org29}Dipartimento di Fisica `E.R.~Caianiello' dell'Universit\`{a} and Gruppo Collegato INFN, Salerno, Italy
\item \Idef{org30}Dipartimento DISAT del Politecnico and Sezione INFN, Turin, Italy
\item \Idef{org31}Dipartimento di Scienze e Innovazione Tecnologica dell'Universit\`{a} del Piemonte Orientale and INFN Sezione di Torino, Alessandria, Italy
\item \Idef{org32}Dipartimento di Scienze MIFT, Universit\`{a} di Messina, Messina, Italy
\item \Idef{org33}Dipartimento Interateneo di Fisica `M.~Merlin' and Sezione INFN, Bari, Italy
\item \Idef{org34}European Organization for Nuclear Research (CERN), Geneva, Switzerland
\item \Idef{org35}Faculty of Electrical Engineering, Mechanical Engineering and Naval Architecture, University of Split, Split, Croatia
\item \Idef{org36}Faculty of Engineering and Science, Western Norway University of Applied Sciences, Bergen, Norway
\item \Idef{org37}Faculty of Nuclear Sciences and Physical Engineering, Czech Technical University in Prague, Prague, Czech Republic
\item \Idef{org38}Faculty of Science, P.J.~\v{S}af\'{a}rik University, Ko\v{s}ice, Slovakia
\item \Idef{org39}Frankfurt Institute for Advanced Studies, Johann Wolfgang Goethe-Universit\"{a}t Frankfurt, Frankfurt, Germany
\item \Idef{org40}Fudan University, Shanghai, China
\item \Idef{org41}Gangneung-Wonju National University, Gangneung, Republic of Korea
\item \Idef{org42}Gauhati University, Department of Physics, Guwahati, India
\item \Idef{org43}Helmholtz-Institut f\"{u}r Strahlen- und Kernphysik, Rheinische Friedrich-Wilhelms-Universit\"{a}t Bonn, Bonn, Germany
\item \Idef{org44}Helsinki Institute of Physics (HIP), Helsinki, Finland
\item \Idef{org45}High Energy Physics Group,  Universidad Aut\'{o}noma de Puebla, Puebla, Mexico
\item \Idef{org46}Hiroshima University, Hiroshima, Japan
\item \Idef{org47}Hochschule Worms, Zentrum  f\"{u}r Technologietransfer und Telekommunikation (ZTT), Worms, Germany
\item \Idef{org48}Horia Hulubei National Institute of Physics and Nuclear Engineering, Bucharest, Romania
\item \Idef{org49}Indian Institute of Technology Bombay (IIT), Mumbai, India
\item \Idef{org50}Indian Institute of Technology Indore, Indore, India
\item \Idef{org51}Indonesian Institute of Sciences, Jakarta, Indonesia
\item \Idef{org52}INFN, Laboratori Nazionali di Frascati, Frascati, Italy
\item \Idef{org53}INFN, Sezione di Bari, Bari, Italy
\item \Idef{org54}INFN, Sezione di Bologna, Bologna, Italy
\item \Idef{org55}INFN, Sezione di Cagliari, Cagliari, Italy
\item \Idef{org56}INFN, Sezione di Catania, Catania, Italy
\item \Idef{org57}INFN, Sezione di Padova, Padova, Italy
\item \Idef{org58}INFN, Sezione di Roma, Rome, Italy
\item \Idef{org59}INFN, Sezione di Torino, Turin, Italy
\item \Idef{org60}INFN, Sezione di Trieste, Trieste, Italy
\item \Idef{org61}Inha University, Incheon, Republic of Korea
\item \Idef{org62}Institute for Nuclear Research, Academy of Sciences, Moscow, Russia
\item \Idef{org63}Institute for Subatomic Physics, Utrecht University/Nikhef, Utrecht, Netherlands
\item \Idef{org64}Institute of Experimental Physics, Slovak Academy of Sciences, Ko\v{s}ice, Slovakia
\item \Idef{org65}Institute of Physics, Homi Bhabha National Institute, Bhubaneswar, India
\item \Idef{org66}Institute of Physics of the Czech Academy of Sciences, Prague, Czech Republic
\item \Idef{org67}Institute of Space Science (ISS), Bucharest, Romania
\item \Idef{org68}Institut f\"{u}r Kernphysik, Johann Wolfgang Goethe-Universit\"{a}t Frankfurt, Frankfurt, Germany
\item \Idef{org69}Instituto de Ciencias Nucleares, Universidad Nacional Aut\'{o}noma de M\'{e}xico, Mexico City, Mexico
\item \Idef{org70}Instituto de F\'{i}sica, Universidade Federal do Rio Grande do Sul (UFRGS), Porto Alegre, Brazil
\item \Idef{org71}Instituto de F\'{\i}sica, Universidad Nacional Aut\'{o}noma de M\'{e}xico, Mexico City, Mexico
\item \Idef{org72}iThemba LABS, National Research Foundation, Somerset West, South Africa
\item \Idef{org73}Jeonbuk National University, Jeonju, Republic of Korea
\item \Idef{org74}Johann-Wolfgang-Goethe Universit\"{a}t Frankfurt Institut f\"{u}r Informatik, Fachbereich Informatik und Mathematik, Frankfurt, Germany
\item \Idef{org75}Joint Institute for Nuclear Research (JINR), Dubna, Russia
\item \Idef{org76}Korea Institute of Science and Technology Information, Daejeon, Republic of Korea
\item \Idef{org77}KTO Karatay University, Konya, Turkey
\item \Idef{org78}Laboratoire de Physique des 2 Infinis, Ir\`{e}ne Joliot-Curie, Orsay, France
\item \Idef{org79}Laboratoire de Physique Subatomique et de Cosmologie, Universit\'{e} Grenoble-Alpes, CNRS-IN2P3, Grenoble, France
\item \Idef{org80}Lawrence Berkeley National Laboratory, Berkeley, California, United States
\item \Idef{org81}Lund University Department of Physics, Division of Particle Physics, Lund, Sweden
\item \Idef{org82}Nagasaki Institute of Applied Science, Nagasaki, Japan
\item \Idef{org83}Nara Women{'}s University (NWU), Nara, Japan
\item \Idef{org84}National and Kapodistrian University of Athens, School of Science, Department of Physics , Athens, Greece
\item \Idef{org85}National Centre for Nuclear Research, Warsaw, Poland
\item \Idef{org86}National Institute of Science Education and Research, Homi Bhabha National Institute, Jatni, India
\item \Idef{org87}National Nuclear Research Center, Baku, Azerbaijan
\item \Idef{org88}National Research Centre Kurchatov Institute, Moscow, Russia
\item \Idef{org89}Niels Bohr Institute, University of Copenhagen, Copenhagen, Denmark
\item \Idef{org90}Nikhef, National institute for subatomic physics, Amsterdam, Netherlands
\item \Idef{org91}NRC Kurchatov Institute IHEP, Protvino, Russia
\item \Idef{org92}NRC \guillemotleft Kurchatov\guillemotright  Institute - ITEP, Moscow, Russia
\item \Idef{org93}NRNU Moscow Engineering Physics Institute, Moscow, Russia
\item \Idef{org94}Nuclear Physics Group, STFC Daresbury Laboratory, Daresbury, United Kingdom
\item \Idef{org95}Nuclear Physics Institute of the Czech Academy of Sciences, \v{R}e\v{z} u Prahy, Czech Republic
\item \Idef{org96}Oak Ridge National Laboratory, Oak Ridge, Tennessee, United States
\item \Idef{org97}Ohio State University, Columbus, Ohio, United States
\item \Idef{org98}Petersburg Nuclear Physics Institute, Gatchina, Russia
\item \Idef{org99}Physics department, Faculty of science, University of Zagreb, Zagreb, Croatia
\item \Idef{org100}Physics Department, Panjab University, Chandigarh, India
\item \Idef{org101}Physics Department, University of Jammu, Jammu, India
\item \Idef{org102}Physics Department, University of Rajasthan, Jaipur, India
\item \Idef{org103}Physikalisches Institut, Eberhard-Karls-Universit\"{a}t T\"{u}bingen, T\"{u}bingen, Germany
\item \Idef{org104}Physikalisches Institut, Ruprecht-Karls-Universit\"{a}t Heidelberg, Heidelberg, Germany
\item \Idef{org105}Physik Department, Technische Universit\"{a}t M\"{u}nchen, Munich, Germany
\item \Idef{org106}Politecnico di Bari, Bari, Italy
\item \Idef{org107}Research Division and ExtreMe Matter Institute EMMI, GSI Helmholtzzentrum f\"ur Schwerionenforschung GmbH, Darmstadt, Germany
\item \Idef{org108}Rudjer Bo\v{s}kovi\'{c} Institute, Zagreb, Croatia
\item \Idef{org109}Russian Federal Nuclear Center (VNIIEF), Sarov, Russia
\item \Idef{org110}Saha Institute of Nuclear Physics, Homi Bhabha National Institute, Kolkata, India
\item \Idef{org111}School of Physics and Astronomy, University of Birmingham, Birmingham, United Kingdom
\item \Idef{org112}Secci\'{o}n F\'{\i}sica, Departamento de Ciencias, Pontificia Universidad Cat\'{o}lica del Per\'{u}, Lima, Peru
\item \Idef{org113}St. Petersburg State University, St. Petersburg, Russia
\item \Idef{org114}Stefan Meyer Institut f\"{u}r Subatomare Physik (SMI), Vienna, Austria
\item \Idef{org115}SUBATECH, IMT Atlantique, Universit\'{e} de Nantes, CNRS-IN2P3, Nantes, France
\item \Idef{org116}Suranaree University of Technology, Nakhon Ratchasima, Thailand
\item \Idef{org117}Technical University of Ko\v{s}ice, Ko\v{s}ice, Slovakia
\item \Idef{org118}The Henryk Niewodniczanski Institute of Nuclear Physics, Polish Academy of Sciences, Cracow, Poland
\item \Idef{org119}The University of Texas at Austin, Austin, Texas, United States
\item \Idef{org120}Universidad Aut\'{o}noma de Sinaloa, Culiac\'{a}n, Mexico
\item \Idef{org121}Universidade de S\~{a}o Paulo (USP), S\~{a}o Paulo, Brazil
\item \Idef{org122}Universidade Estadual de Campinas (UNICAMP), Campinas, Brazil
\item \Idef{org123}Universidade Federal do ABC, Santo Andre, Brazil
\item \Idef{org124}University of Cape Town, Cape Town, South Africa
\item \Idef{org125}University of Houston, Houston, Texas, United States
\item \Idef{org126}University of Jyv\"{a}skyl\"{a}, Jyv\"{a}skyl\"{a}, Finland
\item \Idef{org127}University of Liverpool, Liverpool, United Kingdom
\item \Idef{org128}University of Science and Technology of China, Hefei, China
\item \Idef{org129}University of South-Eastern Norway, Tonsberg, Norway
\item \Idef{org130}University of Tennessee, Knoxville, Tennessee, United States
\item \Idef{org131}University of the Witwatersrand, Johannesburg, South Africa
\item \Idef{org132}University of Tokyo, Tokyo, Japan
\item \Idef{org133}University of Tsukuba, Tsukuba, Japan
\item \Idef{org134}Universit\'{e} Clermont Auvergne, CNRS/IN2P3, LPC, Clermont-Ferrand, France
\item \Idef{org135}Universit\'{e} de Lyon, Universit\'{e} Lyon 1, CNRS/IN2P3, IPN-Lyon, Villeurbanne, Lyon, France
\item \Idef{org136}Universit\'{e} de Strasbourg, CNRS, IPHC UMR 7178, F-67000 Strasbourg, France, Strasbourg, France
\item \Idef{org137}Universit\'{e} Paris-Saclay Centre d'Etudes de Saclay (CEA), IRFU, D\'{e}partment de Physique Nucl\'{e}aire (DPhN), Saclay, France
\item \Idef{org138}Universit\`{a} degli Studi di Foggia, Foggia, Italy
\item \Idef{org139}Universit\`{a} degli Studi di Pavia, Pavia, Italy
\item \Idef{org140}Universit\`{a} di Brescia, Brescia, Italy
\item \Idef{org141}Variable Energy Cyclotron Centre, Homi Bhabha National Institute, Kolkata, India
\item \Idef{org142}Warsaw University of Technology, Warsaw, Poland
\item \Idef{org143}Wayne State University, Detroit, Michigan, United States
\item \Idef{org144}Westf\"{a}lische Wilhelms-Universit\"{a}t M\"{u}nster, Institut f\"{u}r Kernphysik, M\"{u}nster, Germany
\item \Idef{org145}Wigner Research Centre for Physics, Budapest, Hungary
\item \Idef{org146}Yale University, New Haven, Connecticut, United States
\item \Idef{org147}Yonsei University, Seoul, Republic of Korea
\end{Authlist}
\endgroup
  
\end{document}